\documentclass{article}
\usepackage{emulateapj,timesfonts}
\usepackage{float,epsfig,psfig}
\usepackage{pstricks}

\def\hea4{{\it HEAO~A4}}
\def\heaoa2{{\it HEAO~A2}}
\def\heao1{{\it HEAO~1}}

\def\amin{$^\prime$}

\def\eg{{\it e.g.}~}

\def\h0{$H_{\rm o}=50$~km~s$^{-1}$~Mpc$^{-1}$}
\def\q0{$q_{\rm o}$}

\def\msun     {M$_{\odot}$}

\def\etal    {{et~al.}~}

\def\cms3  {~{cm$^{-3}$}}

\begin{document}

\submitted{The Astrophysical Journal, 578, in press, 2002, October 10}
\title{{ASCA observations of groups at radii of low overdensity.}\\
{Implications for the cosmic preheating.}}
\author{A.~Finoguenov$^{1,2}$, C. Jones$^2$, H. B\"ohringer$^{1}$, T.J. Ponman$^{3}$}
\affil{\vspace*{0.5cm}{$^1$ Max-Planck-Institut f\"ur extraterrestrische
    Physik, Giessenbachstra\ss e, 85748 Garching, Germany}\\ 
{$^2$ Smithsonian Astrophysical Observatory, 60 Garden st., MS 2, Cambridge,
  MA 02138, USA}\\ 
{$^3$ School of Physics \& Astronomy, University of Birmingham, Edgbaston,
Birmingham B15 2TT, UK}}
\submitted{Received 2002 April 24; accepted 2002 June 19}

\begin{abstract}
Through a 3d-modeling of ASCA observations, we performed a spatially
resolved X-ray spectroscopic study, extending to radii exceeding 150 kpc,
for a sample of 9 groups of galaxies. Combined with published ROSAT results,
we conclude that these systems generally exhibit a strong temperature
decline at outer radii. In our best case, NGC3268, this corresponds to
a flattening of the entropy profile at a level of $\sim400$ keV cm$^2$. This
value is high compared both to the observed entropy floor of $\sim100$ keV
cm$^2$ and to the expected value from gravitational heating. We suggest that
the observed entropy profile in most groups at densities exceeding 500 times
the critical is purely driven by non-gravitational heating processes. After
comparison with a larger sample of groups and clusters, we conclude that
there is a variation in the level of non-gravitational heating between
$\sim100$ keV cm$^2$ and $\sim400$ keV cm$^2$ within every system. Using
models of cluster formation as a reference frame, we established that the
accreted gas reaches an entropy level of $400$ keV cm$^2$ by redshift
$2.0-2.5$, while such high entropies where not present at redshifts higher
than $2.8-3.5$, favoring nearly instantaneous preheating. Adopting galactic
winds as a source of preheating, and scaling the released energy by the
observed metal abundance, the variation in the preheating could be ascribed
mostly to variation in the typical overdensity of the energy injection,
$\sim30$ for an entropy floor ($100$ keV cm$^2$) and to $\sim5$ for an
entropy of $400$ keV cm$^2$.
\end{abstract}


\section{Introduction}


Comparative studies of the scaling relations in clusters of galaxies
reveal strong deviations of the observed relations from predictions
based on self-similar collapse (Evrard \& Henry 1991; Bower 1996; Loewenstein
2000; Finoguenov, Reiprich, B\"ohringer 2001). These deviations are
best characterized by an entropy floor in the X-ray emitting gas.
While negligible compared to the entropy due to the accretion shock in
large clusters, such preheating\footnote{We define preheating as a
change in the initial adiabat for the gas before its accretion into
the potentials of groups and clusters. The plausible sources of
preheating include galactic winds and AGN activity, as well as
gravitational shocks associated with the formation of large scale
structure.} leads to very extended, low-density gas distributions on
the scales of groups, that cause a steepening of most scaling
relations (Ponman, Cannon, Navarro 1999, hereafter PCN). When the
temperature of the preheated gas reaches the level comparable to the virial
temperature, simulations predict adiabatic accretion of the gas (\eg\
Tozzi, Scharf, Norman 2000). This also implies strong temperature
gradients, proportional to the gas density to the 2/3 power
($\gamma=5/3$). As noted by Loewenstein (2000), the observation of
this phenomena is a critical test for the importance of preheating, as
well as other characteristics of cluster collapse. So far no clear
examples of systems with an adiabatic gas distribution have been
found.

Throughout this {\it Paper} we will define the entropy as
$S={kT_e / n_e^{2/3}}h^{-1/3}$ in keV cm$^2$, following PCN. The
observational data is presented assuming $H_{\rm
o}=50$~km~s$^{-1}$~Mpc$^{-1}$.

\section{Observations}

We select our sample of groups by cross-correlating ASCA (Tanaka,
Inoue \& Holt 1994) and ROSAT (Truemper 1983) archival data on
elliptical galaxies with the group catalogs by Huchra \& Geller (1982)
and Hickson (1982). From an initial sample of 23 groups, we omit 12
because of insufficient coverage by ASCA SIS (Burke \etal 1991) or
availability of only ROSAT HRI data, that after a critical check, does
not allow the study of the surface brightness profile at the radii of
interest. Specifically, our criterium was source extent exceeding 150
kpc. In addition, we exclude HCG92 and NGC3923, because the 'group'
component of these two systems does not appear clearly in the ASCA
data (for HCG92, see also ROSAT HRI analysis of Pietsch \etal 1997).

All observations were screened using FTOOLS version 4.2 with standard
screening criteria. An advantage of ASCA data is its low background,
compared to Chandra and XMM, which is essential for studies of low surface
brightness emission. An obvious drawback is the complexity of the broad ASCA
PSF. Detailed treatment of the PSF effects is therefore a critical part of
our analysis, where we follow the approach described in Finoguenov \etal
(1999), also including the geometrical projection of the three-dimensional
distribution of X-ray emitting gas. Compared to clusters of galaxies,
however, the temperature structure of groups is defined by the Fe L-shell
peak, thus decreasing the importance of energy dependent effects of ASCA PSF
on temperature determination, as they affect primarily the continuum. In
addition, for energies below 2 keV, the energy dependence of the ASCA PSF is
negligible.  Our minimization routines are based on the $\chi^2$ criterion.
No energy binning is done, but an error calculation is introduced as in
Churazov \etal (1996) to properly account for small number statistics. Model
fits to the ROSAT surface brightness profiles of these groups are used as
the input to the ASCA data modeling. The details of our minimization
procedure for ASCA spectral analysis are described in Finoguenov and Ponman
(1999). We adapted the XSPEC analysis package to perform the actual fitting
and error estimation.  The spatially resolved spectral characteristics are
quoted as the best fit solution, plus an estimate of the 90 \% confidence
area for possible parameter variation, based on the regularization technique
(Press \etal 1992; Finoguenov and Ponman 1999). To study the systematic
errors related to the spatially resolved spectroscopic analysis of the ASCA
data, we follow the approach described in Finoguenov, David and Ponman
(2000; hereafter FDP). For all ROSAT imaging analysis, we use the software
described in Snowden \etal (1994) and references therein.

\begin{figure*}

   \includegraphics[width=2.0in]{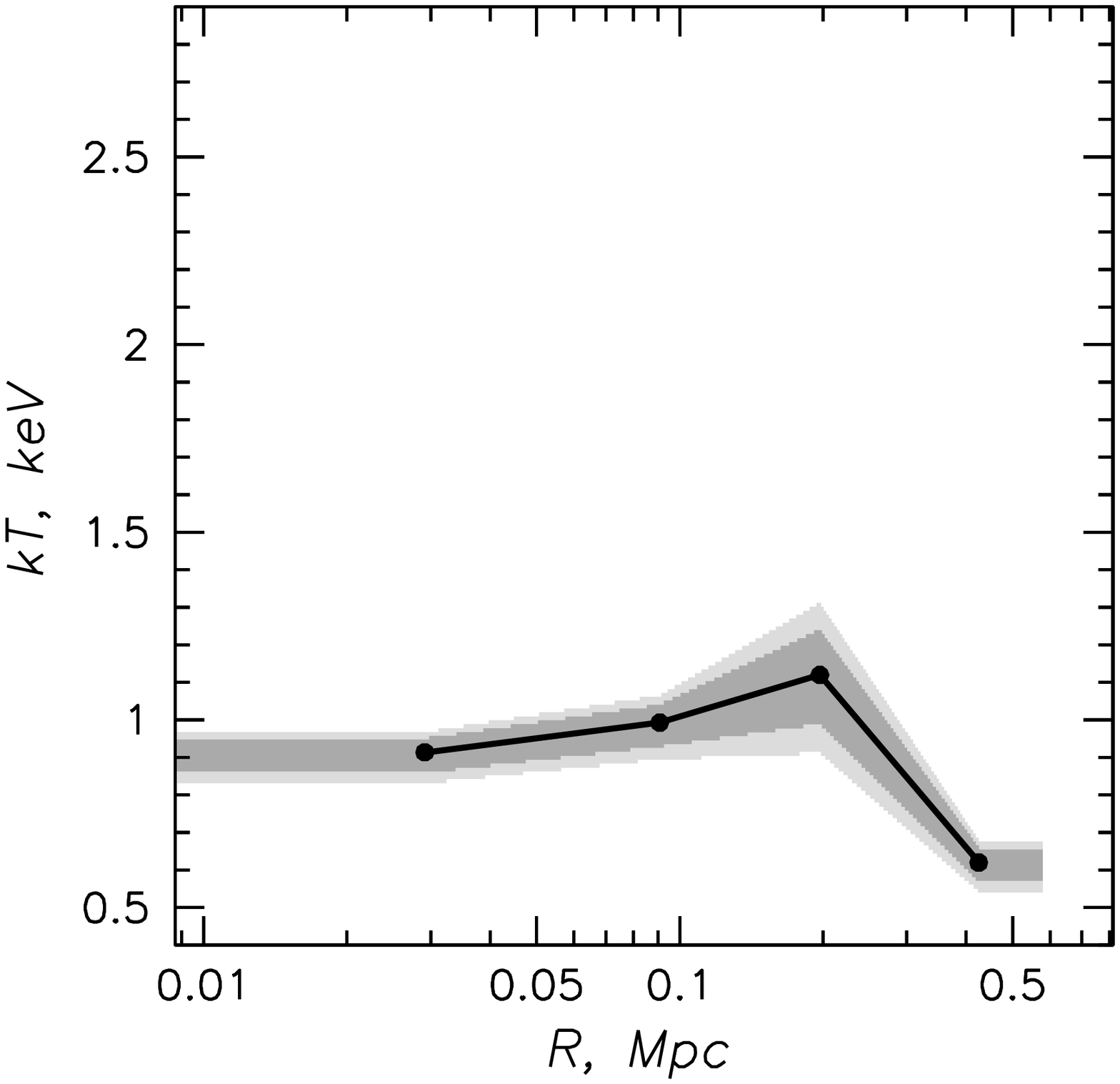} \hfill
   \includegraphics[width=2.0in]{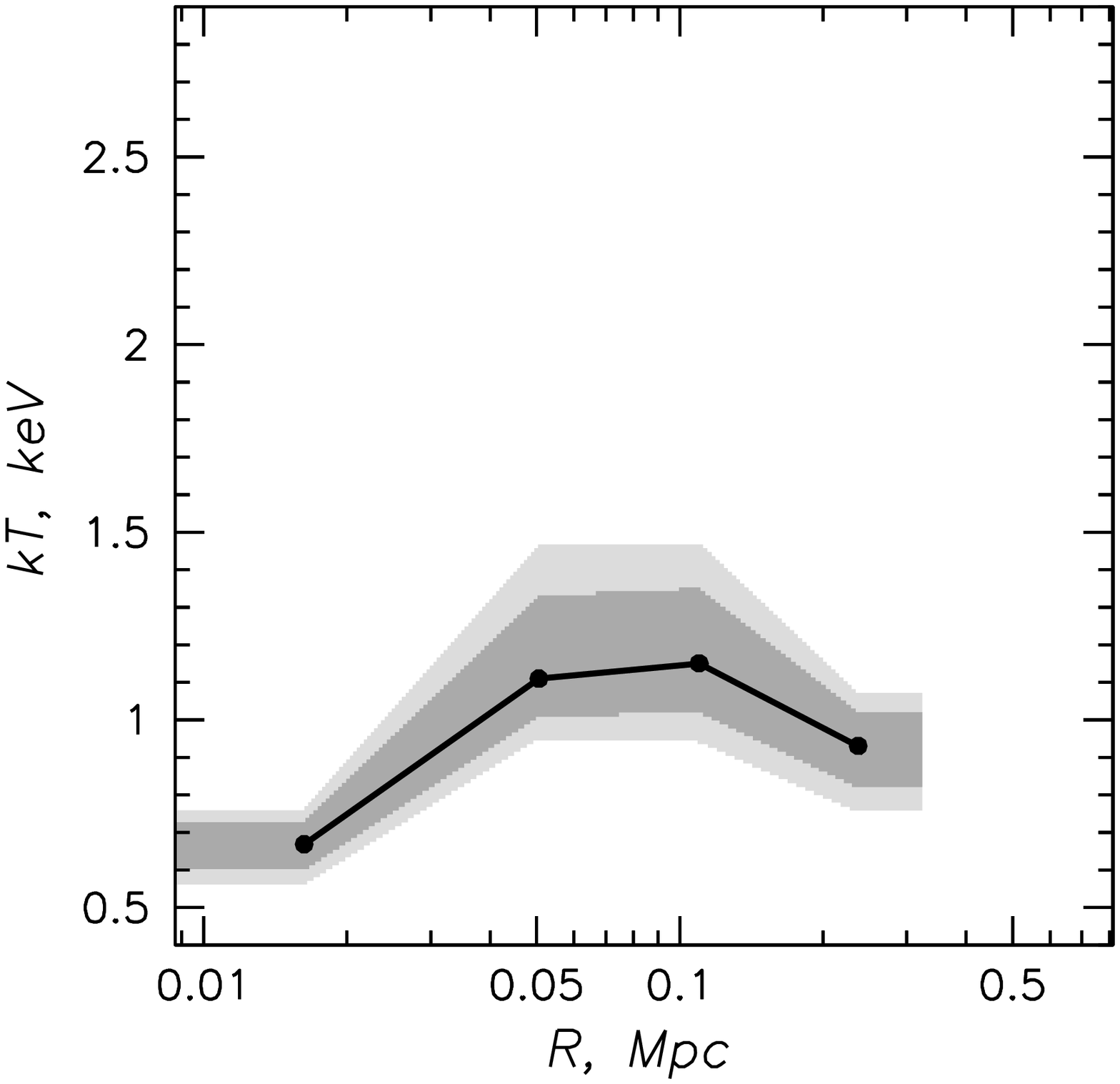} \hfill 
  \includegraphics[width=2.0in]{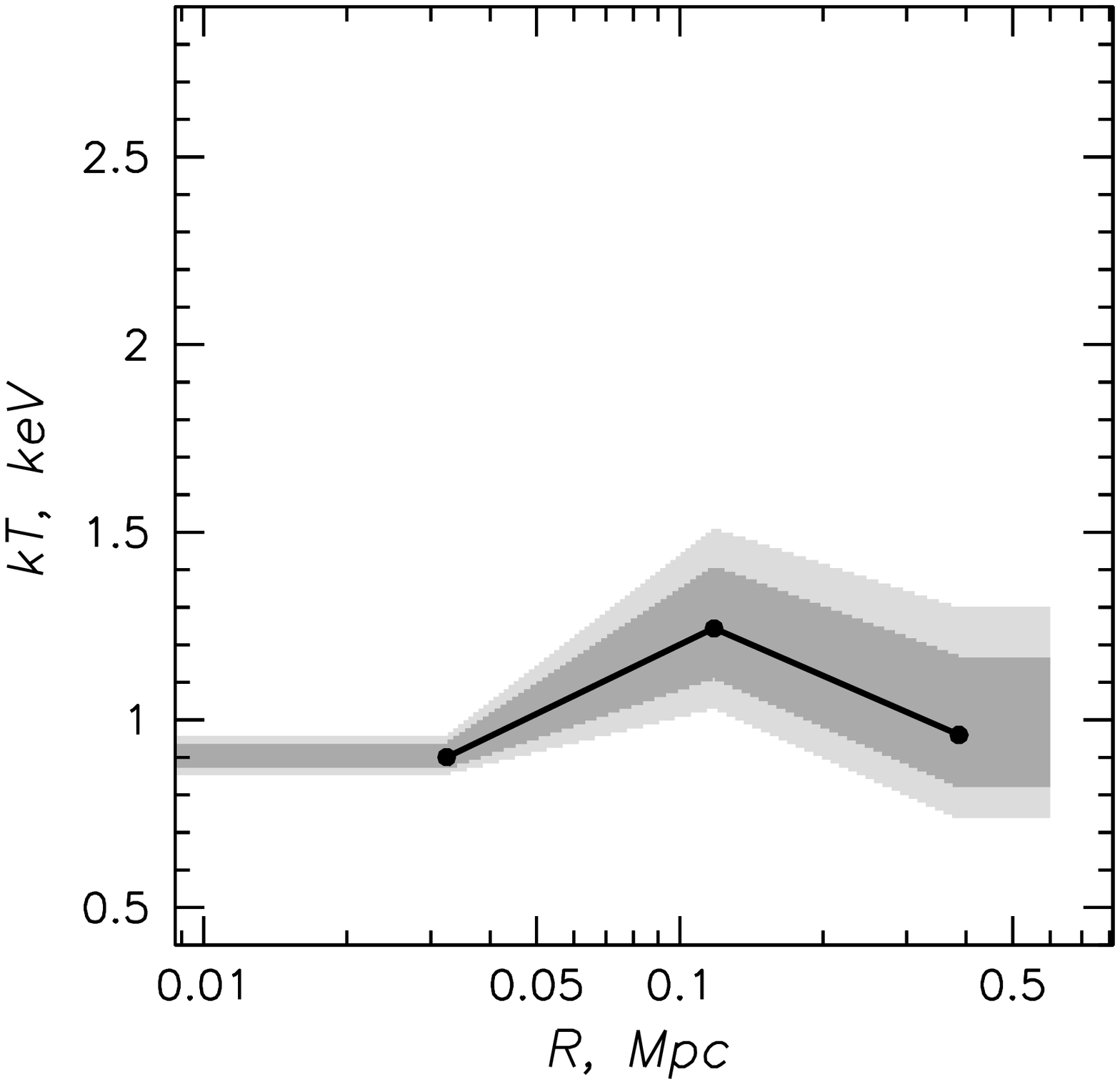} 
 
   \includegraphics[width=2.0in]{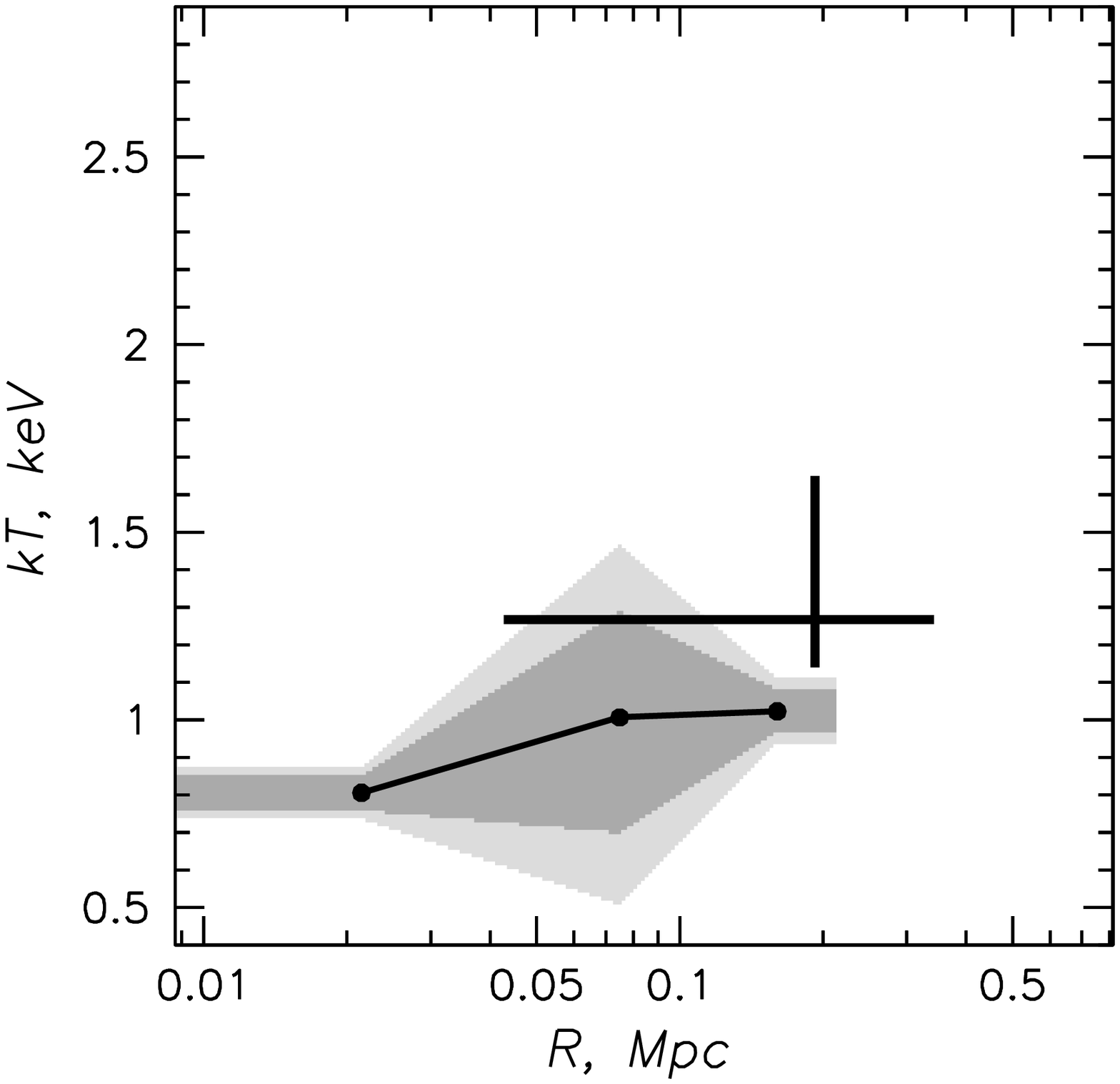} \hfill 
  \includegraphics[width=2.0in]{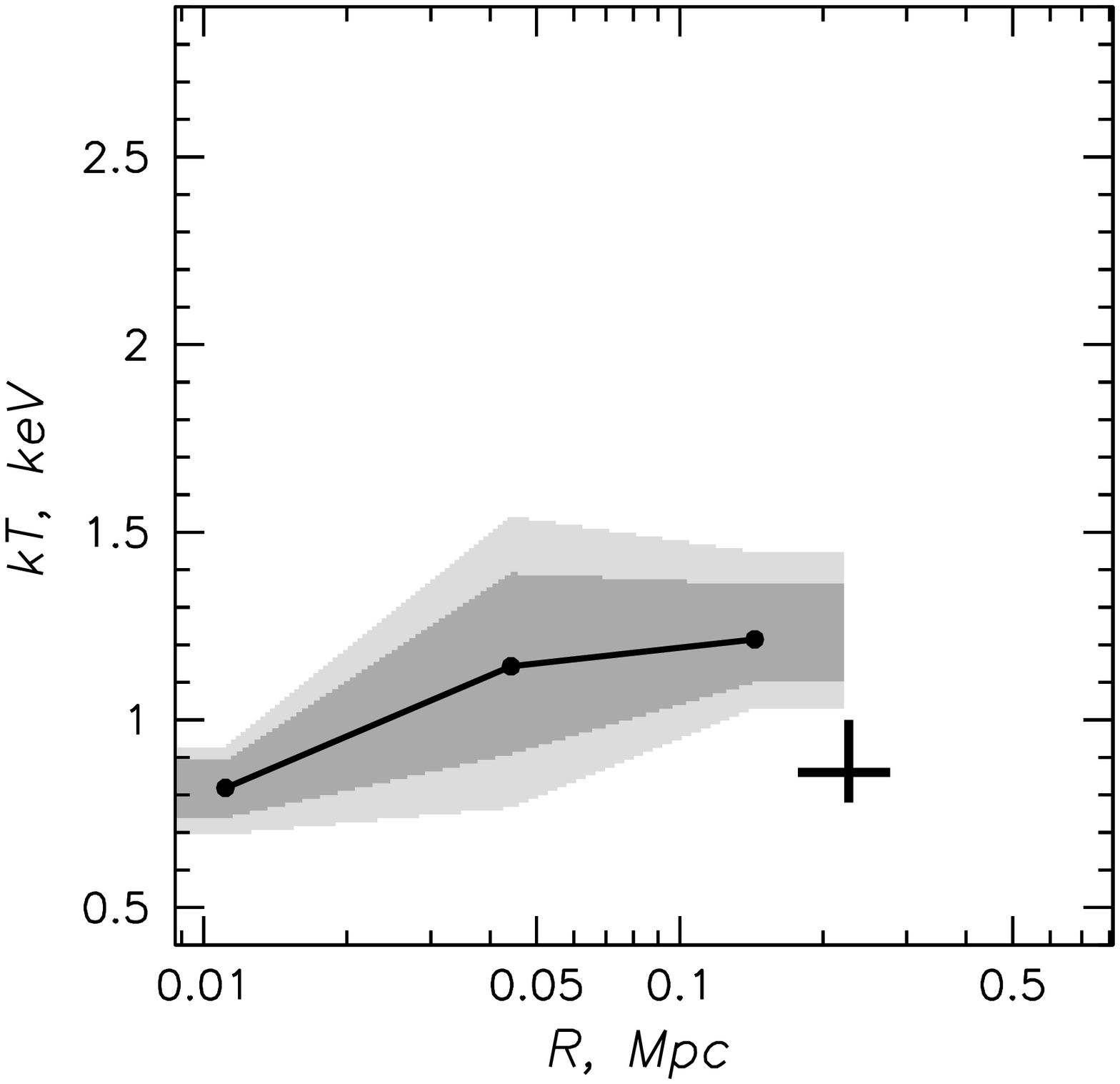} \hfill 
  \includegraphics[width=2.0in]{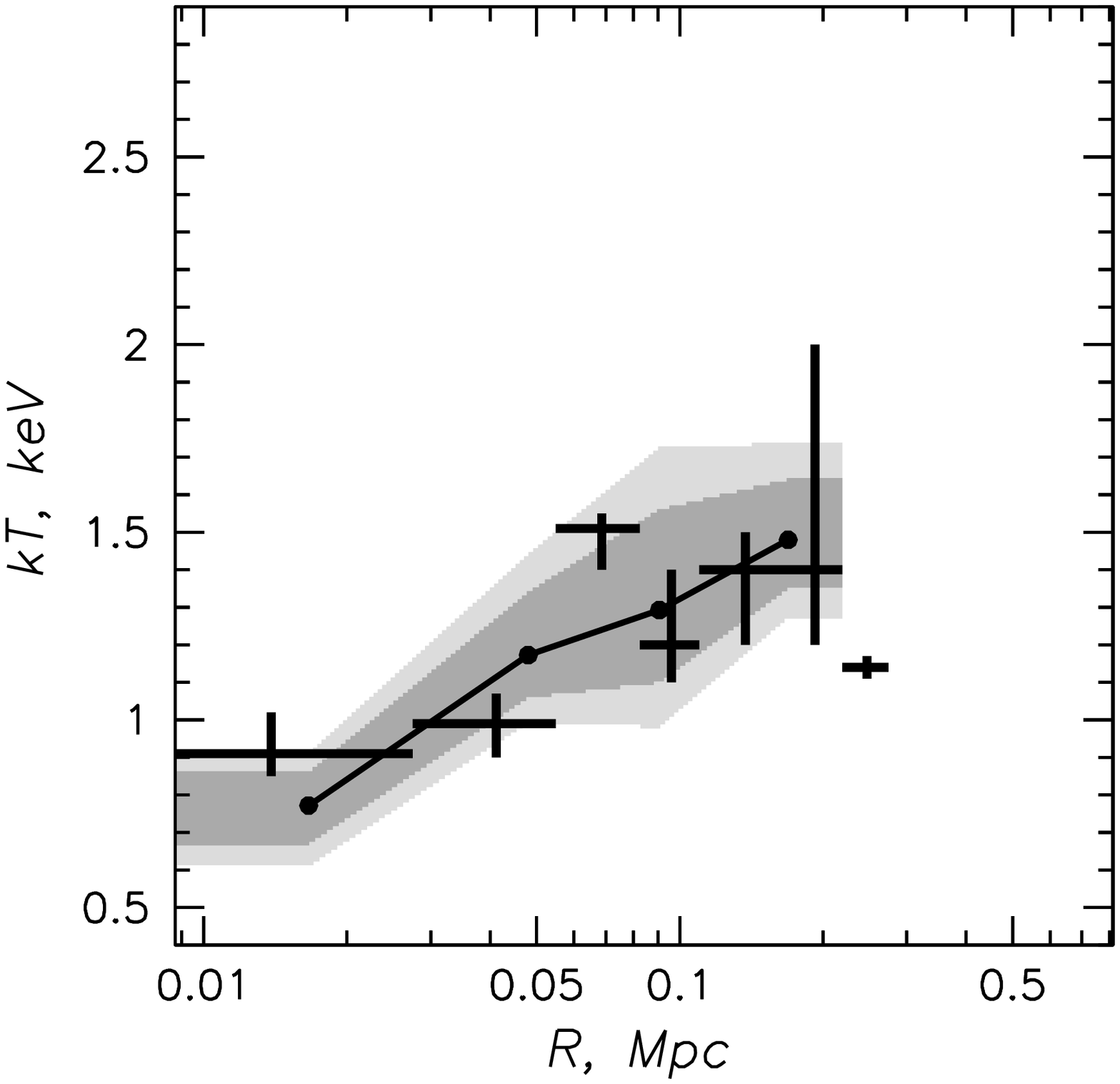} \hfill 

   \includegraphics[width=2.0in]{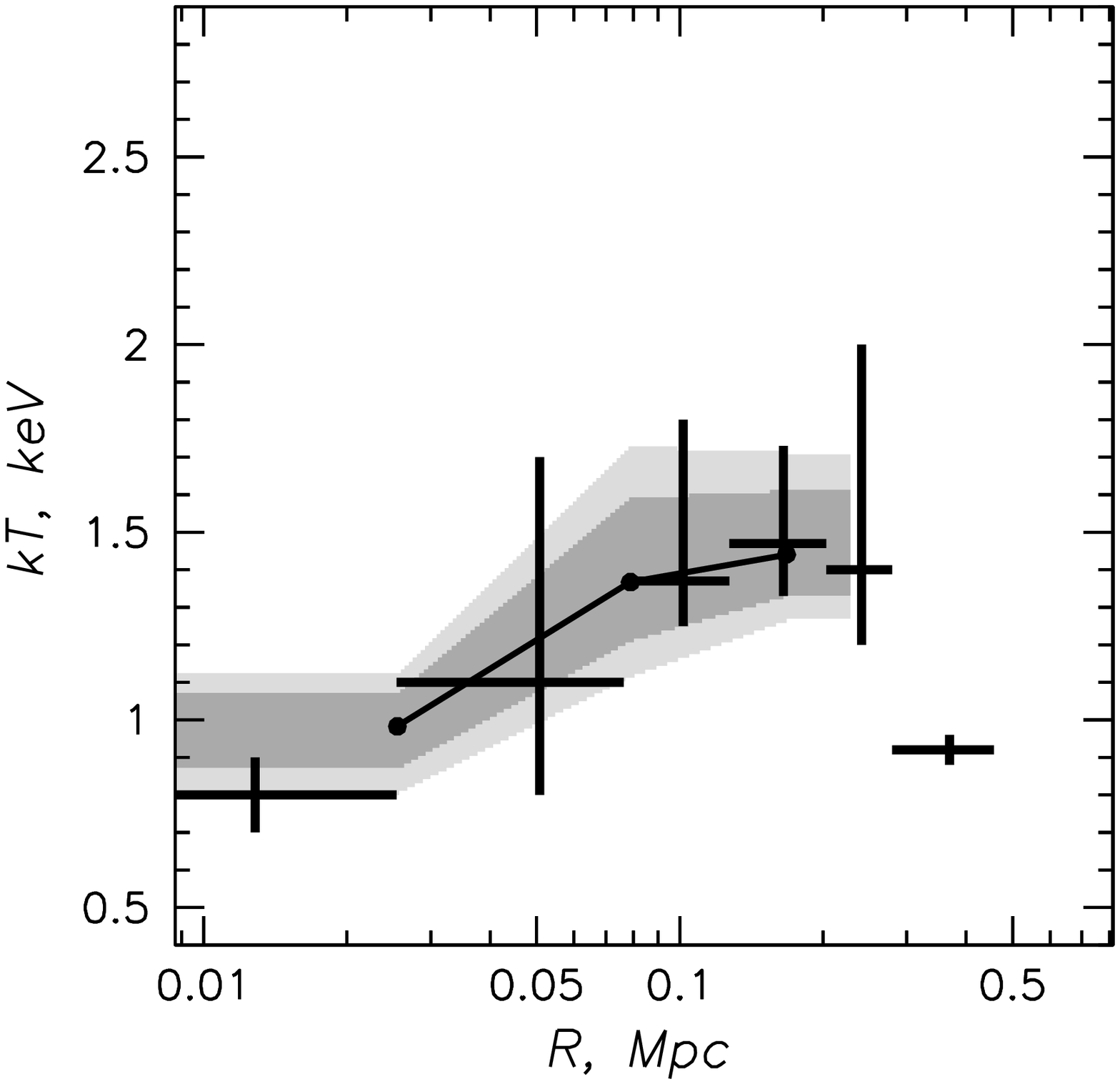}   \hfill 
  \includegraphics[width=2.0in]{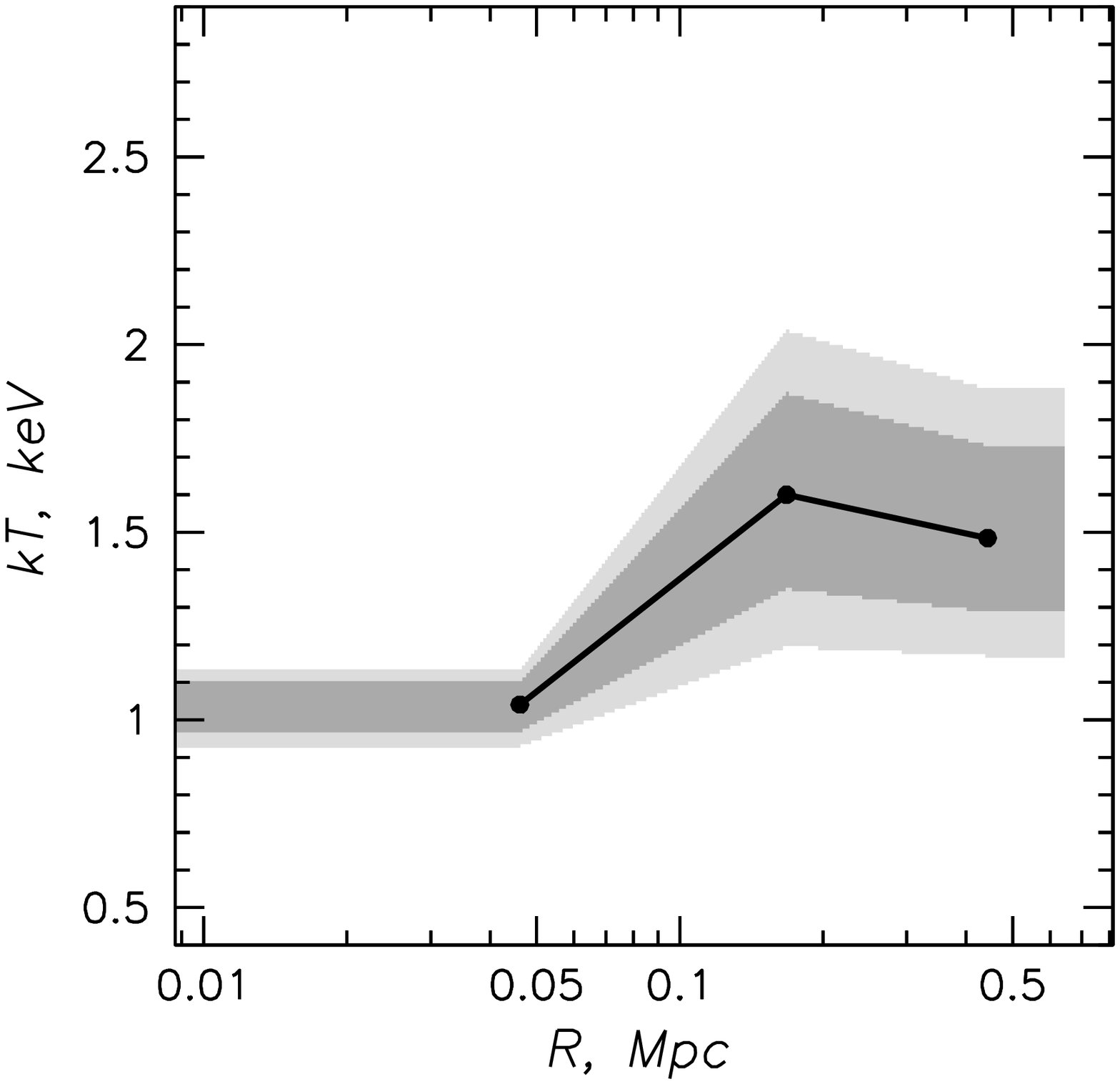} \hfill 
 \includegraphics[width=2.0in]{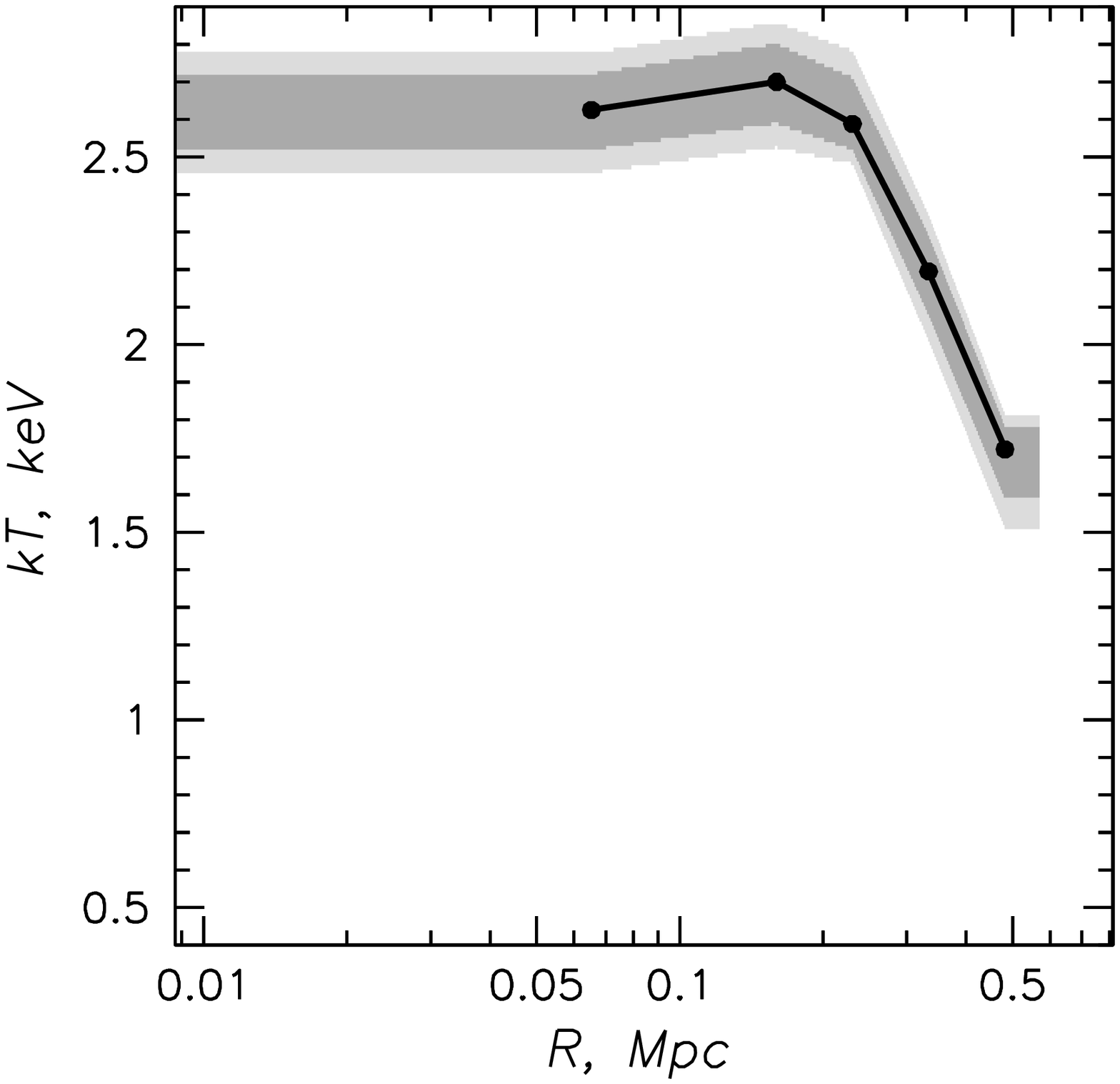}

\figcaption{Deprojected temperature profiles, derived using a single
temperature model to fit the data.  The solid lines correspond to the
best-fit with filled circles indicating the spatial binning used in the
analysis. Dark and light shaded zones around the best fit curves denote the
68 and 90 per cent confidence areas. Black crosses in the NGC507 and NGC2563
panels show the results of a similar deprojection analysis applied to ROSAT
PSPC data by Buote (2000). A black cross in the NGC2300 panel indicates
ROSAT PSPC results for NGC2300 outskirts (Davis \etal 1996) and similarly
for NGC7619 from Trinchieri, Fabiano, Kim (1997). The confidence level for
the cited results corresponds to 68\%.
\label{kt-fig}}
\vspace*{-15.cm}

{\it \hspace*{1.0cm} NGC5129 \hspace*{5.2cm} IC4296 \hspace*{5.6cm} NGC4325}

\vspace*{4.5cm}

{\it \hspace*{1.cm} NGC7619 \hspace*{5.2cm} NGC2300 \hspace*{5.2cm} NGC507 }

\vspace*{4.5cm}

{\it \hspace*{1.cm} NGC2563 \hspace*{5.2cm}  NGC6329 \hspace*{5.2cm}  NGC3268}

\vspace*{6.cm}

\end{figure*}

\begin{figure*}

   \includegraphics[width=2.0in]{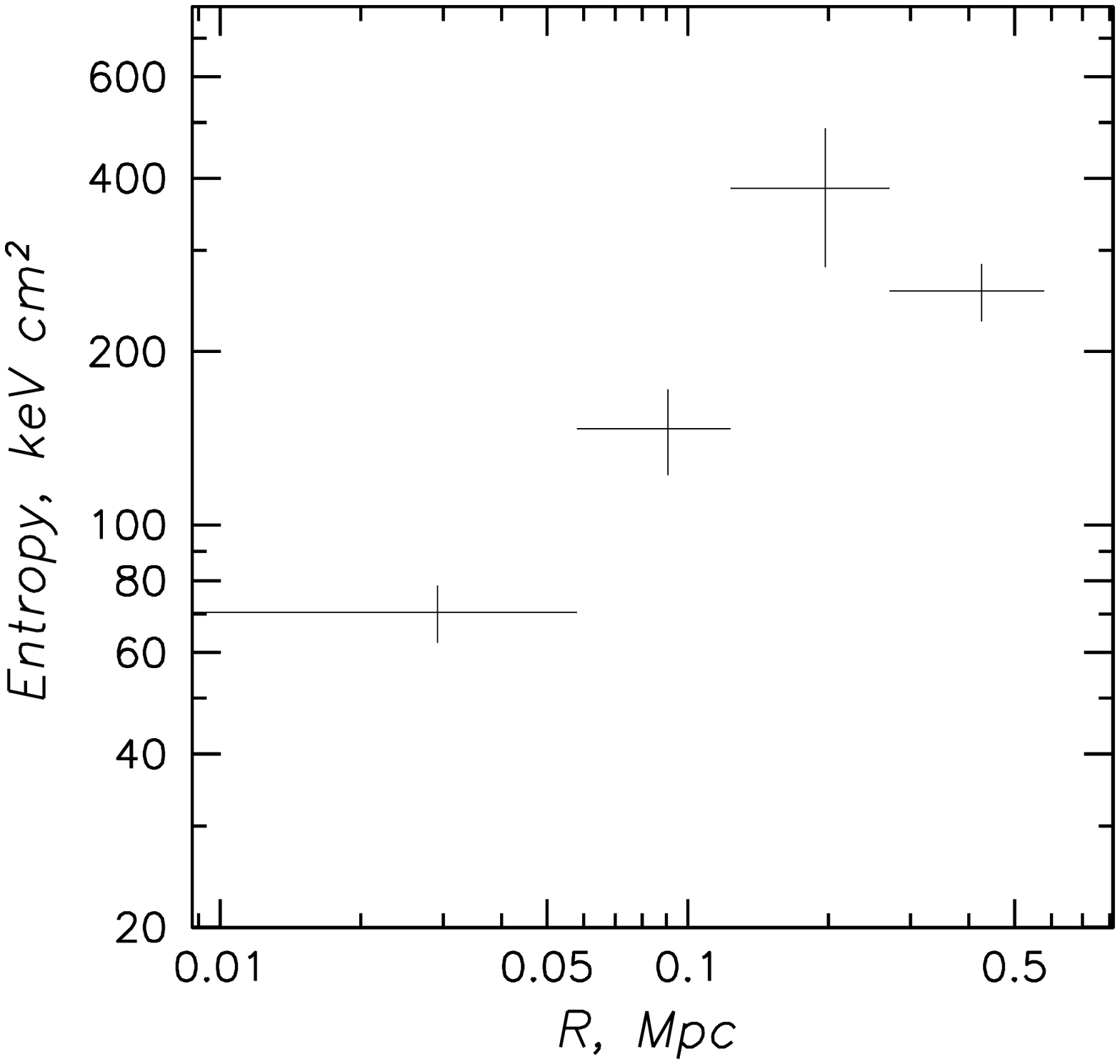} \hfill
   \includegraphics[width=2.0in]{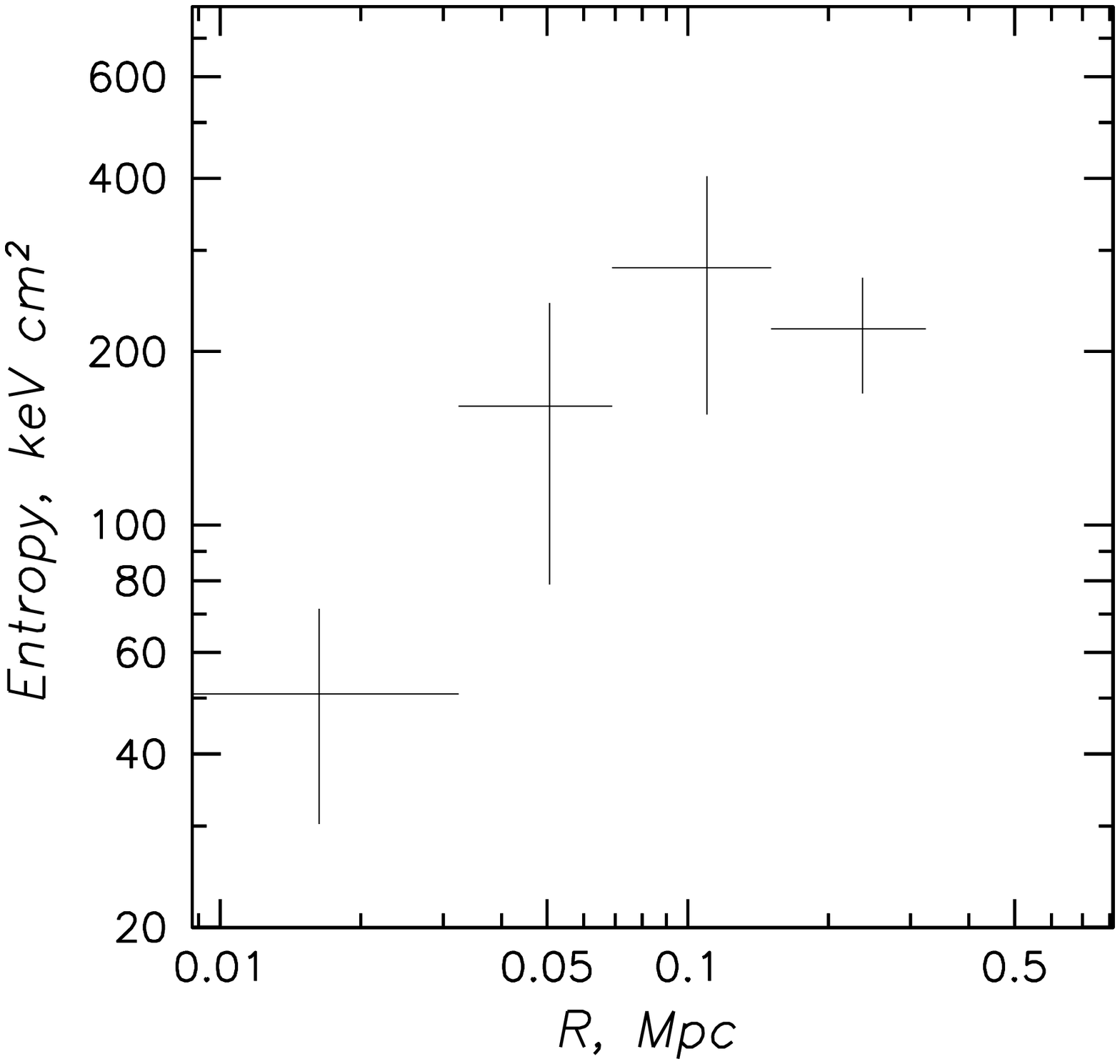} \hfill 
  \includegraphics[width=2.0in]{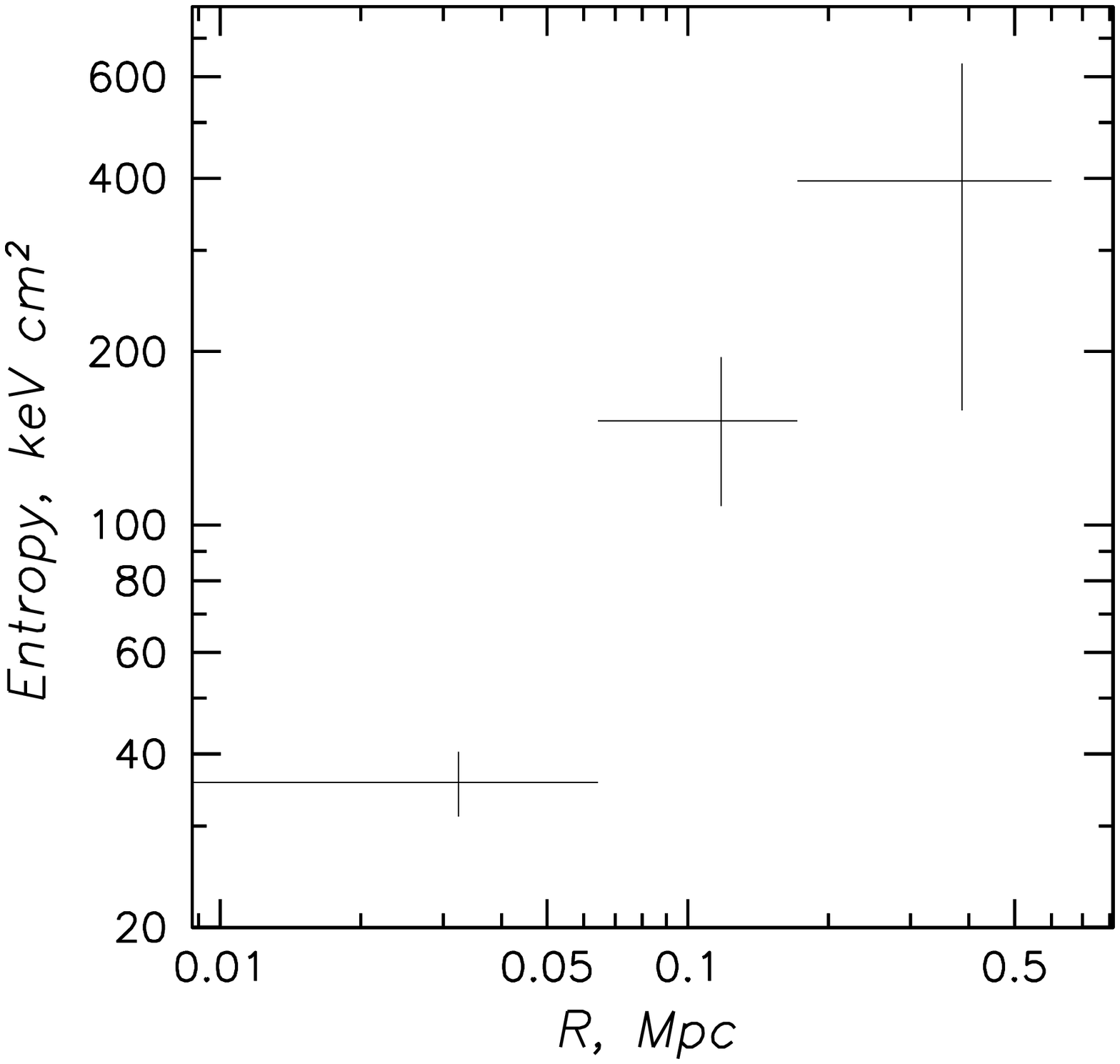} 
 
   \includegraphics[width=2.0in]{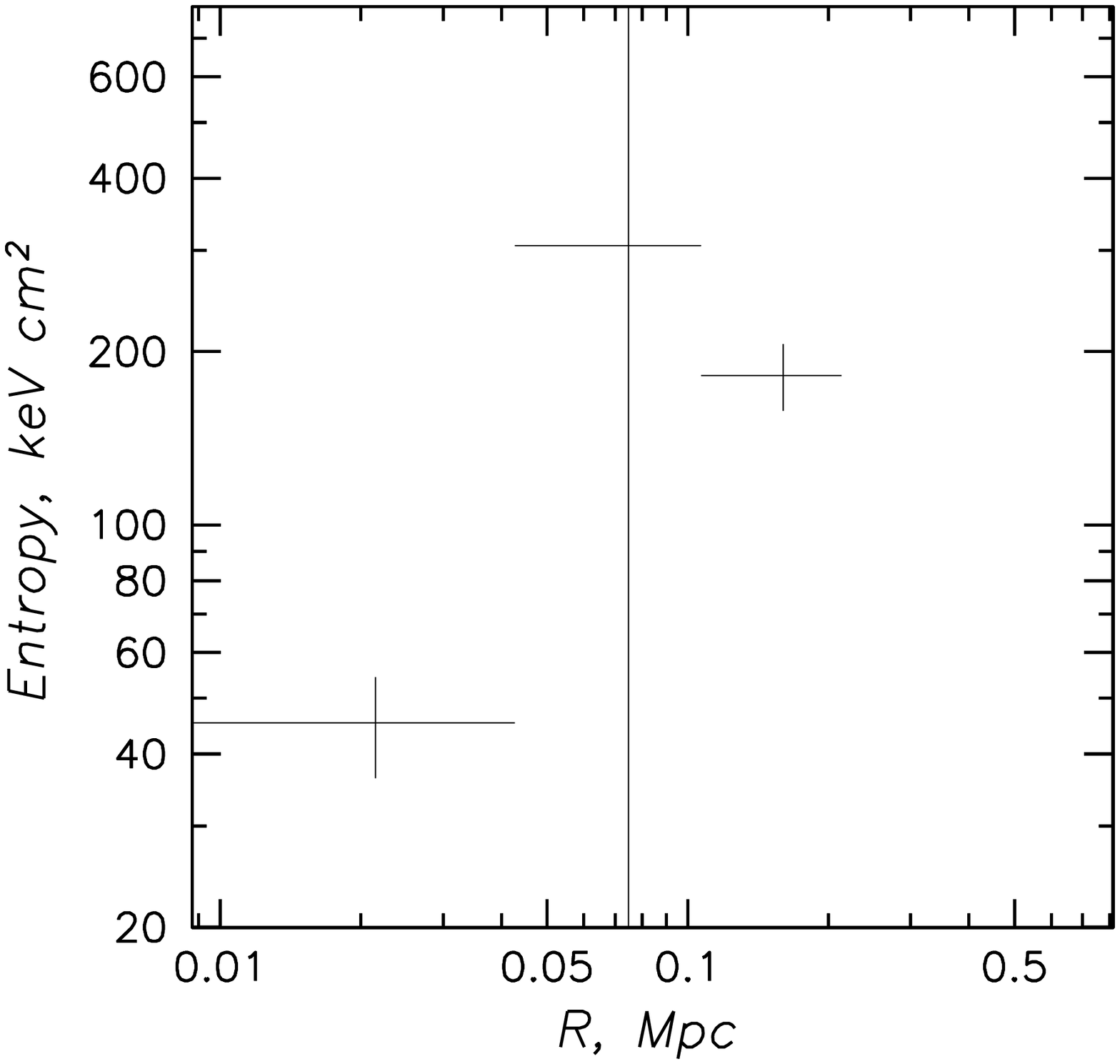} \hfill 
  \includegraphics[width=2.0in]{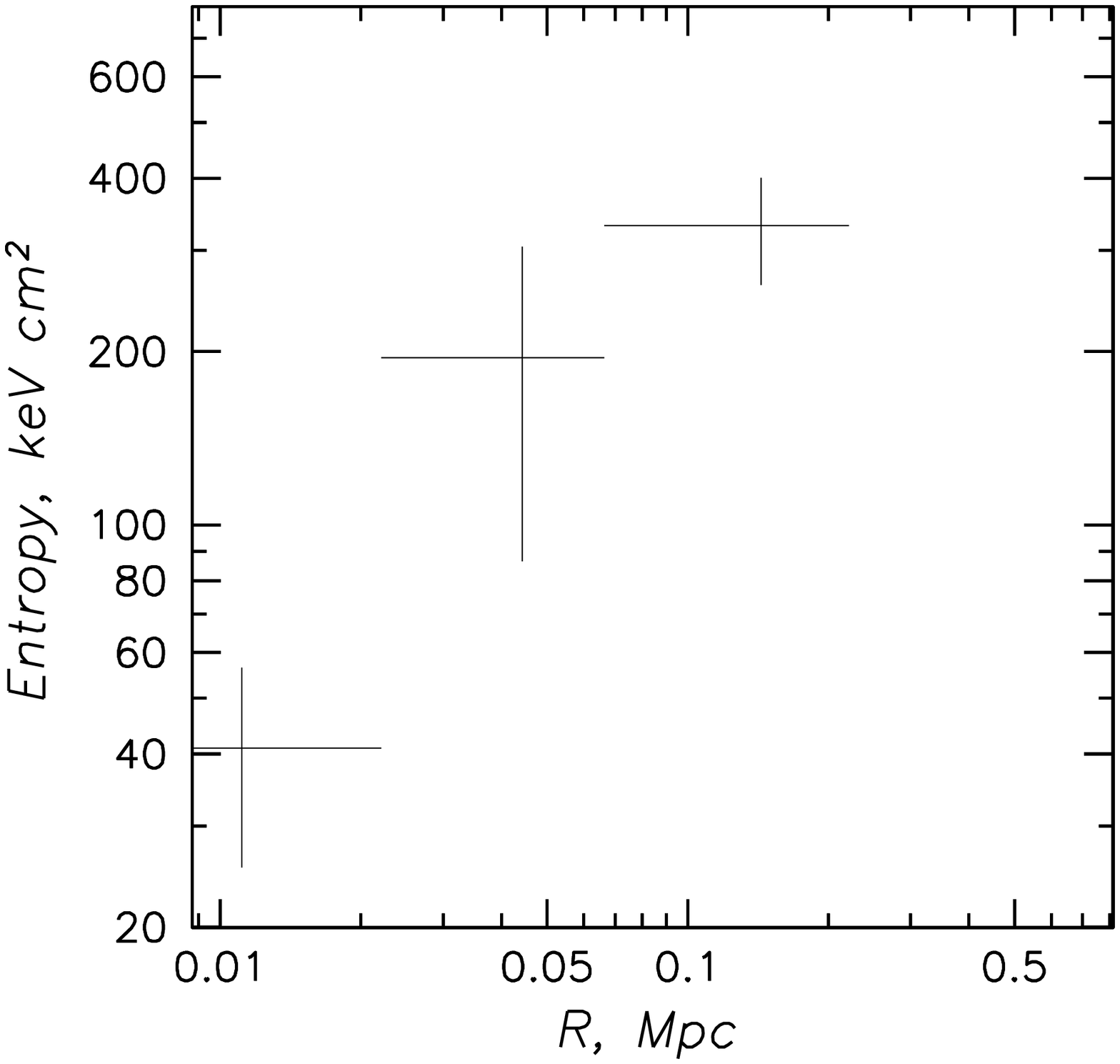} \hfill 
  \includegraphics[width=2.0in]{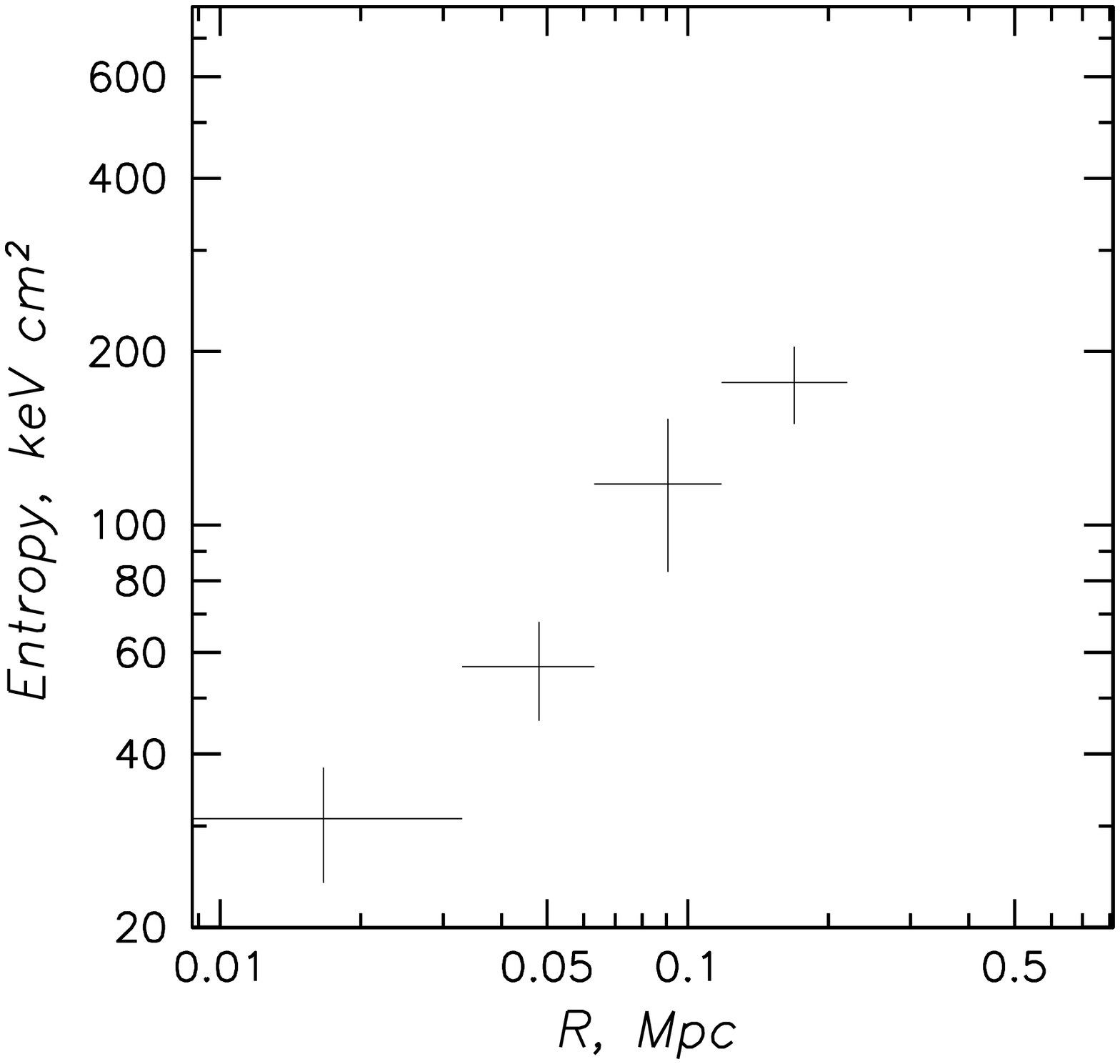} \hfill 

   \includegraphics[width=2.0in]{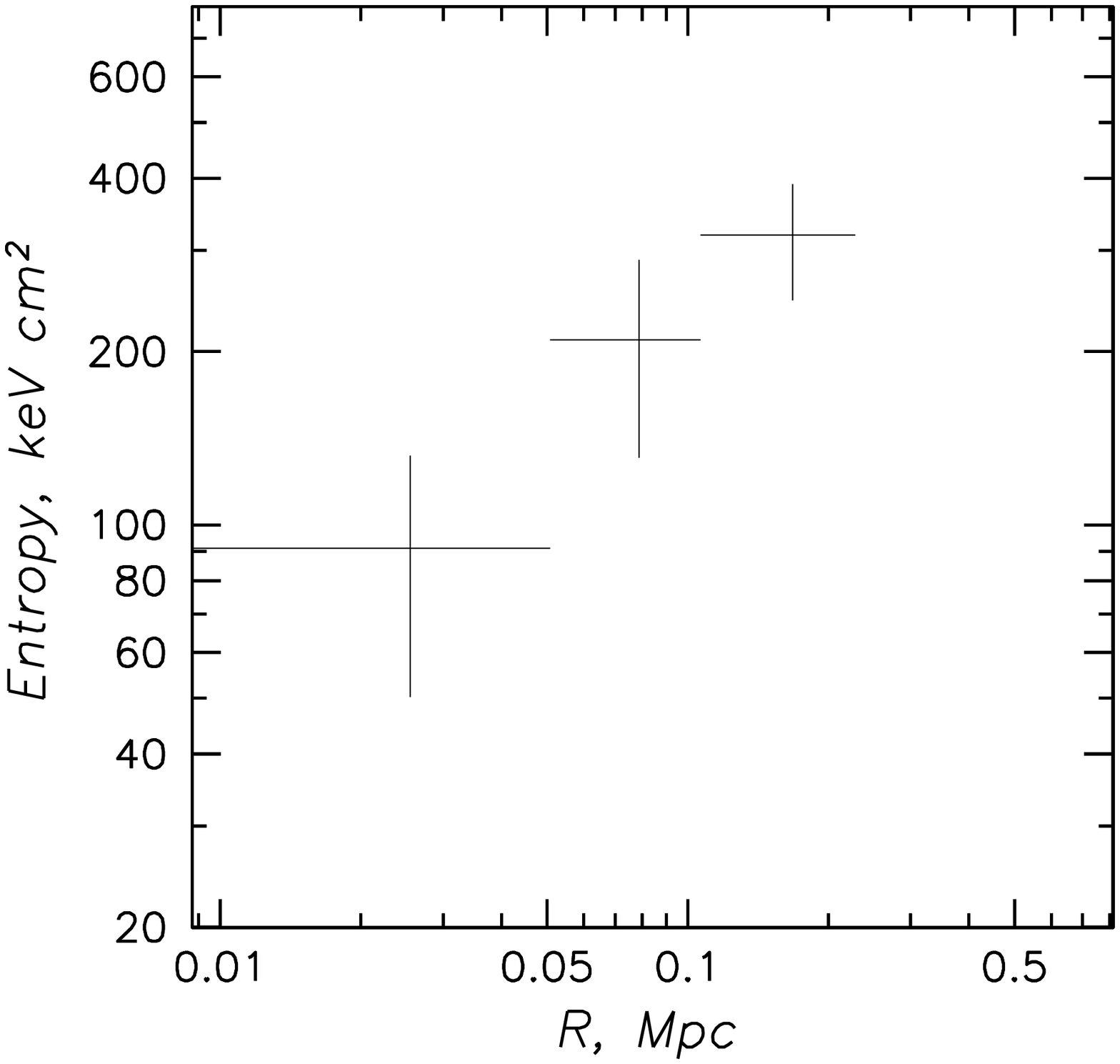}   \hfill 
  \includegraphics[width=2.0in]{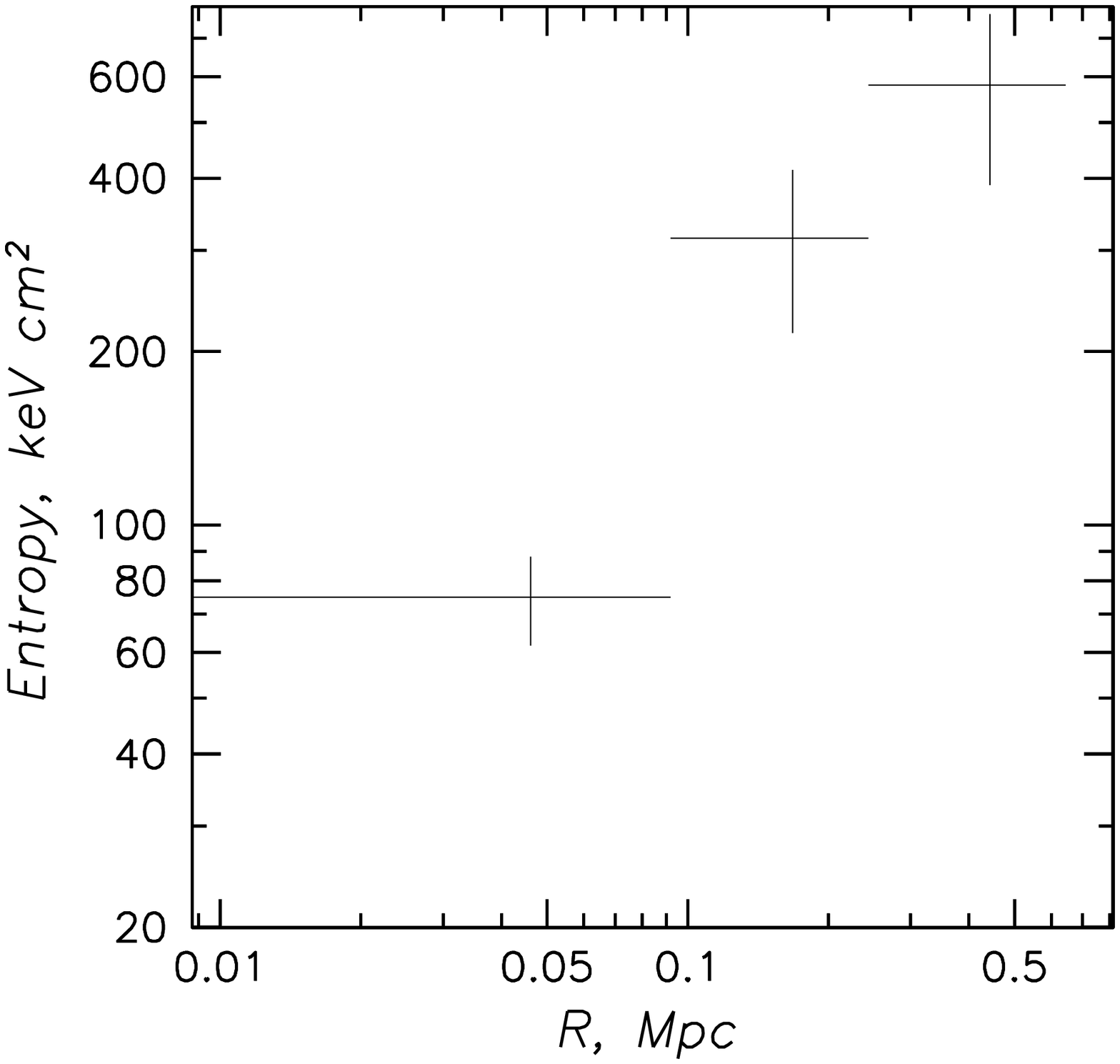} \hfill 
 \includegraphics[width=2.0in]{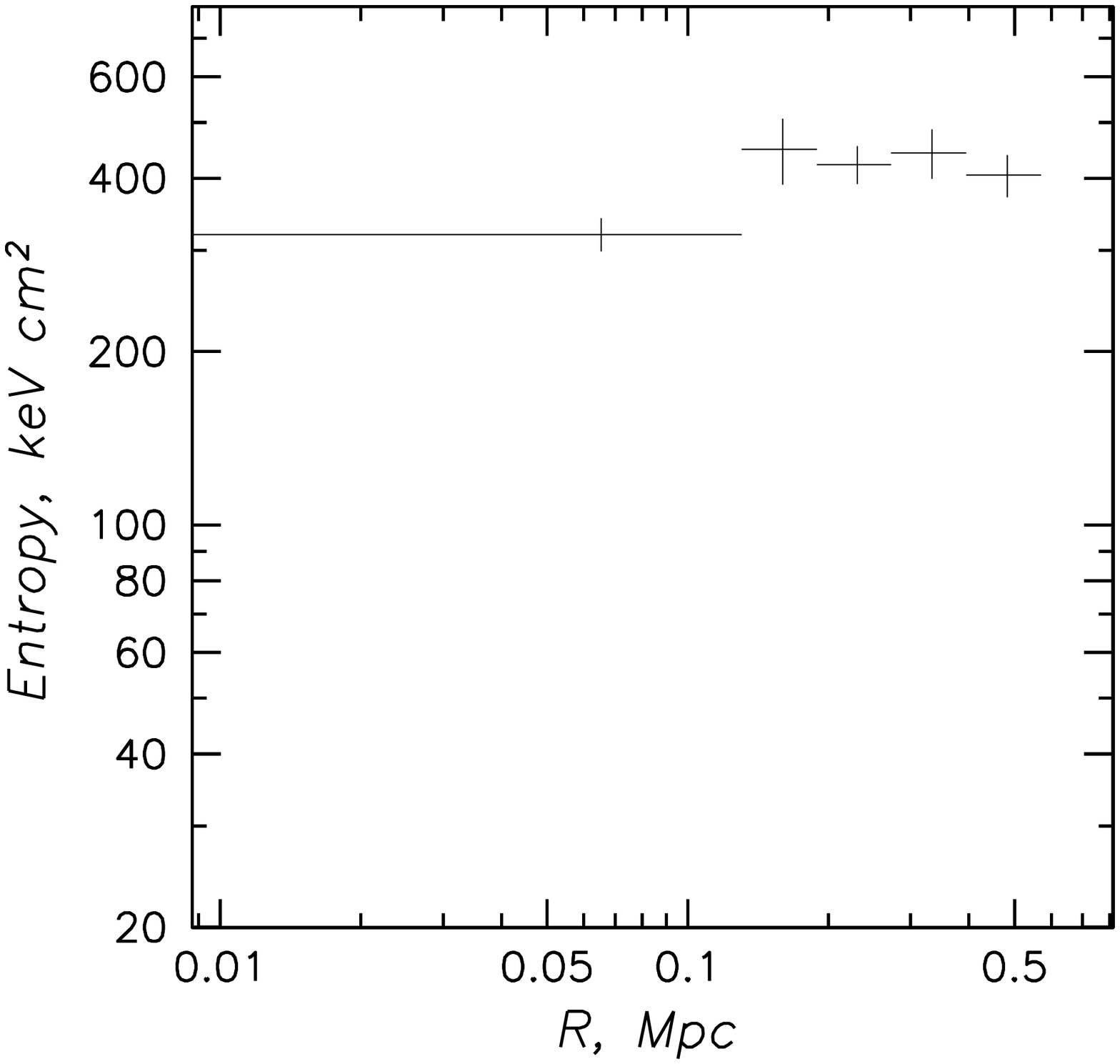}

 \figcaption{Entropy profiles in units of keV cm$^2$, derived using a single
   temperature model to fit the data.  The error bars are shown at 90\%
   confidence level.
\label{ent-fig}}
\vspace*{-14.8cm}

{\it \hspace*{1.0cm} NGC5129 \hspace*{5.2cm} IC4296 \hspace*{5.6cm} NGC4325}

\vspace*{4.5cm}

{\it \hspace*{1.cm} NGC7619 \hspace*{5.2cm} NGC2300 \hspace*{5.2cm} NGC507 }

\vspace*{4.5cm}

{\it \hspace*{1.cm} NGC2563 \hspace*{5.2cm}  NGC6329 \hspace*{5.2cm}  NGC3268}

\vspace*{5.cm}

\end{figure*}

\begin{table*}
{
\begin{center}
\footnotesize
\tabcaption{\centerline{\footnotesize
Characteristics  of  the sample}
\label{tab:opt}}

\begin{tabular}{lrrccccccccc}
\hline
\hline
Name  &D   &\amin & $\beta$&$r_{c}$& $kT_e$& $\delta kT_e$ & $R_{v}$& $L_x$$^{\natural}$&  $f_{line}{\dag}$ & $L_x$$^{\natural}$&  $f_{line}{\dag}$ \\
      &Mpc & kpc  &        & kpc   & keV   & keV           & Mpc    & \multicolumn{2}{c}{total}  & \multicolumn{2}{c}{galaxy} \\
\hline                                                                    
NGC5129& 139. & 39.& 0.60 & 96.& 0.75& 0.07& 1.07&18.4&0.69&5.5&0.66\\
 IC4296&  76. & 22.& 0.31 & 58.& 0.95& 0.16& 1.20&8.3&0.42&1.4&0.92\\
NGC4325& 155. & 43.& 0.59 & 14.& 0.98& 0.07& 1.22&54.8&0.61&33.3&0.62\\
NGC7619&  75. & 21.& 0.33 & 48.& 1.00& 0.02& 1.23&6.2&0.82&2.2&0.64\\
NGC2300&  38. & 11.& 0.33 & 34.& 1.01& 0.23& 1.24&0.8&0.63&0.4&0.56\\
 NGC507&  98. & 27.& 0.44 & 16.& 1.34& 0.05& 1.43&28.6&0.89&9.2&0.85\\
NGC2563&  90. & 25.& 0.40 & 86.& 1.36& 0.10& 1.44&4.5&0.86&0.8&0.60\\
NGC6329& 167. & 46.& 0.53 &120.& 1.45& 0.12& 1.49&19.4&0.89&3.8&0.88\\
NGC3268$^{\flat}$                                   
       &  57. & 16.& 0.32 & 62.& 1.82& 0.06& 1.66&40.1&0.86&$<6$&$-$\\
\hline
\end{tabular}
\end{center}
\hspace*{3cm} $^{\natural}$ \hspace*{0.3cm}{\footnotesize Deprojected luminosity in units of
  $10^{41}$ ergs/s, in the 0.5--2 keV band}

\hspace*{3cm} $^{\flat}$ \hspace*{0.3cm}{\footnotesize also referred to as
the Antlia cluster or NGC3258, the second brightest galaxy}

\hspace*{3cm} $^{\dag}$ \hspace*{0.3cm}{\footnotesize fraction of line emission, based on the iron abundance and assuming the solar abundance ratio}
}
\end{table*}

We use the MEKAL plasma code (Mewe \etal 1985, Mewe and Kaastra 1995,
Liedahl \etal 1995) in all of our spectral analysis.  Spectral fitting was
twofold. First we use the energy range 0.8--7 keV and determine the
intensity of the hard component, arising from the unresolved point sources
in the central galaxy, fixing its spectral shape to a bremsstrahlung
temperature of 6.5 keV (Finoguenov \& Jones 2000). This is mostly important
for heavy element abundance determination (Matsushita 1998) and is a
specific of the analysis of X-ray spectra of early-type galaxies, where the
gas-to-light ratio is an order of magnitude lower than in clusters of
galaxies.  Later we perform the analysis in the 0.8--3 keV band, where the
group's emission is clearly seen over the background. The exception is
NGC3268, which is much hotter (2.5 keV) than other groups. For it we use the
0.8--7 keV energy band. Below 1.5 keV, we add a 10\% systematic error at the
68\% confidence level to the spectra to account for systematics in spectral
modeling.

The groups, centered on NGC2300, IC4296, NGC5129, NGC6329 and NGC2563 have
in the ASCA fields of view point-like background sources, which are brighter
than the 'blank-sky' threshold chosen for ASCA background files and thus
should be removed individually. In these cases we extract the spectra of
point sources using both the ROSAT/PSPC and the ASCA/SIS, subtract the
background, analyze the spectra and then, at all energies of the ASCA
response matrix, estimate the contribution from these point sources to the
regions of diffuse emission selected for the analysis.

Basic characteristics of the sample are listed in Tab.\ref{tab:opt}. Column
(1) identifies the system, (2) is the adopted luminosity distance, (3) the
corresponding scale length.  Columns (4--5) give the results of the surface
brightness fitting outside the central region using a $\beta$ model.  Column
(6) gives the best fit cluster emission-weighted temperature, (7) the
corresponding 90\% error, and (8) an estimate of the virial radius of the
system ($r_{180}=1.23 T_{keV}^{0.5}h_{50}^{-1}$ Mpc; Evrard, Metzler,
Navarro 1996) using $kT_e$. Estimates of the virial radii using X-ray
observations by Finoguenov, Reiprich and B\"ohringer (2001) show
that the actual virial radii are 20\% lower than the estimation of Evrard et
al. (1996). Columns (9--12) list the luminosity in the 0.5-2 keV band and a
fractional contribution from the line emission for the system in total and
the central galaxy (excluding the hard component). The deprojected
luminosity is calculated using the profiles of temperature, element
abundance and normalization, obtained in our analysis of ASCA SIS data. The
luminosity is not corrected for differences in normalization between ASCA
SIS and other experiments.  Since line emission in our energy band is
dominated by iron, in calculating the fractional contribution of the line
emission, we use the derived iron abundance profile and assume a solar
abundance ratio for other elements. We attribute to the central galaxy the
X-ray emission of temperature matching the stellar velocity dispersion,
which is typically one-two central bins in our analysis. 

\section{Radial profiles of temperature and entropy} \begin{figure*}
\hfill\includegraphics[width=3.6in]{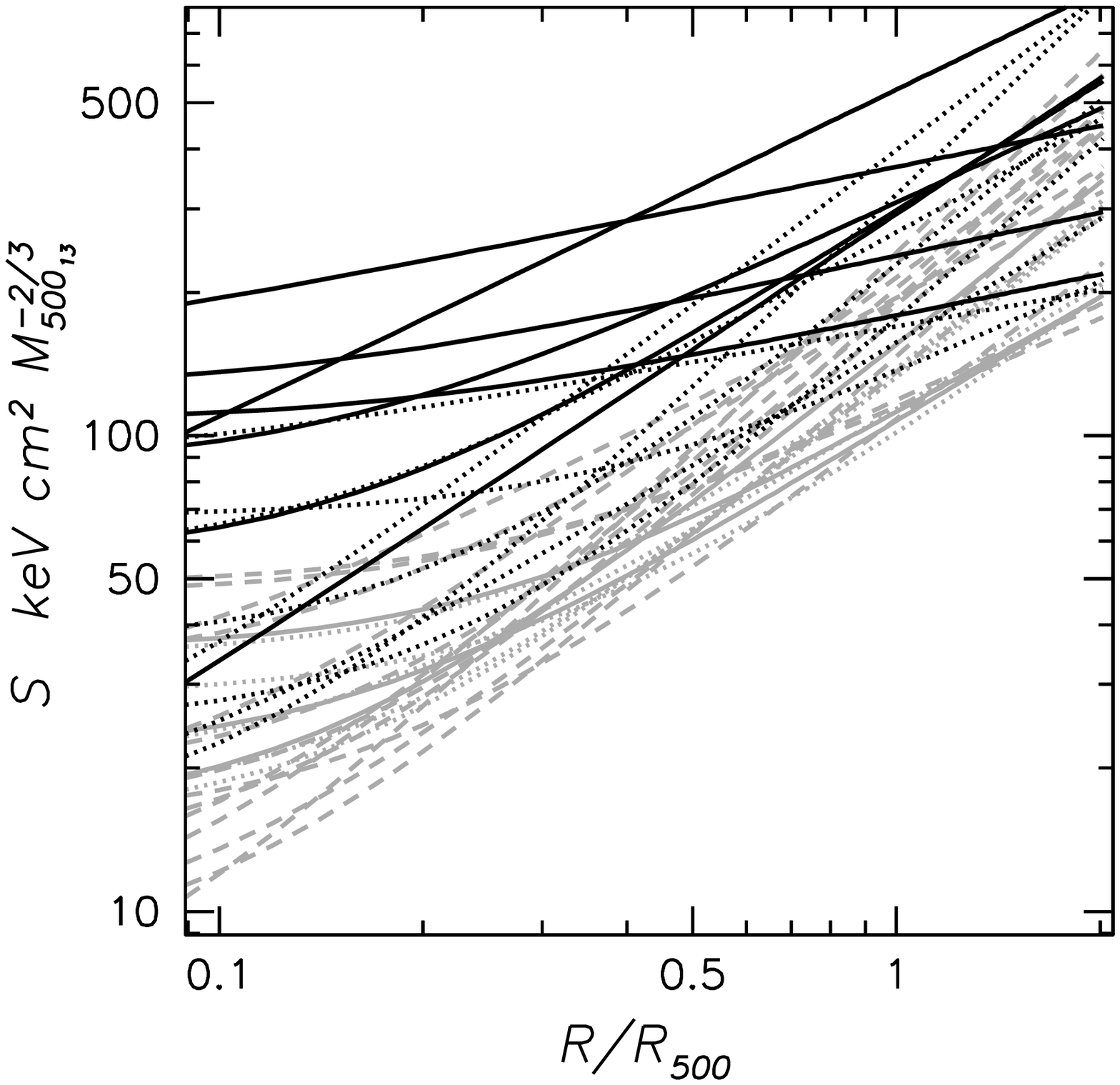}\hfill\hfill\includegraphics[width=3.6in]{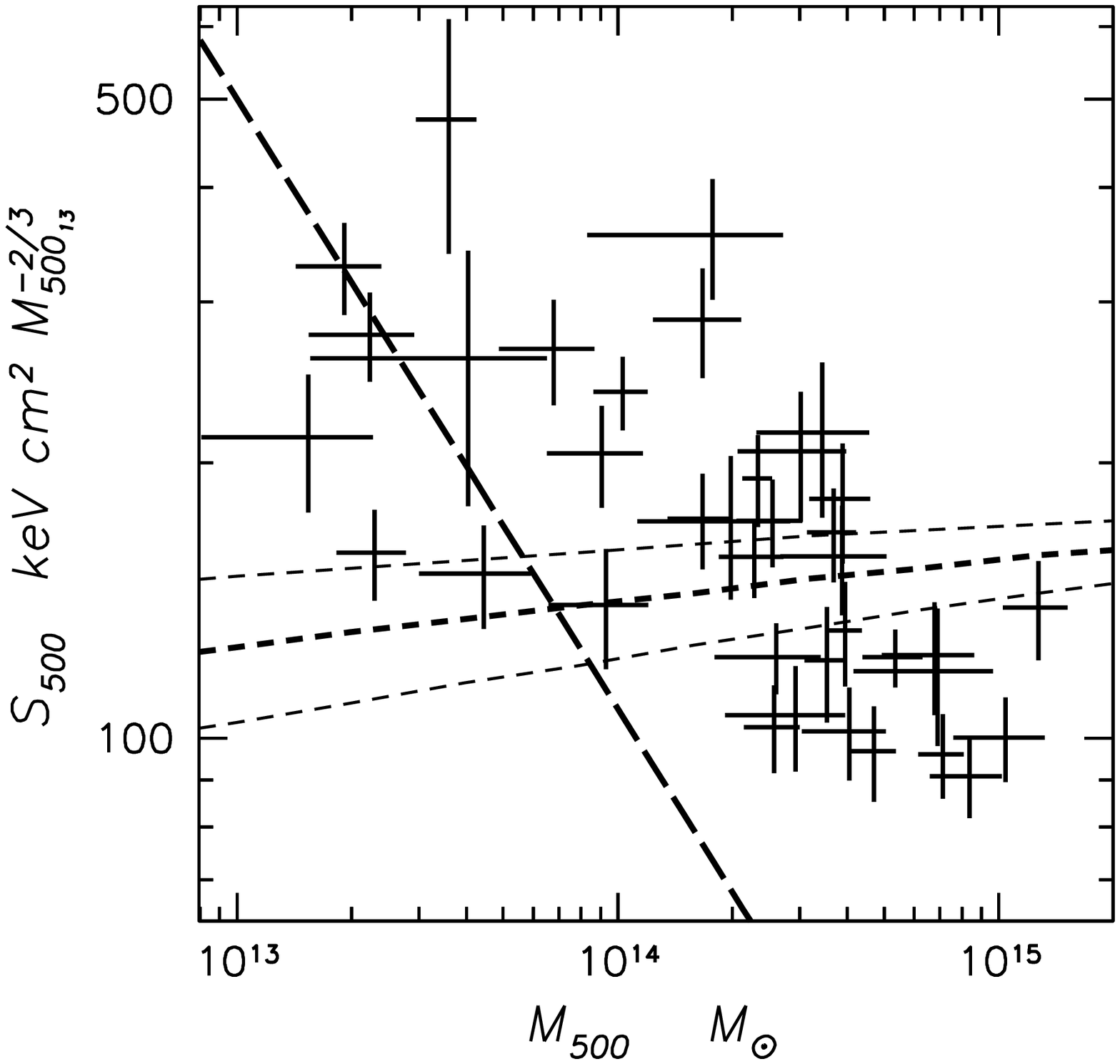}\hfill
\figcaption{{\it Left panel.} Entropy profiles scaled by the total
mass in units of $10^{13}$ \msun, measured at an overdensity of 500
and plotted against a scaling radius measured as a fraction of
$r_{500}$. Black solid lines denote systems with kT$<1.5$ keV
(NGC5129, IC4296, NGC4325, HCG62, NGC5044, HCG51, NGC6329), black
dotted lines 1.5 keV $<$kT$<3$ keV (MKW4, NGC3268, A262, MKW4S,
A539, AWM4, MKW9, A2634), gray dashed 3 keV $<$kT$<5$ keV (A1060,
A4038, 2A0335, HCG94, A2052, AWM7, A2657, MKW3S, A2063, Hydra A,
A3112, A4059, A496, A2199), gray dotted lines 5 keV $<$kT$<7.5$ keV
(A3558, A119, A1651, A3571) and gray solid lines 7.5 keV $<$kT$<11$ keV
(A401, A3266, A2029). {\it Right panel.}  Scaled entropy at the
overdensity of 500 vs the total gravitational mass of the system,
measured within an overdensity of 500. The short-dashed line indicates
the prediction from shock heating, corrected for the mean formation
redshift, normalized to fit the intermediate-mass clusters. Thinner
dashed lines indicate a scatter induced by the expected variation in
the formation redshifts on the 75\% level. The long-dashed line
indicates the effect of the preheating value of 500 keV cm$^2$.
\label{fig:ent} } 
\end{figure*}

Temperature and entropy profiles are presented in Figs.\ref{kt-fig}
and \ref{ent-fig}. Heavy element abundances are of secondary
interest to this paper and are discussed in the appendix.
The data on temperature, entropy, Fe
and Si abundance is given in Tab.\ref{t:data}. The general behavior of
the temperature profiles is defined by kinematics of the stellar mass
loss at the center (Matsushita 2001), characterized by the velocity
dispersion of stars in the central galaxy.  
Beyond radii of 150 kpc, the galaxy contribution is small and the
emission is dominated by the intragroup gas, which exhibits a gradual
decrease in temperature, extending to the observational limits. In our
sample, the best examples of such temperature behavior are the
NGC3268, NGC5129 and IC4296 groups. A central drop in temperature is
not seen in NGC3268, suggesting that the emission is dominated by the
ICM gas at radii as small as 50 kpc. ROSAT observations, extending to
larger radii, show that the temperature also drops in NGC2563, NGC507
(Buote 2000), and possibly in NGC2300 (Davis et al. 1996). A decrease
of the temperature profile at radii exceeding 150 kpc is commonly seen
in groups (Trinchieri \etal 1997; Finoguenov \etal 1999; Finoguenov \&
Ponman 1999, Buote 2000, Finoguenov \& Jones 2000).  Total
gravitational mass estimates for the five best observed systems
(NGC3268, NGC6329, NGC4325, IC4296 and NGC5129) are given in
Finoguenov, Reiprich, B\"ohringer (2001).

In deriving the temperature profiles, we were particularly interested in
revealing any isentropic zones in groups. As one can see from
Fig.\ref{kt-fig}, where temperature profiles are shown, several groups show
sharp temperature declines at large radii, typically exceeding 200 kpc. Thus
the detection of isentropic behavior is limited by the spatial extent of the
data.

\subsection*{\normalsize\it \center I\lowercase{sentropic outskirts in the} NGC3268 \lowercase{group}}

The X-ray emission from the Antlia (NGC3268) group is extended, filling the
ROSAT PSPC field of view. The determination of the surface brightness with
the ROSAT PSPC has a large uncertainty, due to the source extent and
uncertain level of the ROSAT background, caused by the short exposure time
(6 ksec). Inclusion of ASCA data for the surface brightness analysis
constrains the $\beta$ parameter, because ASCA has a lower background at the
radii of interest, due to additional pointings that cover the group
outskirts. With four pointings, about 30 ksec each, the ASCA data cover the
central 35\amin\ of the cluster. The results of the analysis are listed in
Tab.\ref{tab:opt}.

Given the flat gas density profile, projection effects are important. ASCA
SIS projected temperature is $\sim2$ keV, in agreement with the GIS results
of Nakazawa \etal (2000), while the deprojected temperature at the center is
$\sim2.7$ keV (Fig.\ref{kt-fig}). We note that the temperature drop from 0.2
to 0.6 Mpc in the Antlia cluster is detected in two off-set ASCA pointings,
that cover the south-west part of the cluster. From the constant entropy
level seen at outer radii in the Antlia cluster (Fig.\ref{ent-fig}), we
conclude that the temperature drop is primarily due to the cluster emission
and not to a contribution from NGC3258, located south-west of NGC3268. Our
arguments are based on a factor of 3 lower entropy, typical of the group
centers (Lloyd-Davies \etal 2000 and this work), compared to the adiabat of
the gas at the place of NGC3258. In addition, enhancement in the surface
brightness would be seen in the ROSAT map, as well as more than just one
galaxy (NGC3258) ought to be associated with the putative subgroup. Finally
a coincidental match of the temperature close to the center of the Antlia
cluster and at larger distances, reaching 600 kpc, would require fine
tuning. For another group, AS1101, first XMM-Newton results reveal a
temperature profile in remarkable correspondence to NGC3268 (Kaastra et
al. 2001).

From radii of 150 kpc to 600 kpc, the entropy profile of the Antlia cluster
is flat at the level of 430 keV cm$^2$. In many groups the temperature drop
is very sharp, corresponding to the leveling off of the entropy at a value
of $350\pm50$ keV cm$^{2}$ (see Fig.\ref{ent-fig}).  While the plateau in
the entropy profile is predicted by the preheating scenario, the observed
entropy level is 3 times higher than the measurement of PCN. In the
following we consider the reasons for this phenomenon.

\section{A ceiling in the preheating value for entropy}

We compare the observed entropy level at the group's outskirts to that
predicted from scaling relations, choosing an overdensity
($\overline{\rho}/\rho_{crit}-1$) of 500, $\Delta_{500}$, where
$\overline{\rho}$ is the mean density of the cluster integrated out to
$r_{\Delta}$. In Finoguenov, Reiprich and B\"ohringer (2001) we estimated
radii for $\Delta_{500}$ (hereafter $r_{500}$), the polytropic index and the
total mass $M_{500}$, while in Sanderson \etal (2002) we fit a $\beta$-model
to the observed ASCA de-projected density. The employed $\beta$-model fits
to surface brightness profiles and polytropic fits to the temperature
profiles are only valid outside the cores of the systems, typically larger
than $0.1r_{500}$. The resulting entropy profiles for samples of groups and
clusters are plotted in Fig.\ref{fig:ent} vs the fraction of $r_{500}$,
scaled by $M^{-2/3}_{500_{13}}$, the total enclosed mass measured within
$\Delta_{500}$ in units of $10^{13}$ \msun. Such scaling is equivalent to
scaling by virial temperature and is introduced to remove the dependence of
the shock strength on the cluster mass. We use the mass instead of the
measured temperature to exclude the effect of adiabatic compression. Entropy
profiles produced by the accretion shock exhibit the strongest rise with
radius, as seen for rich clusters (also David et al. 1996). Entropy profiles
of groups are much shallower (solid and dotted lines in Fig.\ref{fig:ent})
and, especially at smaller radii, are higher than found for clusters.

The flatter entropy profiles found for groups are in good agreement with
simulations that introduce an entropy floor (Borgani \etal 2001). However,
at low overdensities groups still deviate from clusters. These deviations
are opposite to the prediction of structure formation, since groups form
earlier than clusters, resulting in lower entropy. These deviations are
larger than expected due to the entropy floor (PCN) and the effect of gas
cooling (Voit \& Bryan 2001).


\subsection{Effects of gravitational heating and substructure}

Clusters do not appear as uniform entities. As Fig.\ref{fig:ent} (left)
shows, at $0.5r_{500}$ the entropy profiles of clusters start to diverge, so
the scatter in the entropy among clusters at a given overdensity becomes
similar to the differences between clusters and groups. A critical radius at
which such deviations occur corresponds to $r_{500}$, matching an earlier
suggestion of Evrard et al. (1996) that X-ray determinations of cluster mass
should generally be trusted to such radii only. We produce a cut in the
entropy profiles at $r_{500}$ and in Fig.\ref{fig:ent} (right) plot the
resulting entropy values vs the gravitational mass of the system, measured
at overdensity of 500. Preheating at the level discussed in this paper
(shown as a long-dashed line) can only affect systems with mass ($M_{500}$)
below $10^{14}$ \msun, and is successful in describing the entropy behavior
in groups.

In Fig.\ref{fig:ent} (right) we also present the rescaled theoretical
prediction for shock heating (Eke, Navarro, Frenk 1998; PCN), with
details presented in the appendix. As is seen from the Figure, 
clusters exhibit a large spread around the theoretical prediction,
compared to the effect of variation in the cluster formation epochs.

In this paragraph we consider the possible reasons for the large spread in
the entropy among clusters. Incomplete merger shock heating predicts a sharp
temperature and density changes, arising from a final time of shock
propagation through the cluster. Since Chandra cluster images often show
surface brightness edges (Markevitch 2002), the temperature behavior
generally suggests cold fronts rather than shocks (Markevitch et al. 2000;
Vikhlinin, Markevitch \& Murray 2001). Another possibility is that merger
shocks start in the cluster core and decrease in strength with radius (due
to the flat distribution of the gas density). Finally, survival of the gas
associated with the infalling subcluster results in a decreased entropy
(e.g. Borgani et al. 2002). Since formation of large clusters is still in
progress, compared to poor clusters (Kauffmann 1995), hotter clusters tend
to have lower entropy, resulting in a larger range around the entropy
prediction which includes the effect of cluster formation epoch.  In
addition, an observed scatter in the entropy could result from our
simplified approximations to both temperature and density profiles.  At least
in one example of the hot cluster (although not from our sample), A2163,
where a decreasing temperature profile from ASCA is confirmed by
Chandra and XMM-Newton (Markevitch, Vikhlinin 2001; Pratt, Arnaud, Aghanim
2001), a steepening of the density profile is also suggested at large radii
(Vikhlinin, Forman, Jones 1999a). If such behavior is present in other
systems, the resulting entropy would be underestimated.


\subsection{Leaning over the gravitational threshold}

At radii beyond $\sim0.1r_{500}$, entropy profiles in clusters may be the
result of purely gravitational heating. Preheating will increase the scatter
in the entropy profiles between clusters and groups. One can see from
Fig.\ref{fig:ent} (left) that the scatter between cluster and group
profiles, compared to the scatter within the cluster profiles, decreases
with increasing radius. For example at half $r_{500}$, the range among all
the systems is a factor of 10, while it is a factor of 2 for clusters. At
$r_{500}$ the scatter is a factor of 3 for all systems and a factor of 2 for
clusters, while at yet larger radii the scatter does not change when groups
are excluded. This indicates that $r_{500}$ is a limiting radius where the
scatter arising from the non-gravitational effects can still be seen above
the scatter resulting from the gravitational heating. This corresponds to a
typical mass for the groups in our sample of around $3\times10^{13}$ \msun.

Let's compare the entropy level measured for radii of equal enclosed mass
for groups and clusters. Observations show that there is no significant
trend in gas fraction with system mass at radii defined this way (Sanderson
et al. 2003), which is important in estimating the entropy produced by a
given shock strength, as explained in \S\ref{s:imp}. Consideration of the
entropy at similar enclosed mass is fruitful because the implied strength of
the accretion shock is thought to be similar, unless the epoch of cluster
formation plays a dominant role. We choose a mass of $3\times10^{13}$ \msun\
for this comparison, which is still within the analyzed radii for the
groups, while outside the 150 kpc radii for clusters in our sample (and thus
is free from complications due to BCG potential, cooling flow and BCG AGN
heating).

\begin{figure*}
\hfill\includegraphics[width=3.6in]{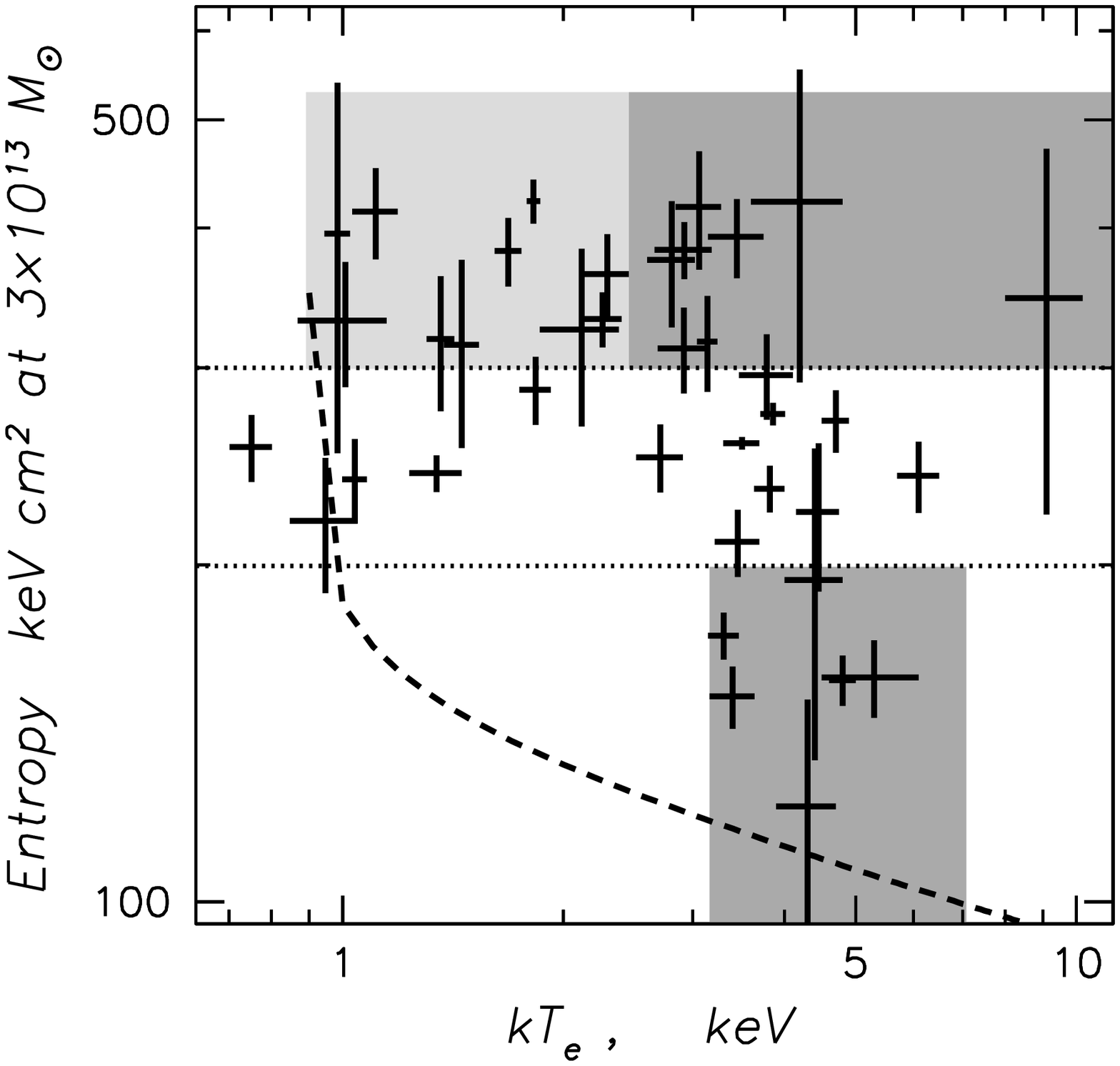}\hfill\hfill\includegraphics[width=3.6in]{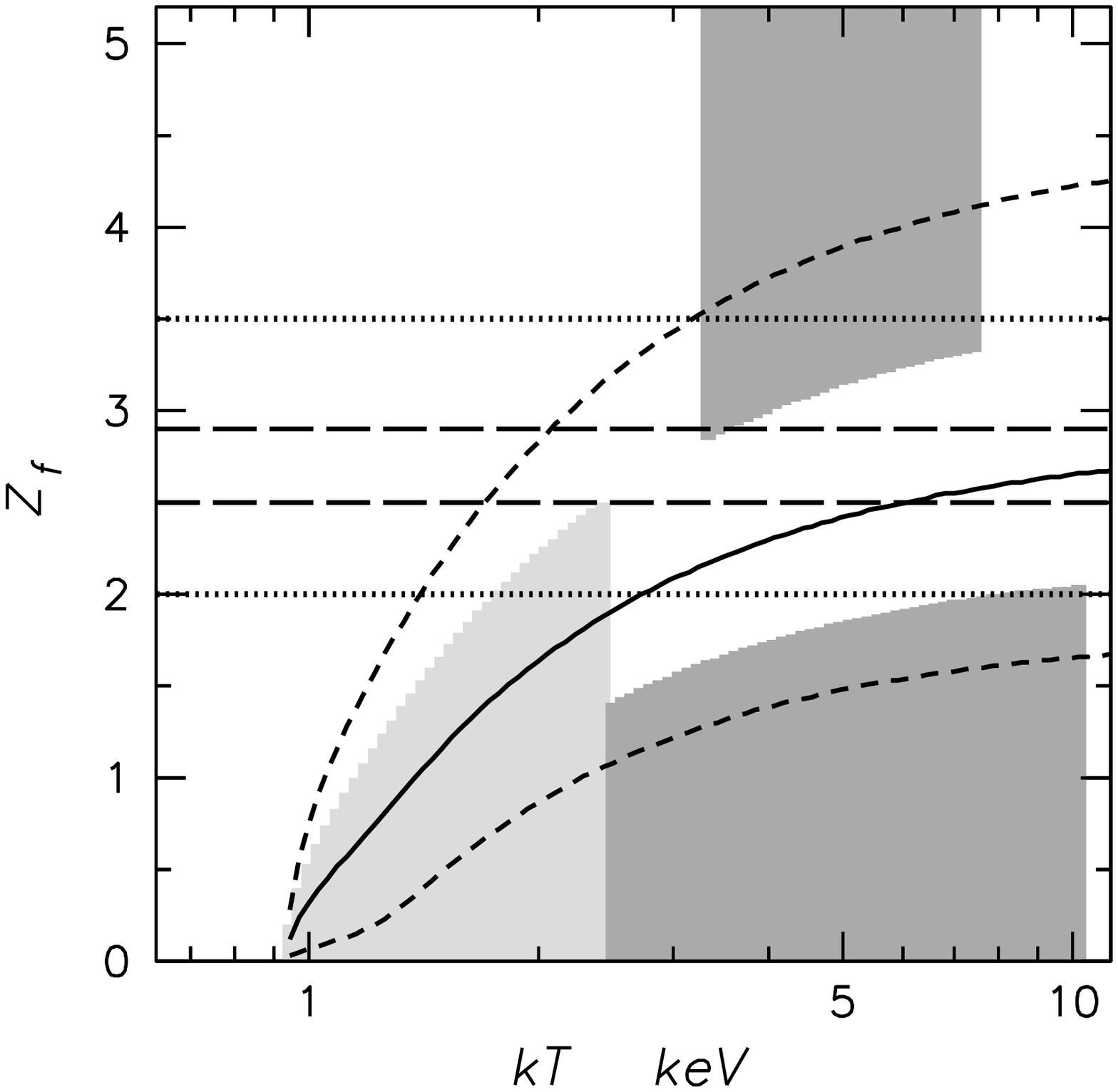}\hfill

\figcaption{ {\it Left panel.} Entropy of the system derived at equal
enclosed mass ($3\times10^{13}$\msun) plotted against the system
temperature. Dashed line represents a prediction from shock heating
artificially scaled up (by a factor of 2) to reproduce the group
measurements. Two dotted lines separate the sample into three
groups. Systems with entropy exceeding 300 keV cm$^2$ are assumed to achieve
the highest preheating level, the systems with entropy below 200 keV cm$^2$
are counted as forming before the high entropy level is achieved and the
rest of the systems are considered unconstrained.  {\it Right panel.}
Virialization redshift of the first $3\times10^{13}$\msun in clusters vs the
observed temperature. Solid-line curve shows the location of 50\% of the
systems from numerical simulations of structure formation (Lacey \& Cole
1993), calculated using the formulae in Balogh \etal (1999) for $\Omega_{\rm
m,\,o}=0.3$. Short dashed-line curves show the location of formation
redshift for 10\% and 90\% of systems. Shaded zones on both panels indicate
whether or not the high level of preheating has been achieved. Light
shadowing indicates that 75\% of groups has been formed after and two dark
shadows indicate that 25\% of cluster cores form before and another 25\%
form after the high level of preheating has been achieved. Two long-dashed
lines show a variation in the redshift consistent with the data, while two
dotted lines indicate the limits where the model prediction becomes strongly
inconsistent with the data.
\label{fig:entm}\label{fig:zf}
}
\end{figure*}

A comparison of the entropy for this enclosed mass is presented in
Fig.\ref{fig:entm} (left panel) for each\footnote{In this approach we are
limited to samples analyzed by Finoguenov (FDP, FAD and this paper). This
excludes a few hot clusters, introduced in Fig.\ref{fig:ent}: A3558, A119,
A3571, A401, A3266.} of the clusters and groups (38 systems in total) shown
in Fig.\ref{fig:ent}. An entropy level of 350 keV cm$^2$ is typical for
systems with temperatures below 3 keV. A dramatic change in the behavior of
entropy vs temperature occurs above 3 keV, exhibiting a trend towards lower
entropy for hotter systems. The dashed line in the left panel of
Fig.\ref{fig:entm} shows the entropy attributed to shock heating, calculated
for the mean formation redshift ($z_f$) of groups and clusters (the solid
line on the right panel of Fig.\ref{fig:entm}) and up-scaled by a factor of
two (it is off the graph when the normalization of PCN is adopted). It is
clearly seen that the observed trend is different from the entropy behavior
predicted from shock heating.

If we assume a similar history of gas preheating in groups and clusters,
then the trend in Fig.\ref{fig:entm} (left) requires a growth in the gas
entropy with (Hubble) time, with the entropy increasing from cluster cores
($z_f\sim2.7$) to the outskirts of groups ($z_f\sim0.$). The entropy in poor
clusters (kT$\sim3$ keV; $z_f\sim2.0$ ) reaches a level, comparable to that
found in groups. This picture implies both prompt preheating at high-z and a
lack of preheating at low-z. Thus we suggest that the preheating level,
observed through the entropy, will not increase in other observational tests
and corresponding therefore to the {\it entropy ceiling}.

An entropy floor has been suggested based on the observation of excess
entropy in cores of groups compared to the gravitational heating
(PCN). Introduction of the term "ceiling" would represent the upper limit on
the preheating, which is a characteristic of any realistic model of the
preheating, based on the limited energetics provided by putative sources of
preheating, such as SNe. Term entropy ceiling is also appropriate for
scenarios ignoring the preheating, where one can reformulate a key
assumption of Voit \& Bryan (2001), to read 'a ceiling in the preheating
value for entropy is much lower than the entropy floor'. Since we prove
otherwise, the scenario of Voit \& Bryan (2001) in its original formulation
is dismissed. Below, we present a more detailed consideration of the
establishment of the entropy ceiling, after first noting other supporting
observations.
 
The entropy level for the gas accreting onto the outskirts of clusters can
be estimated from the entropy level of filaments connecting the
clusters. Durret et al. (1998) found that a galaxy filament close to A85
exhibits an elongated X-ray emission in the 0.4--2 keV band. From the
reported flux of $2.7\times 10^{-13}$ ergs cm$^{-2}$ s$^{-1}$, estimating
the flux collecting area as $60^{\prime}\times 10^{\prime}$, and taking the
short axis for the third dimension, we obtain a density of the X-ray
emitting electrons of $\sim3\times10^{-5}h_{50}^{1/2}$ cm$^{-3}$. For
emission to be detected in the 0.4--2 keV band, the gas temperature should
exceed 0.3 keV, corresponding to an entropy of $>310$ keV cm$^{2}$. Briel
and Henry (1995) used the ROSAT all sky survey data to combine the
measurements of regions connecting physically close clusters of galaxies,
identified in Huchra et al. (1990). They obtain a $2\sigma$ limit on the
emission of $3.9\times 10^{-16}$ ergs cm$^{-2}$ s$^{-1}$. With the
assumption of a 0.5 keV temperature for the gas in filaments, justified by
observation of metals in clusters of galaxies (Finoguenov, Arnaud, David
2001; hereafter FAD), Briel and Henry (1995) derived a limit on the gas
density\footnote{assuming an iron abundance of 0.3 solar} of
$7.4\times10^{-5}h^{1/2}$ cm$^{-3}$, which for $h=0.5$
translates into a lower limit on the entropy of 360 keV cm$^{2}$. Thus we
conclude that also in rich clusters, the entropy of the accreting gas is
higher than the entropy floor.

Observations of a newly discovered class of objects, fossil groups or OLEGs
(Ponman \etal 1994; Vikhlinin \etal 1999b) also show a high level of
preheating (entropy is 350 keV cm$^2$ at radii of the enclosed mass of
$3\times10^{13}$ \msun, Vikhlinin, A. 2001, private communication).

Interpretation of the $M_{gas}-T$ relation for clusters of galaxies also
requires a higher preheating level of $\sim400$ keV cm$^2$ (Babul et
al. 2002, McCarthy et al. 2002). The issue of the level of preheating in the
theoretical models is still unsettled, as e.g. model of Dos Santos and
Dor\`e (2002) does not require such a high entropy.

\subsection{Dating the preheating using a prescription for cluster formation}

With recent advances in high-resolution numerical simulations, cosmic
microwave background observations, the SN Ia Cosmology project, and surveys
of clusters of galaxies, studies of the evolution of the Universe have
achieved a convincing picture with a large degree of detail, involving dark
energy and dark matter. These constrain the geometry of the Universe, its
dynamics and the large-scale structure of the dark matter distribution.
Thus, from the difference in the formation epochs for the outskirts of
groups compared to cluster cores, we can infer the redshift over which the
entropy ceiling has been maintained.

In order to estimate the redshift when most of the preheating occurs, we
show in Fig.\ref{fig:zf} (right) the analytical and numerical results for
the formation of clusters (Lacey \& Cole 1993), following the formulae
presented in Balogh \etal (1999) for $\Omega_{\rm m,\,o}=0.3$. With a solid
line we indicate the mean of the cluster formation redshift distribution and
with dashed lines the corresponding 10\% and 90\% values. To represent the
measurements on the left panel, we divide the systems into three groups
according to the entropy as low, intermediate and high, $<200$, $200-300$,
$>300$ keV cm$^2$, respectively. We attribute the gas with entropy lower
than 200 keV cm$^2$ to that accreted before the establishment of the entropy
ceiling, while the gas with entropy higher than 300 keV cm$^2$ is considered
to achieve the entropy ceiling before accretion.  Analysis of the left panel
reveals that at least 75\% of the low temperature systems have achieved a
high level of entropy and therefore have formed later than the entropy
ceiling was established. This is shown by the light-grey shaded zone in the
right panel.  The dark-shaded zone at higher temperatures indicates the
presence of both high (at least 25\% of clusters) and low (at least 25\% of
clusters) entropy systems. Low entropy systems are treated as formed before
the entropy ceiling was established.


\begin{figure*}
\begin{center}
\includegraphics[width=3.6in]{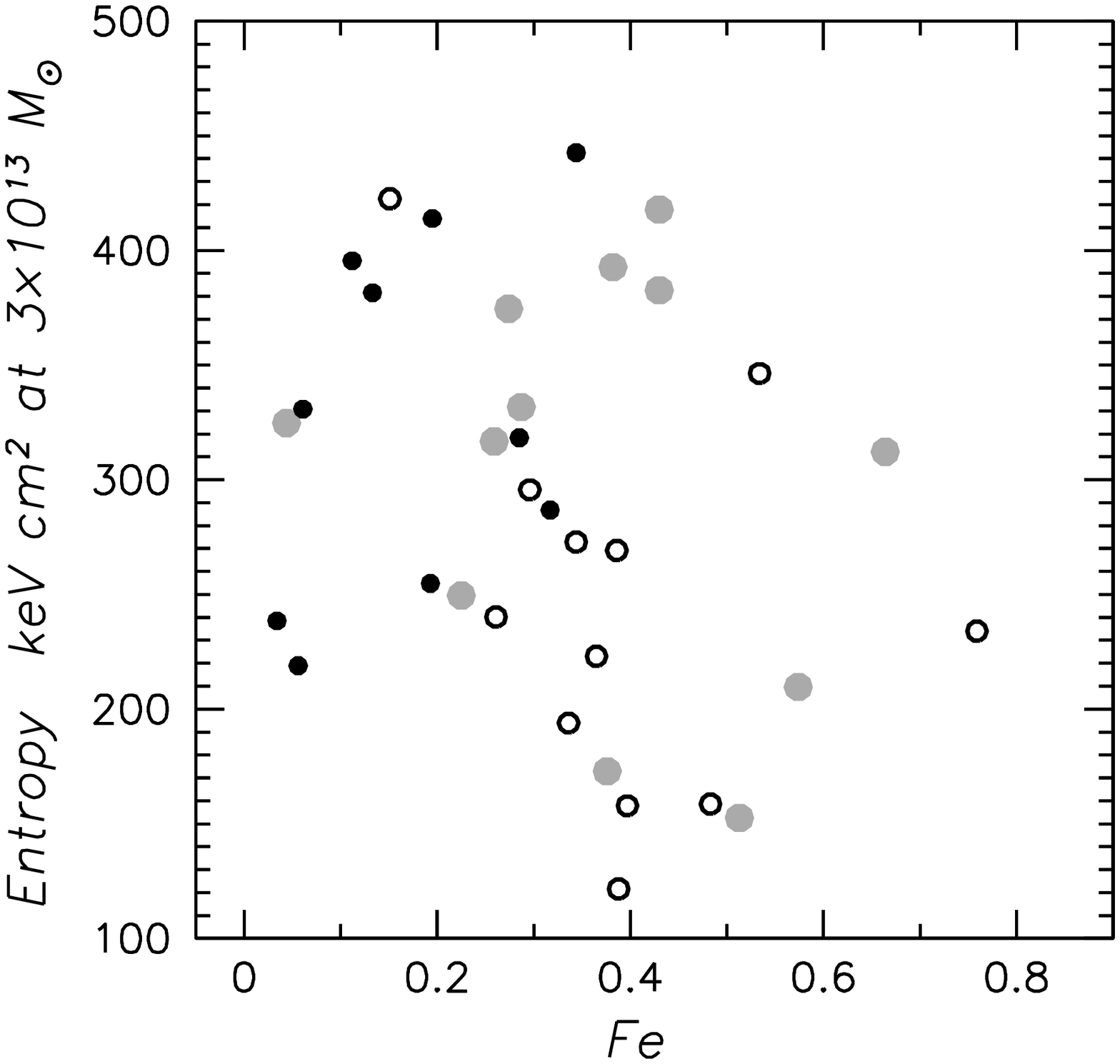}\hspace*{-1.5cm}\includegraphics[width=3.6in]{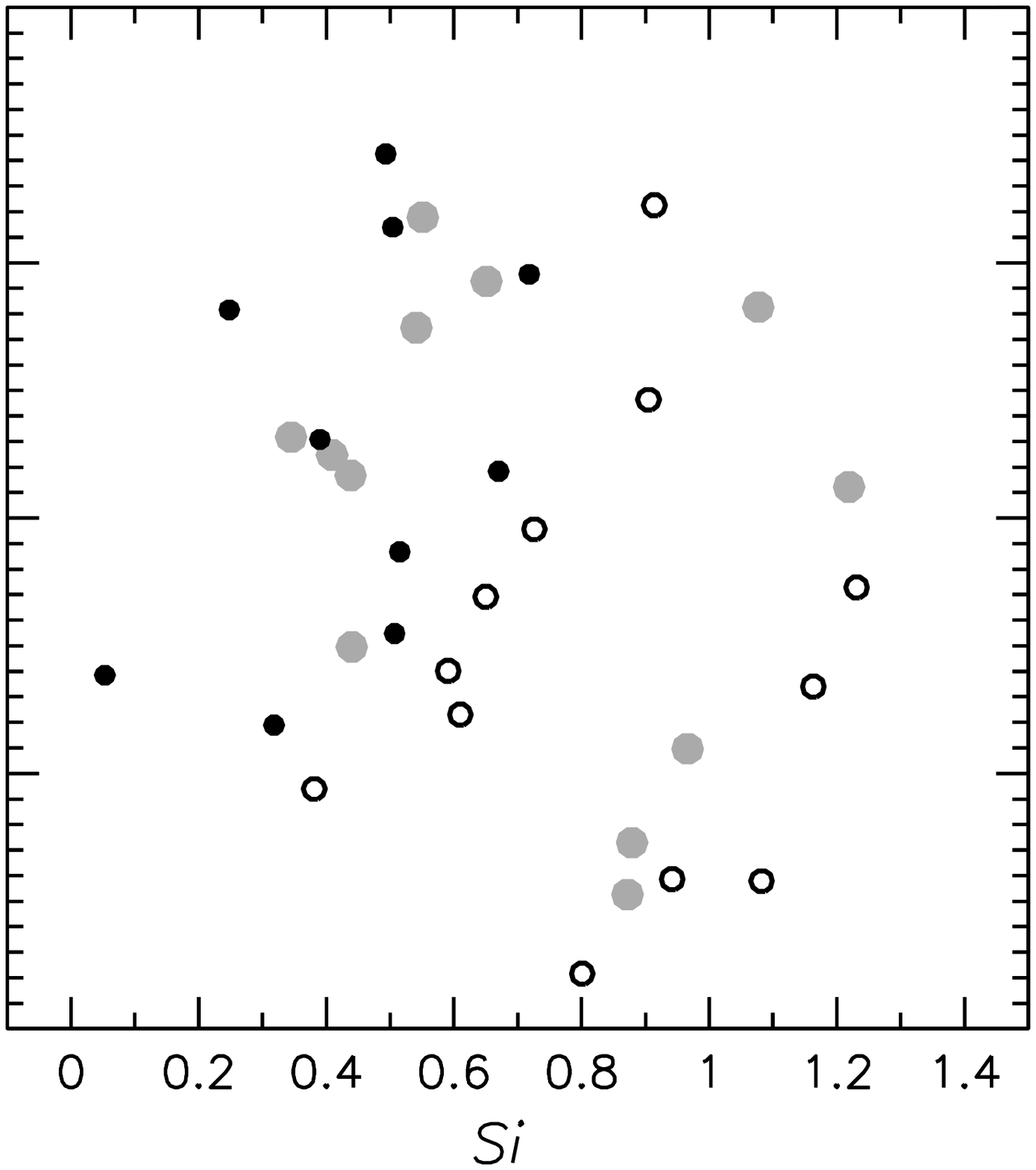}
\end{center}
\figcaption{Search for a correlation of the entropy measured at radii of
equal enclosed mass ($3\times10^{13}$ \msun) vs Fe {\it (left)} and Si {\it
(right)} abundance. Groups are shown with filled black circles, cool
clusters (2 keV $< kT < 3.5$ keV) with filled grey circles and hot clusters
with open circles.
\label{fig:ent-ab}
}
\end{figure*}

As is shown in Fig.\ref{fig:zf} (right), at redshifts greater than 2.8--3.5
the accreted gas was characterized by low entropy, while starting at
redshifts 2--2.5, the accreted gas is on a high adiabat. Duration of this
transition period is very short, $0.3-1.4$ Gyr.  The limits on the epoch for
feedback processes, based on the entropy measurements, crucially depend on
the way such feedback occurs. However, a lower limit of $z_f>2$ on the
feedback redshift should hold in any scenario for external (pre) heating.
Yet, the requirement of high redshift for the feedback processes responsible
for the preheating further supports the dominant role of bulges and is an
important input to modeling galaxy formation in high-density
environments. This conclusion is in agreement with observational evidence of
no star-formation in early-type galaxies since a redshift of 2 (Ellis et
al. 1997; Bender et al 1998).


\subsection{Relating enrichment to the preheating}\label{Z2S}

In this section we attribute the establishment of the preheating level for
the gas to action of the galactic winds. Within this scenario, the metals
found in the hot intracluster medium can serve as an additional constraint
on the preheating, as they are directly related to the amount of energy
released into the ICM (FAD).

In the previous section we concluded that a high preheating level
corresponding to an entropy of about 350 keV cm$^2$ is reached at large
radii in all systems. On the other hand, the observed level of heavy element
enrichment varies between groups and clusters leading to the suggestion that
galaxy winds deposit more energy in clusters (FAD). Moreover, such
enrichment was found to be constant over the cluster (FDP), while a simple
interpretation of the results of the previous section leads to the
expectation of an abundance profile rising with radius. To better illustrate
this point, we show in Fig.\ref{fig:ent-ab} that the entropy at a radius of
equal enclosed mass does not correlate with either the Fe or Si abundance.

\begin{figure*}
\hfill\includegraphics[width=3.6in]{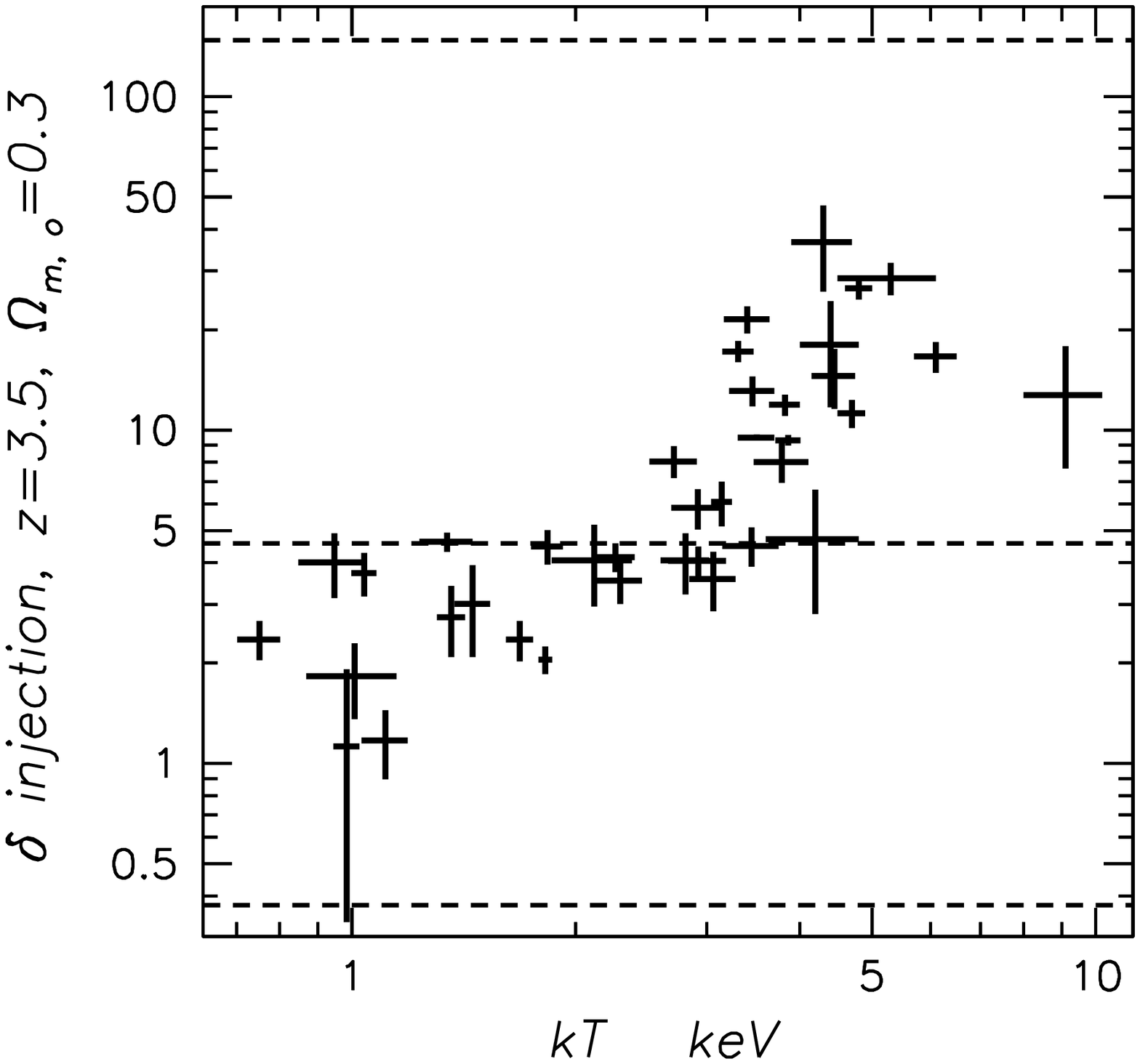}\hfill\hfill\includegraphics[width=3.6in]{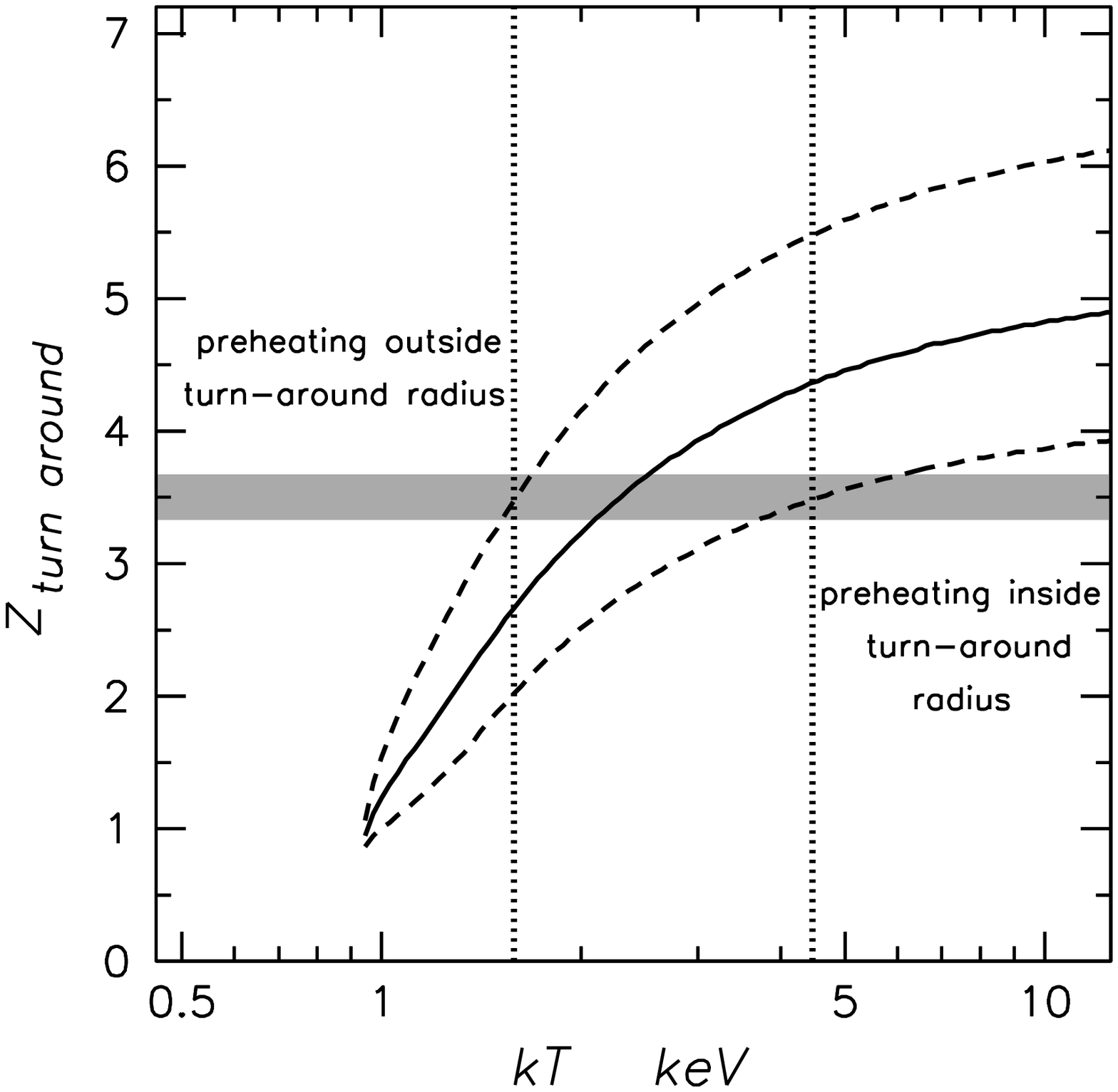}\hfill

\figcaption{ {\it Left panel.} Overdensity at which galaxy winds have been
ejected, calculated from the requirement to match both amount of energy
associated with metal enrichment as well as the observed entropy level (at
radii of $3\times10^{13}$ \msun\ enclosed total mass), assuming redshift of
star formation of 3.5, and using the scaling appropriate for the
$\Omega_{\rm m,\,o}=0.3$. The dramatic difference in the entropy between
groups ($kT<2$ keV) and clusters, translates into the statement that
preheating of groups is complete at the turn around radius, while preheating
of cluster cores occurred at higher overdensities (yet outside the virial
radius). Three dashed lines mark overdensities indicating three stages of
cluster formation, linear growth of initial density fluctuation
($1.69/(1+z)=0.38$), turn around (4.6) and virialization (150).  {\it Right
panel.}  Redshift of turn around, derived by taking half of the formation
time from Fig.\ref{fig:zf}. Grey line shows a consistency with the
assumption in the left panel on the redshift of energy injection. Dotted
lines indicate the temperature range for systems at which the transition in
the derived overdensities occurs. It matches well the analytical prediction,
shown by solid line. Dashed lines show the expected spread in turn around
redshifts, drawn at 75\% confidence level.
\label{fig:ent-rv-ab}
}
\end{figure*}

Apparently, at least part of the enrichment is not associated with the
preheating (\eg\ SN Ia enrichment occurs at later times, which determines
the Fe abundance in clusters at radii corresponding to
Fig.\ref{fig:ent-ab}). Furthermore, Chandra and XMM-Newton observations do
not significantly change the picture of the distribution of SN II products
(David \etal 2001; Matsushita \etal 2002; Molendi \& Gastaldello 2001;
Tamura \etal 2001).

We have concluded that preheating of the gas reaching the virial radius
should be complete by a redshift of $2-3.5$. Since it takes some time before
the preheated gas will be accreted, the redshift of the feedback is somewhat
higher.  Let's choose a redshift of 3.5 for the energy injection and
calculate the overdensity at which energy injection takes place in order to
reproduce the observed entropy for radii corresponding to $3\times10^{13}$
\msun\ enclosed total mass (it will be possible for us to check the validity
of a particular assumption for the redshift). We account only for the
effects of star-formation, estimating the amount of energy from direct
observations of the chemical enrichment, as calculated by FAD. (For
simplicity we will take an average for systems with similar total mass,
which is 0.4 keV/particle for groups, reaching 1.2 keV/particle for rich
clusters, and analytically expressed as $0.44\times (kT/keV)^{0.5}$
keV/particle). In Fig.\ref{fig:ent-rv-ab} (left) we determine the
overdensity where this amount of heat would lead to the observed entropy
level. We assume the electron density at critical density of
$4.1\times10^{-7}$ cm$^{-3}$ (taking $n_e/n_{ion}=1.09$, $m_{ion}=1.27m_p$,
$f_{baryon}=0.17$). For the chosen redshift of 3.5, the scaling suitable for
$\Omega_{\rm m, o}=0.3$ results in $\overline{n_e}=2.1\times10^{-5}$
cm$^{-3}$ corresponding to the critical density. There is a slight
underestimate of the overdensity since we take the density at which we eject
the heat as the overdensity. We can take into account at least galaxies
embedded in the preheated gas, if we introduce an increase in the
overdensity by 20\% due to a gas-to-bulge ratio of 5. Variation in the
assumption for the redshift changes both left and right panels of
Fig.\ref{fig:ent-rv-ab} such that matching between the temperature of the
system at turn-around on both panels is maintained for a broad range of
redshifts. A requirement for the minimum derived overdensity to be higher
than linear approximation to the growth of primordial density fluctuations
($1.69/(1+z_{injection})$) rules out redshifts $>4$.

Fig.\ref{fig:ent-rv-ab} (left) shows that many systems, including the
outskirts of groups, are characterized by SN enrichment at turn-around,
while cluster cores are characterized by enrichment occurring primarily at
higher overdensities. In Fig.\ref{fig:ent-rv-ab} (right) we present a
calculation for the turn-around redshift that corresponds to the
virialization epoch, shown in Fig.\ref{fig:zf} (right). The turn-around time
is calculated as half the collapse time, defined in Fig.\ref{fig:zf}
(right).  This plot allows us to check our assumption for the star-formation
redshift. For the chosen redshift of 3.5, the gas presently accreting onto
the outskirts of groups should not necessary be inside the turn around
radius. Thus, there is no contradiction in our estimate that injection
overdensities are less than 5. In our approach, the preheating level in
groups is defined by the overdensity of the material destined to collapse
($0.4<\delta<4.6$). Constant entropy behavior at radii of equal enclosed
mass is therefore a consequence of a narrow range of possible overdensities
and a small scatter in the metal production. For the cores of clusters with
temperatures above 4.5 keV, the high overdensity shows that the metal
enrichment occurred during the advanced stages of cluster core formation. A
more refined approach, would allow for the expectation that most star
formation takes place at larger redshifts in the regions which now form
cluster cores, compared to the outskirts of groups.  This will reduce the
range in derived overdensities at which preheating is inferred to happen,
and in particular it will increase the very low overdensities inferred for
group outskirts. Another effect arises from the importance of SN Ia in the
production of metals in groups, generally understood as enrichment at high
density regions, thus reducing the level of preheating. Both effects just
mentioned are of the same factor of few, compensating each other. Two other
uncertainties related to our calculation are the energy-to-temperature
transformation, as discussed in Lloyd-Davies et al. (2001), which we assume
to be $3/2 kT$ and the energy released per SNe, which we take as
$1.5\times10^{51}$ ergs in our calculation.

In summary, a gradual increase in the entropy of the accreting gas in {\it
any} system could be characterized by a single preheating epoch, with the
cores of systems being already at an advanced stage of collapse. Flat
$\alpha$-element abundance profiles observed in clusters of galaxies (FDP)
are naturally reproduced, since at a derived range of overdensities there is
no segregation between gas and galaxies. A smoking-gun to define the epoch
and level of preheating would be an observation of the thermodynamic state
of baryons in the infall regions of clusters, either in emission or
absorption.

\subsection{Comparison to existing approaches}

Borgani et al. (2002) have carried out simulations where various preheating
entropy levels are reproduced. One difference between their approach and
ours lies in the attitude regarding the high density regions. Borgani et
al. (2002) rises the entropy of all particles in the system to the same
level, which requires a much higher level of energy injection, as well as
preferential deposition of metals into high-density regions. The highest
overdensity where we introduce preheating is 30. Since in our approach, we
do not add energy to high density regions, we allow star-formation to
continue in Ly break galaxies. Still, we are able to preheat the bulk of the
gas mass, which resides at lower overdensities. As a consequence, our
approach is different to that of Borgani et al. (2002) by requiring the
dense regions to cool, either by forming galaxies outside the virial radius
or via early cooling flows, as proposed by Voit and Bryan (2001). An
advantage of our approach lies in lowering the requirements for the total
energy required to achieve the observed preheating level.


One firm conclusion that can be derived from our work is that for
star-formation to be responsible for the preheating in groups and clusters,
the energy ejection should occur at an early epoch and preferentially in a
low-density environment. As star-formation itself usually peaks in
high-density regions, it might seem that such a requirement could rule out
this scenario. However, examples of local star-formation show that galaxies
are very good in building chimneys expel most of the energy and metals
(self-regulating star-formation; Larson \& Dinerstein 1975). Also if energy
is equally deposited at various densities, the low-entropy gas experiences
large radiative losses and falls back onto the galaxy and is consumed in
subsequent star-formation. Since high overdensities surrounding the galaxy
may not become part of the ICM (Larson \& Dinerstein 1975), it is quite
logical to assume that the energy associated with metals in the ICM comes
from the high-entropy tail of the energy ejection (FAD).

Injecting the energy into overdensities lower than 5 requires that the
infall of the gas into the group horizon occurs after the gas is
preheated. Also, when we calculate the amount of metals in groups, we only
account for the filaments that fall in and not those that do not. This
explains why it is reasonable to assume no losses due to the removal from
galaxies by galactic winds of the part of metals found in clusters.

\subsection{Alternative sources of preheating}\label{s:imp}

To estimate the uncertainties in using entropy to constrain the feed-back
epoch, one can consider the influence of cooling flows and AGN heating, as
well as segregation between mass and gas inside the cluster virial radius.

The high level of entropy, associated with the entropy ceiling is beyond the
reach of models that invoke gas cooling (e.g. Voit \& Bryan 2001) to explain
the 100 keV cm$^2$ entropy floor. We have shown that the high entropy level
in the groups' outskirts is related to recently accreted gas, while lower
entropy levels would be more typical of an earlier epoch. One effect of gas
cooling is to reduce of the entropy of the gas, due to entropy carried out by
emitting photons. However, most of the cooling gas is expected to drop out at
the same location as the cooling, so part of the information originally
recorded in the entropy at centers of groups and clusters is lost due to
the conversion of the X-ray emitting gas into stars. Regarding the systematics
of using only the cores of clusters as probes, note that our systems are
sampled at a radial range outside the regions, where cooling is appreciable
and in addition we use only the hotter component in our analysis of cluster
cores.

AGNs are proposed as a source of the preheating, as they may provide much
higher energies per particle, compared to SNe (Wu, Fabian \& Nulsen
2000). Using the AGN luminosity function with a proper scaling by cluster
volume and overdensity, and an increase by a factor of 80 in the AGN
activity with the redshift (Miyaji, Hasinger and Schmidt 2000), the energy
radiated in X-rays over a Hubble time for all AGN corresponds to 0.2
keV/particle. AGN still could play a dominant role if, according to Fabian
\& Iwazawa (1999), most of them are absorbed and most of the emission is
re-radiated in the infrared. Assuming an efficient accretion mode forming
100\% of the local volume density of black holes, a corresponding estimate
of the radiated energy is 10-15 keV/particle. Observationally there is a
lack of such luminous absorbed AGNs. In particular the optical follow-up of
the XMM Lockman-Hole field and Chandra Deep Field South reveal that absorbed
AGNs are located at low redshifts (Hasinger 2002, Rosati et al. 2002).
Although the identification of AGNs is not yet complete, the unidentified
sources are too faint to account for efficient mode of accretion growth of
the black holes. Ciotti \& Ostriker (1999) also argue that feedback
processes prevent AGN from effective accretion.

Another problem, related to the suggestion of Fabian (1999) on AGN
importance is related to the conversion of radiated energy to kinetic
energy. The most effective mechanism, based on radiative pressure (Silk \&
Rees 1998), requires galactic column densities of $10^{24}$ cm$^2$, while
the bulk of baryons at high redshift have column densities less than
$10^{21}$ cm$^2$ (Ellison et al. 2001). Yamada \& Fujita (2001) considered
heating by AGN jets and concluded that jets are capable of providing about
0.3 keV/particle at redshifts below 3.  They assume that 1\% of normal
galaxies are radio galaxies and limit the redshift to be less then 3 using
the prediction for the deformation of the cosmic microwave background
(Sunyaev-Zel'dovich effect). One can elaborate this scenario by considering
the association of radio-loud quasars with elliptical galaxies (e.g. Laor
2000), which formally increases the preheating in clusters by up to an order
of magnitude (3 keV/particle).  However, Ho (2002) proposed that the
association of radio-loud sources with ellipticals is driven by a tendency
of local galaxies to have nuclei that are both underluminous and radio-loud,
in qualitative agreement with predictions of accretion disk theory. Thus,
such an association would not hold when ellipticals were star-forming and
the galactic wind was driven by star-formation. Indeed, as
acknowledged by Laor (2000), Stern et al. (2000) find no evidence that the
radio-loud fraction depends on optical luminosity for $-25>M_B>-28$ at
$z=2-4$.

Another aspect of AGN activity is related to the robustness of our approach,
where we use the cluster cores. We can consider Hydra-A as a
counter-example, where the AGN activity is among the strongest (McNamara et
al. 2000) at the present time, yet its entropy is amongst lowest ($134\pm33$
keV cm$^2$). As discussed in McNamara et al. (2000), no significant shock
heating is seen in Hydra-A. The reason is that the typical kinetic energy
associated with the observed AGN activity in clusters centers is on the
level of $10^{44}-10^{45}$ ergs/s (B\"oringer et al. 2002), while to
substantially change the entropy of the central gas mass of $10^{12}$ \msun,
typical of cluster cores, it requires kinetic input from AGN of $10^{48}$
ergs/s. Since the abundance of hot clusters is $10^{-8}$ Mpc$^{-3}$
(Reiprich \& B\"ohringer 2002), to systematically change the entropy of the
cluster cores it would require an AGN activity three orders of magnitude
higher than the $10^{-11}$ Mpc$^{-3}$ maximal volume density of AGNs with
$10^{47}$ ergs/s, as observed at high redshift. Note that AGNs of such
luminosity are capable of completely ionizing the torus and could not be
hidden from the observer (Nandra et al. 1995).

The last possible source of systematics we consider is the segregation of
mass and gas. In fact the gas fractions are similar between groups and
clusters at radii of the same enclosed mass and are $0.05h_{50}^{-1.5}$
(Sanderson et al. 2003) at radii of total mass of $3\times10^{13}$
\msun. Thus the radii of equal enclosed gas mass and equal total mass agree
in this case.

Observation of a systematic increase in the gas mass fraction with the mass
of the system (David et al. 1995; Jones, Forman 1999; Sanderson \etal 2002),
could be self-consistently explained within the preheating scheme (Babul et
al. 2002). However, if reduction in the gas fraction occurs before the shock
heating, the entropy resulting from the gravitational heating will be
higher. This possibility does not change the consideration of the entropy at
radii of equal enclosed mass, since the gas fractions are similar. The
gravitational heating fails to reproduce the trend in the entropy at radii of
equal enclosed mass, independent of the normalization, thus disfavoring the
leading role of the reduction in the gas fraction before the shock in the
entropy profiles.

\section{Conclusions}

We present the temperature, entropy and heavy element abundance measurements
for nine groups of galaxies and compare their entropy profiles to a larger
sample of groups and clusters of galaxies. We conclude

\begin{itemize}

\item Analysis of groups at radii exceeding 150 kpc generally reveals a
decrease with radius in the gas temperature. In the best-resolved case,
NGC3268, this corresponds to a flat entropy profile at the level of $400$
keV cm$^2$.

\item Comparison of the scaled entropy at an overdensity of 500 between
groups and clusters, still reveals deviations, that require a high level of
preheating ($400-500$ keV cm$^2$).

\item Comparing the entropy in groups and clusters at radii with a similar
enclosed mass (chosen as $3\times10^{13}$ \msun), we conclude that the
entropy of the preheated gas does not exceed $400$ keV cm$^2$. Thus, we
establish a ceiling for the preheating entropy.

\item Comparing our results to the prediction of structure formation,
we find that accretion of the gas preheated to $400$ keV cm$^2$ did
not occur before a redshift of 3.5, but is typical of accretion for
redshifts below 2. Thus, not only the metal enrichment of clusters is
associated with bulges, but also the timing of the preheating suggests
the dominant role of bulges in the feed-back process.

\item If an increase in the level of preheating is due to secular
star-formation activity, duration for energy ejection is constrained to a
narrow time interval (redshifts 2--3.5). Distribution of the level of
preheating cannot result from the progressive release of SN energy from
galaxies, but can be produced by the nearly instantaneous release of SN
energy into a density gradient.

\item Assuming an injection (or star-formation) redshift of 3.5 and scaling
the feed-back by the observed metal content in groups and clusters of
galaxies, (0.4 keV/particle in groups, 1.2 keV/particle in clusters), the
typical overdensity of gas receiving the preheating must be below 5 for
outskirts of groups and 5--30 for cluster cores.

\item Using the luminosity function of AGNs, we show that AGNs cannot be the
dominant source of the preheating. A concentration of obscured AGNs at low
redshift does not support the suggestion of an unseen growth in black-hole
masses.

\item The "energy crisis", related to the preheating with a limited
source of energy, such as SNe, is solved in our approach by allowing the
high-density regions (overdensities higher than 50 at redshift of 3.5)
to cool out and form stars, while we are
able to preheat the dominant gas fraction which resides at low densities.

\end{itemize}

\section*{Acknowledgments}

AF thanks Andrey Doroshkevich for communicating the idea of using
Lagrangian units in entropy studies, Paolo Tozzi for helpful referee
report, Stefano Borgani, Richard Bower, Michael Loewenstein and Chris
Metzler for discussion of effects of star formation on the observed
entropy level, Arif Babul for useful comments on the manuscript,
Guenther Hasinger for enlightening discussion on AGNs. Simulations
required for ASCA data reduction were performed using the computer
facilities of Astrophysikalisches Institut Potsdam. AF acknowledges
receiving the Max-Plank-Gesellschaft Fellowship. The authors
acknowledge the devoted work of the ASCA operation and calibration
teams, without which this paper would not be possible. This work was
partially supported by NASA grant NAG5-3064 and the Smithsonian
Institution.

\appendix{}

\section*{A. Rescaling the entropy from gravitational heating}

The scaling of the entropy, chosen in Fig.\ref{fig:ent} (right)
accounts for the increasing strength of the shock with the cluster
temperature. To scale the theoretical prediction for the shock heating (Eke,
Navarro, Frenk 1998; PCN), stated at $0.1r_{v}$, we rescale the entropy to
the radius of the overdensity of 500, using the $S\sim r^{1.1}$ (such entropy scaling on radius is also suggested in Tozzi \& Norman 2000),
$\rho_{m}\sim r^{-2.4}$, found in the same simulations (Eke, Navarro, Frenk
1998) and use the $M-T$ relation of Evrard et al. (1996) to consistently
remove the entropy scaling on temperature. We scale the prediction to
$h=0.5$, and account for the fact that in the low density Universe, the
shock occurs at lower overdensity and is even lower when lambda is taken
into account (100 compared to 178, Pierpaoli, Scott, White 2001), so the
shock temperature is lower, but the density is even lower. We obtain

\begin{equation}
{S(r_{500})\over M_{500}^{2\over3}} \approx
189\left({f_{gas}\over 0.17}\right)^{-{2\over 3}}\left({1+\delta\over
    100}\right)^{-{2\over3}}\; {\rm cm}^2
\end{equation}

where $M_{500}$ is in units of $10^{13}$ \msun.

Compared to the original

\begin{equation}
S(0.1r_{v}) \approx 45 \left({T\over keV}\right) \left({f_{gas}\over 0.06}\right)^{-{2\over3}} h^{-{4\over 3}}\; keV\; cm^2
\end{equation}

Contribution of various scaling are following
\begin{equation}
S(r_{500})=8.3 S(0.1 r_{v})
\end{equation}

\begin{equation}
\left({f_{gas}\over 0.06}\right)^{-{2\over3}} h^{-{4\over 3}} \equiv 1.26 
\left({f_{gas}\over 0.17}\right)^{-{2\over3}} h_{0.5}^{-{4\over 3}}
\end{equation}

where the PCN scaling of $h^{-4/3}$ resulting from $\rho_{crit}\sim h^2$
is balanced by the scaling coming from $f_{gas}\sim h^{-3/2}$.

\begin{equation}
\left({1+\delta\over 178}\right)^{-{2\over 3}}\equiv 1.47\left({1+\delta\over 100}\right)^{-{2\over 3}}
\end{equation}

In Fig.\ref{fig:ent} (right) we show the prediction for shock heating and
include the influence in the redshifts of cluster formation
($S\sim(1+z_f)^{-1}$ in CDM, while we use a prescription for density and
virial overdensity evolution in $\Lambda$CDM).

\section*{B. Heavy element abundance determination}

All abundances
are given relative to the solar photospheric values in Anders \&
Grevesse (1989).  The abundances of He and C are fixed to their solar
values.  The remaining elements are combined into four groups for
fitting: Mg; Si; S and Ar; and Ca, Fe, and Ni. Large systematic
uncertainties in the ASCA SIS at low energies prevent us from
determining the O abundance. 

A specific problem in the analysis of this sample, compared to the
cluster samples of FAD and FDP, is the low signal-to-noise ratio for
most of these groups. Given the spectral shape for 1 keV plasma, in
most of the systems, no continuum is detected above 2 keV in the outer
annuli. This limits the S abundance determination to only a few
systems (NGC2300, NGC507 and NGC3268) and increases the uncertainties
in the measurement of the Si abundance. Low signal-to-noise also makes
the Mg abundance determination difficult, although temperature and Fe
abundance determinations are robust.

\begin{figure*}

   \includegraphics[width=2.0in]{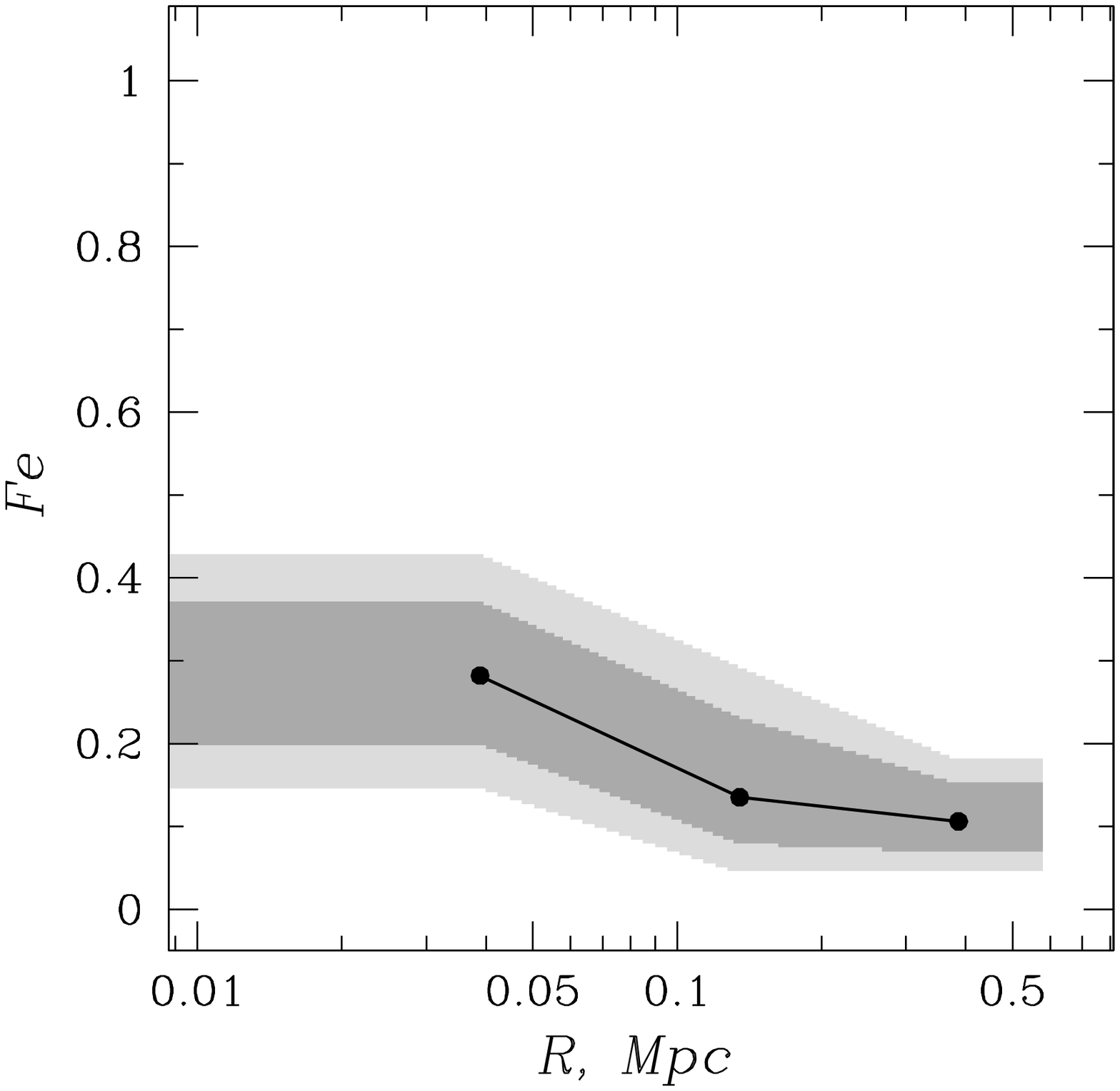} \hfill
   \includegraphics[width=2.0in]{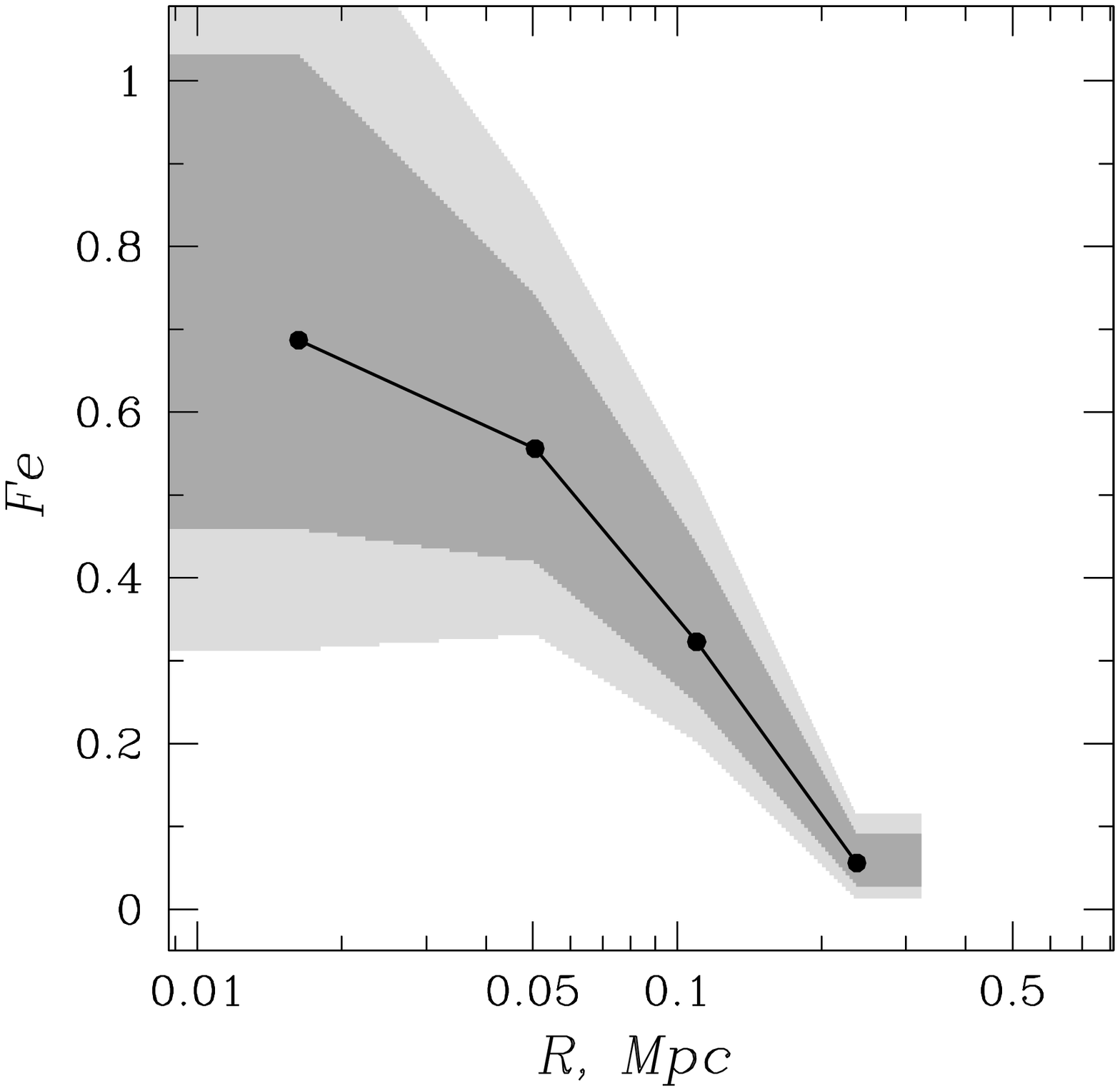} \hfill 
  \includegraphics[width=2.0in]{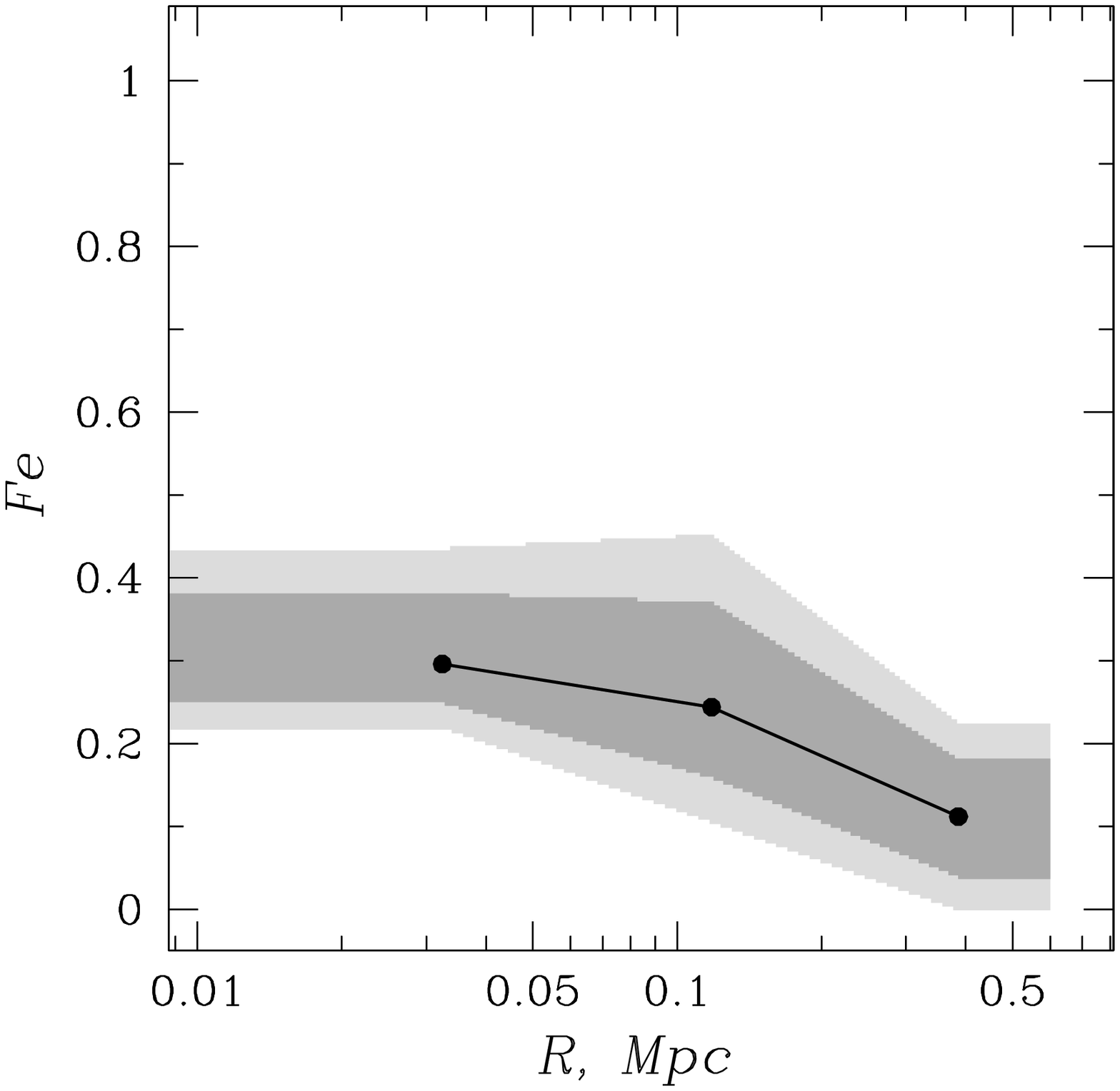} 
 
   \includegraphics[width=2.0in]{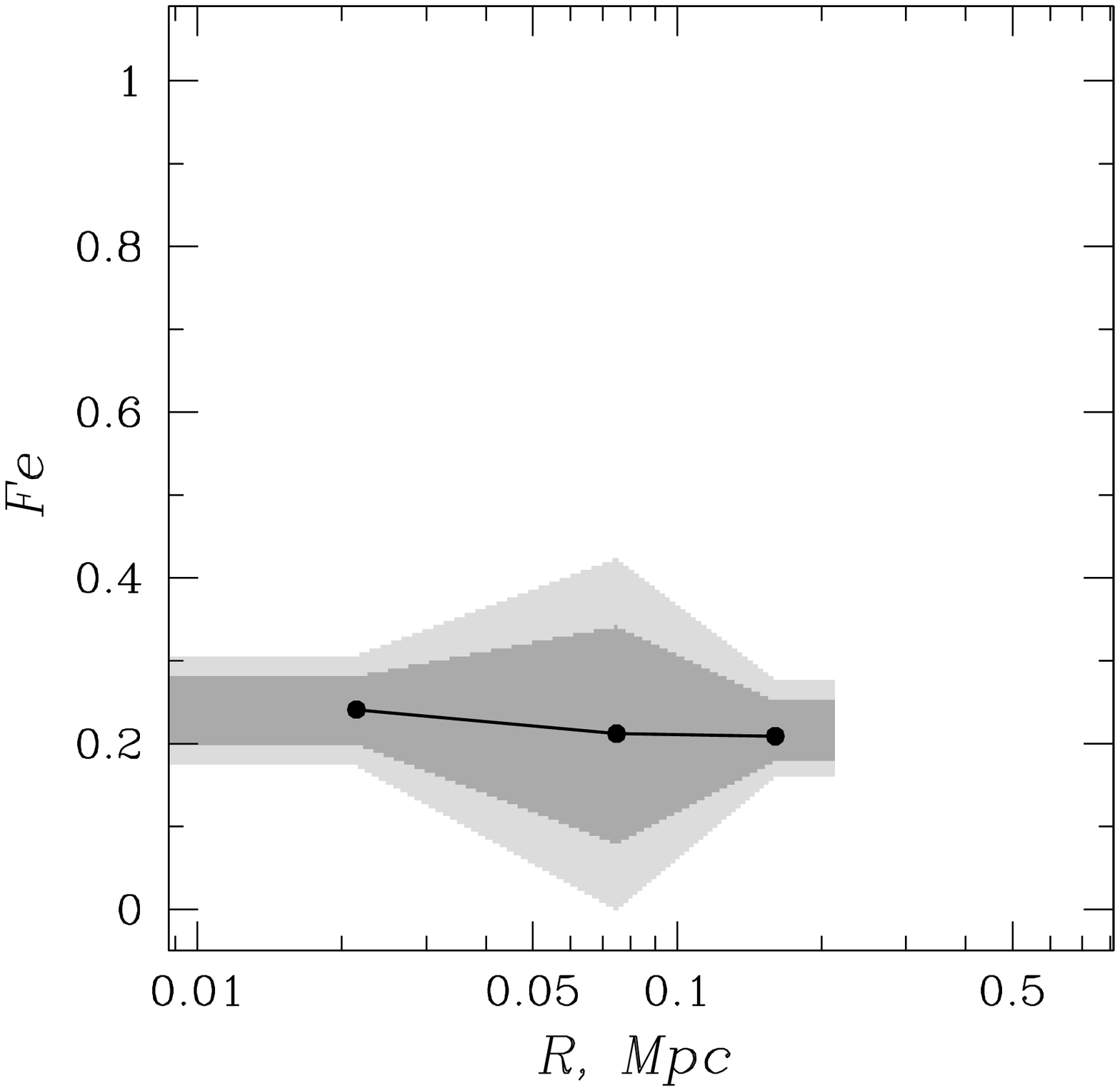} \hfill 
  \includegraphics[width=2.0in]{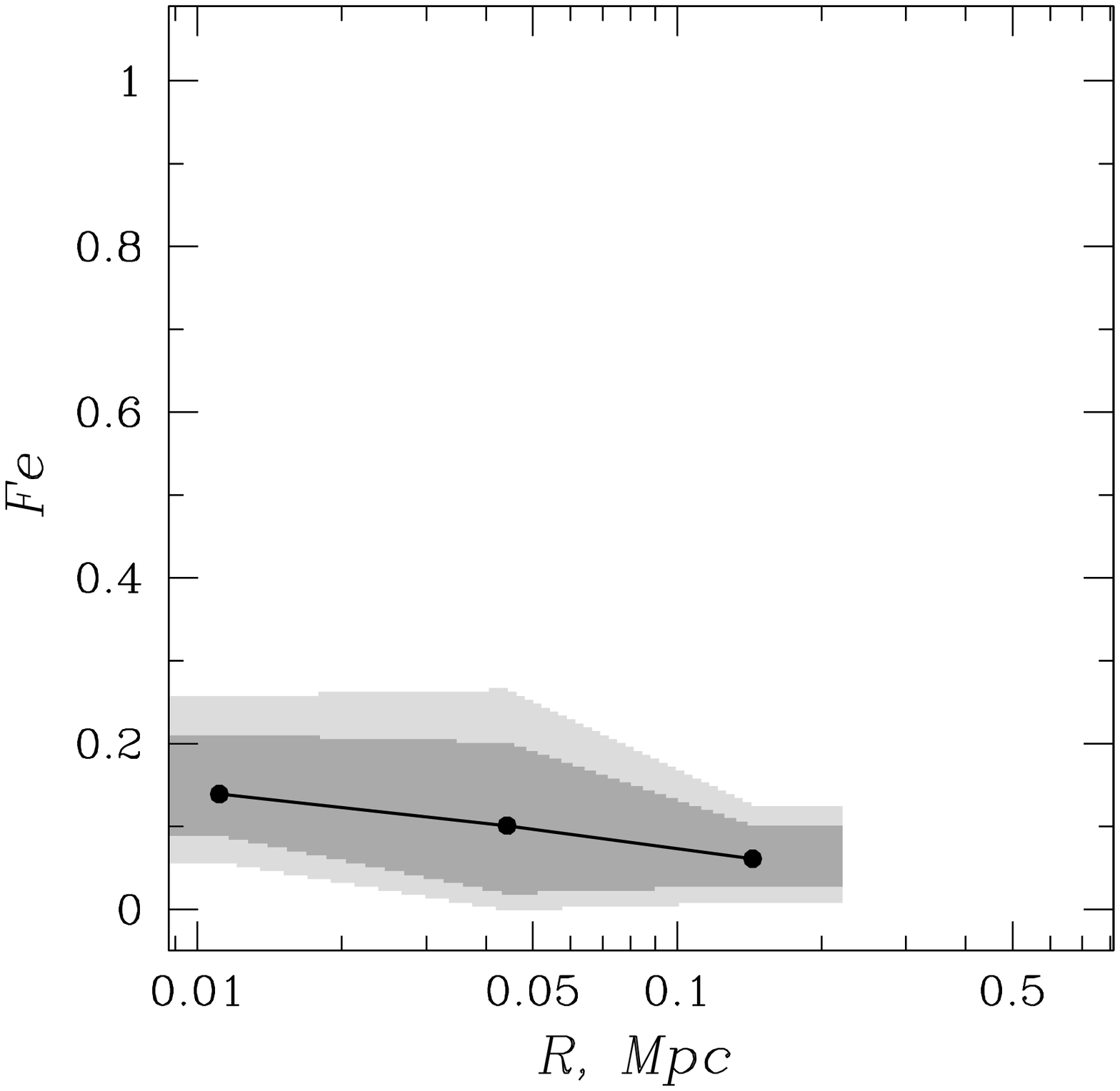} \hfill 
  \includegraphics[width=2.0in]{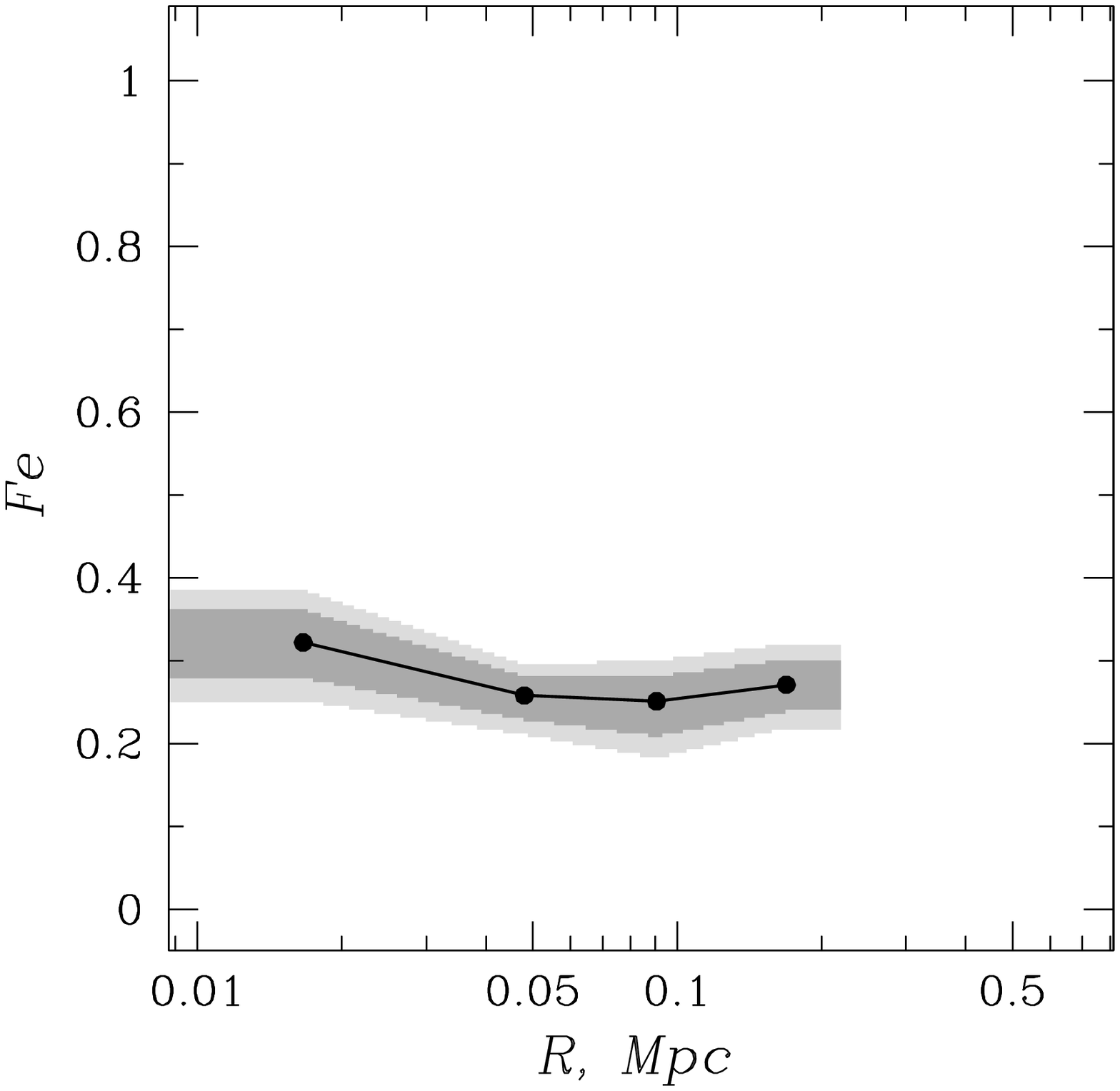} \hfill 

   \includegraphics[width=2.0in]{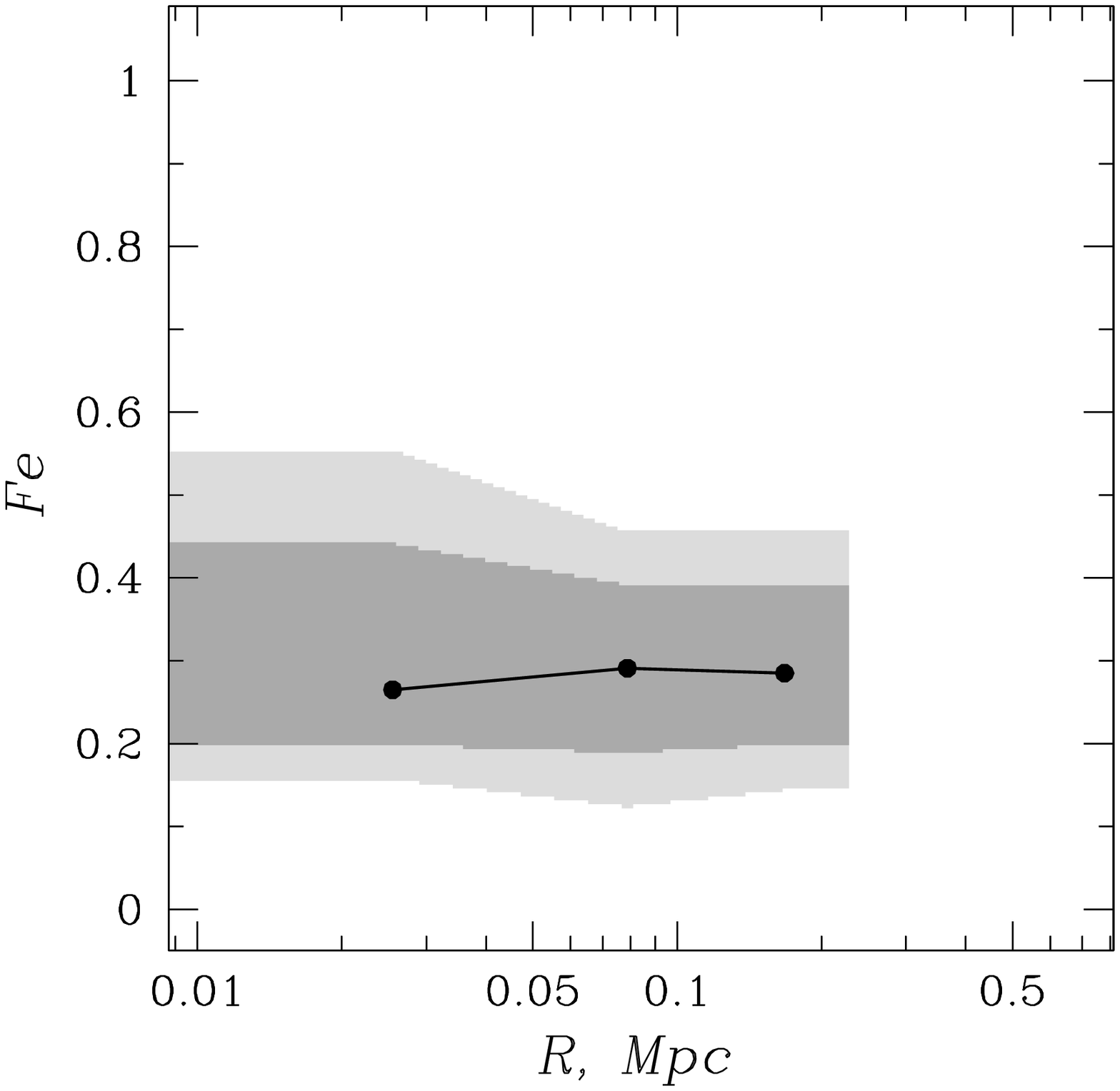}   \hfill 
  \includegraphics[width=2.0in]{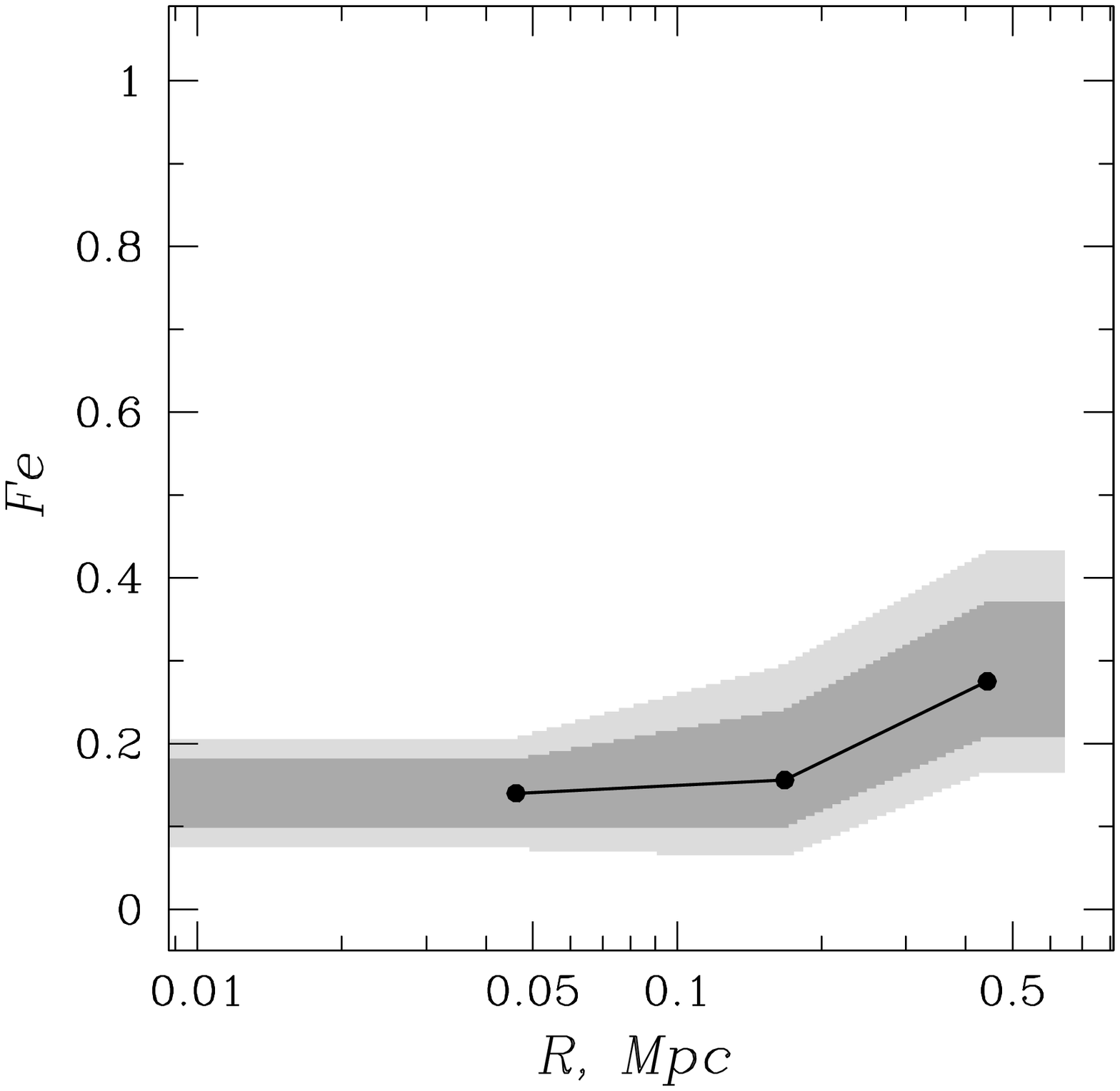} \hfill 
 \includegraphics[width=2.0in]{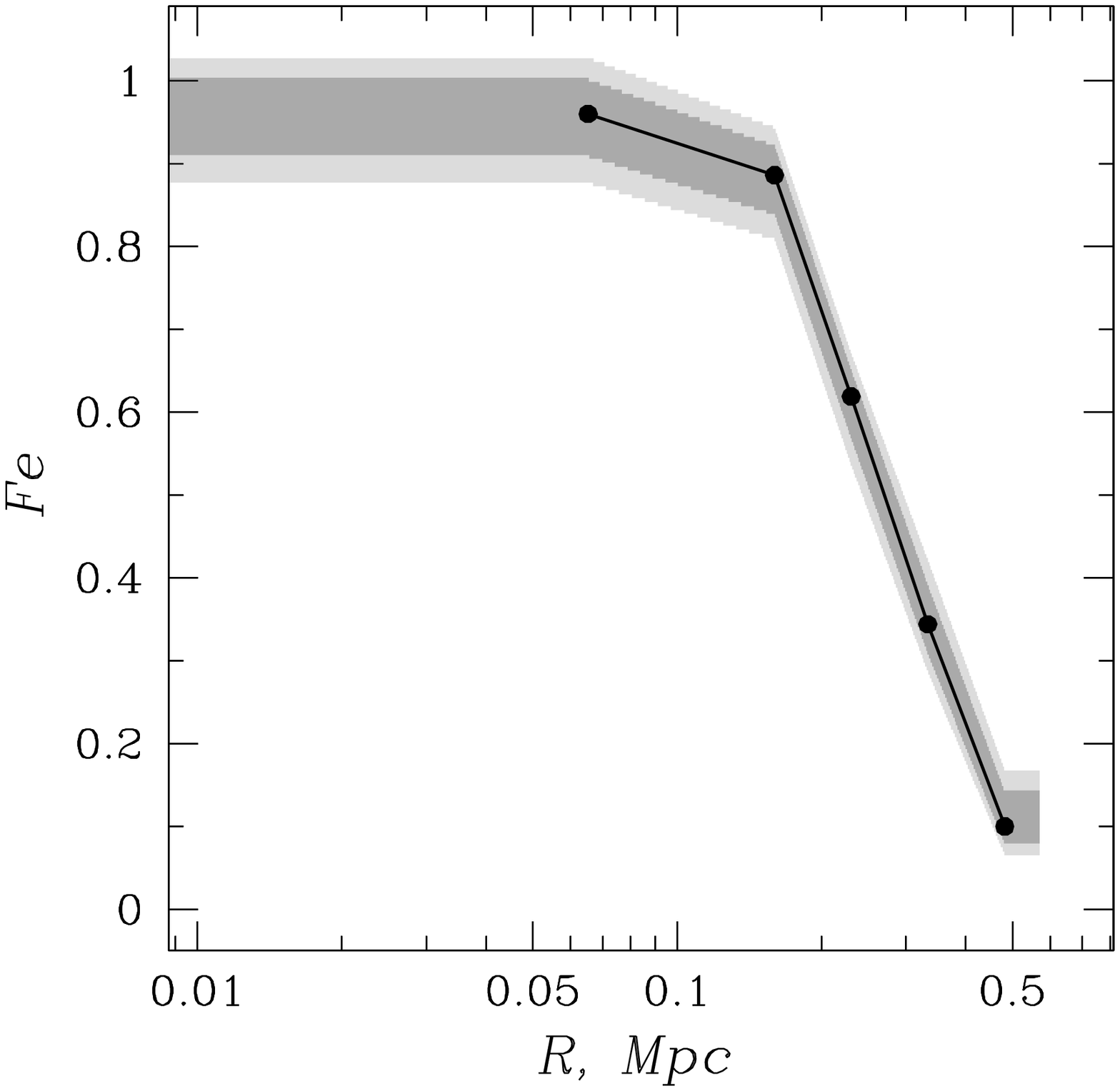}

\figcaption{Derived iron abundances (in units of 4.68 $10^{-5}$ for iron
number abundances relative to H). The solid lines correspond to the best-fit
Fe abundances derived from the ASCA data. The filled circles indicate the
spatial binning used in the analysis. Dark and light shaded zones around the
best fit curves denote the 68 and 90 per cent confidence areas. 
\label{fe-fig}}
\vspace*{-15.cm}

{\it \hspace*{3.cm} NGC5129 \hspace*{5.2cm} IC4296 \hspace*{5.6cm} NGC4325}

\vspace*{4.5cm}

{\it \hspace*{3.cm} NGC7619 \hspace*{5.2cm} NGC2300 \hspace*{5.2cm} NGC507 }

\vspace*{4.5cm}

{\it \hspace*{3.cm} NGC2563 \hspace*{5.2cm}  NGC6329 \hspace*{3.2cm}  NGC3268}

\vspace*{5.cm}

\end{figure*}

\begin{figure*}

   \includegraphics[width=2.0in]{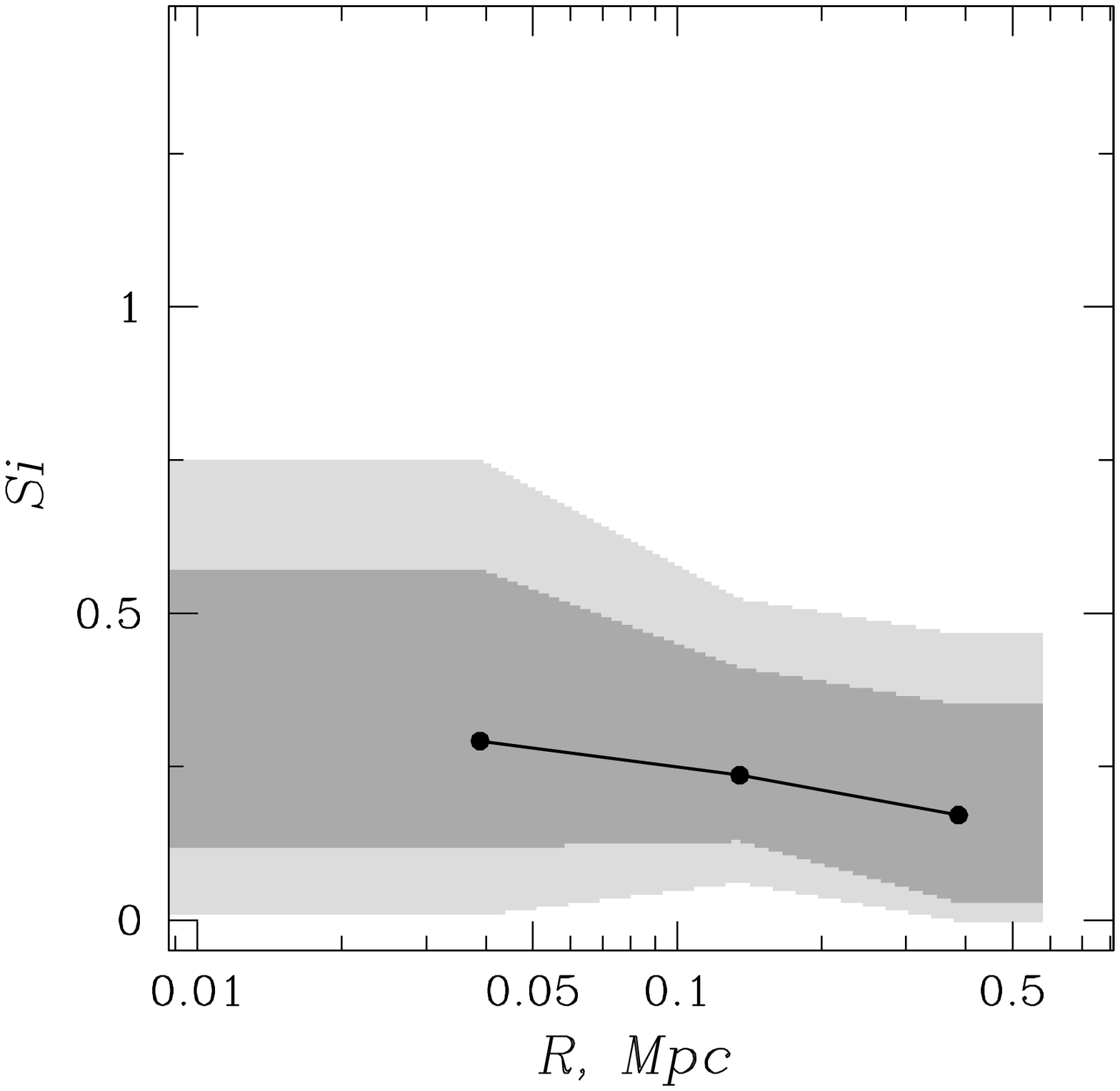} \hfill
   \includegraphics[width=2.0in]{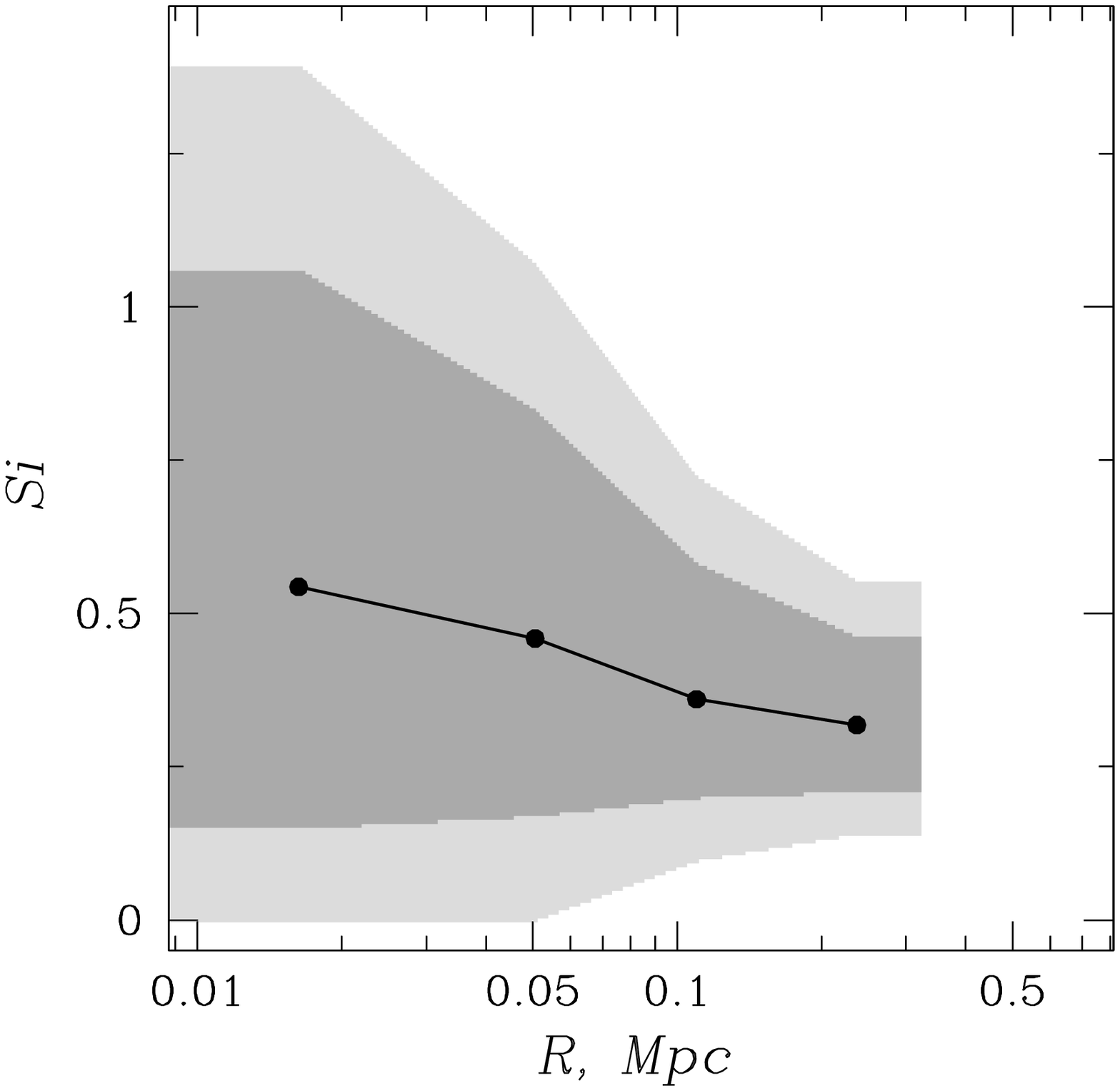} \hfill 
  \includegraphics[width=2.0in]{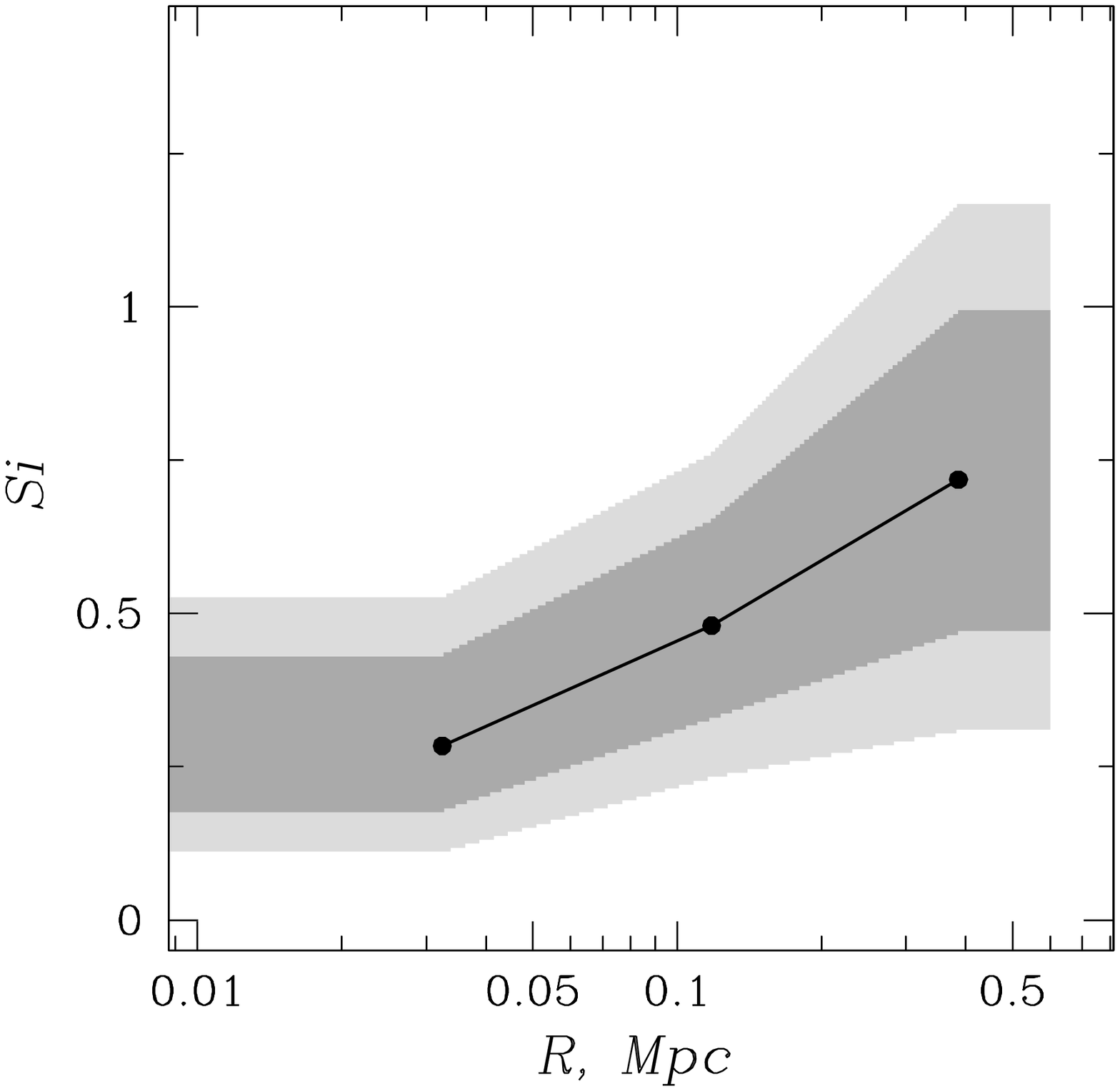} 
 
   \includegraphics[width=2.0in]{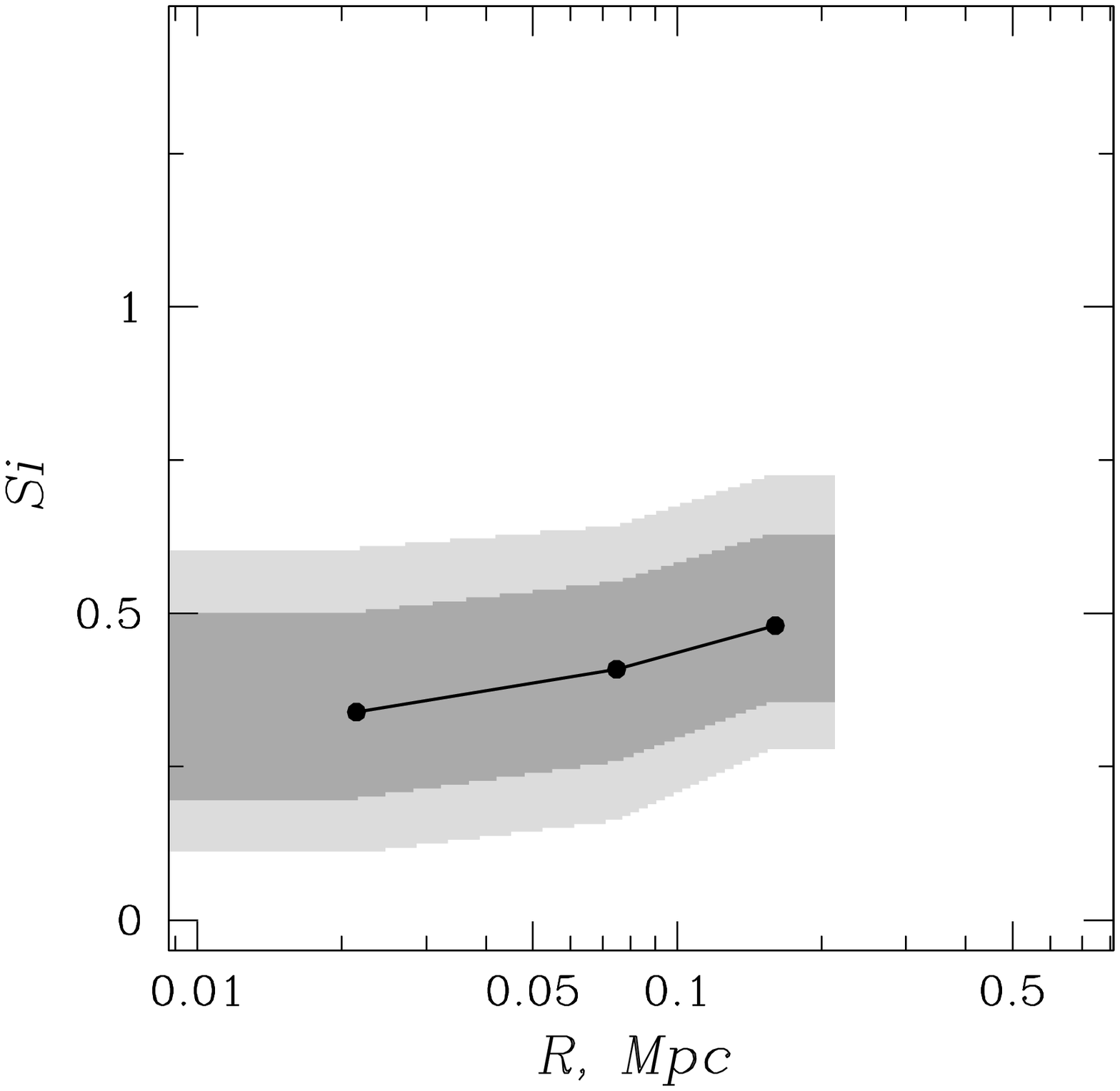} \hfill 
  \includegraphics[width=2.0in]{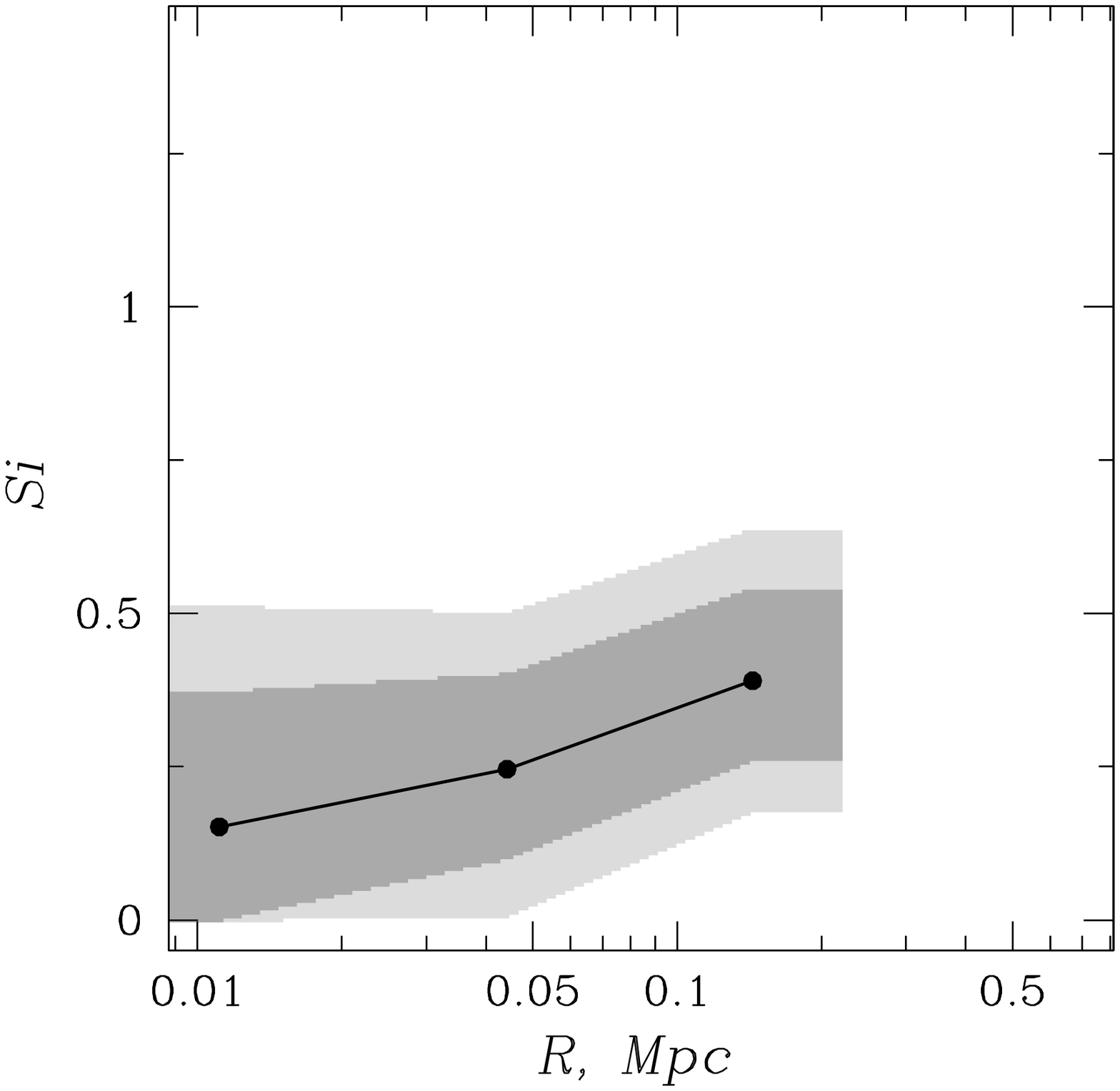} \hfill 
  \includegraphics[width=2.0in]{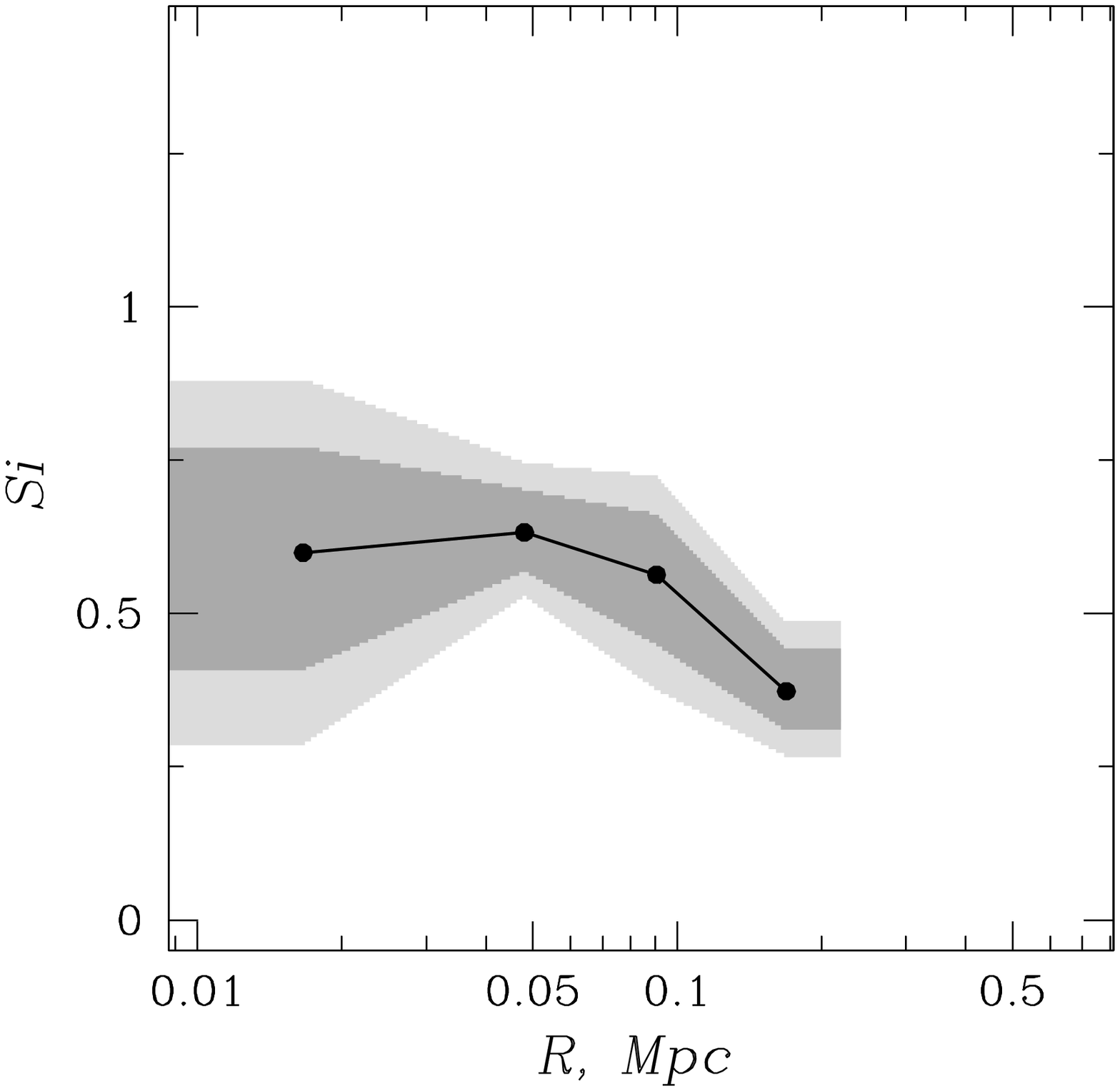} \hfill 

   \includegraphics[width=2.0in]{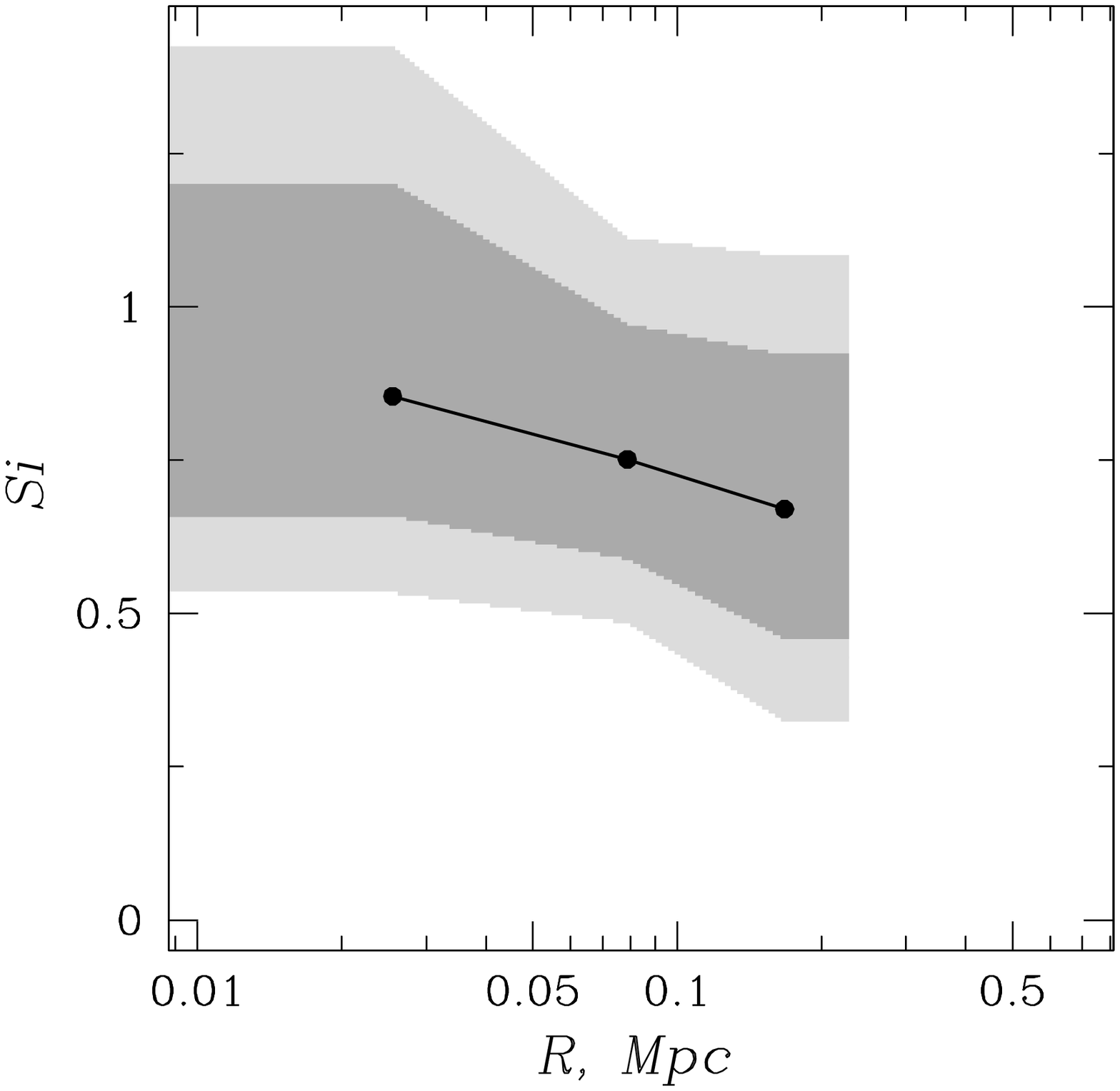}   \hfill 
  \includegraphics[width=2.0in]{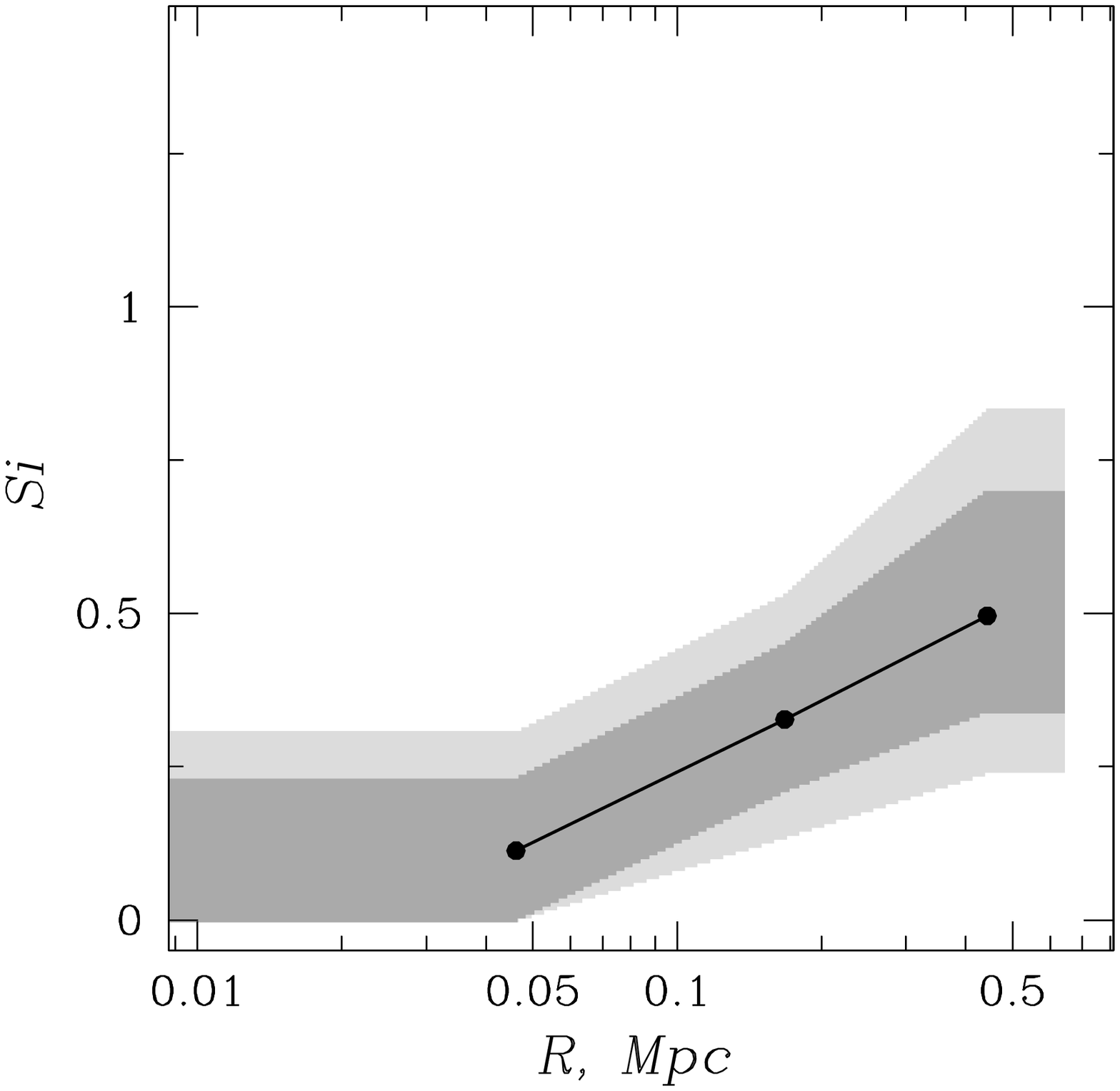} \hfill 
 \includegraphics[width=2.0in]{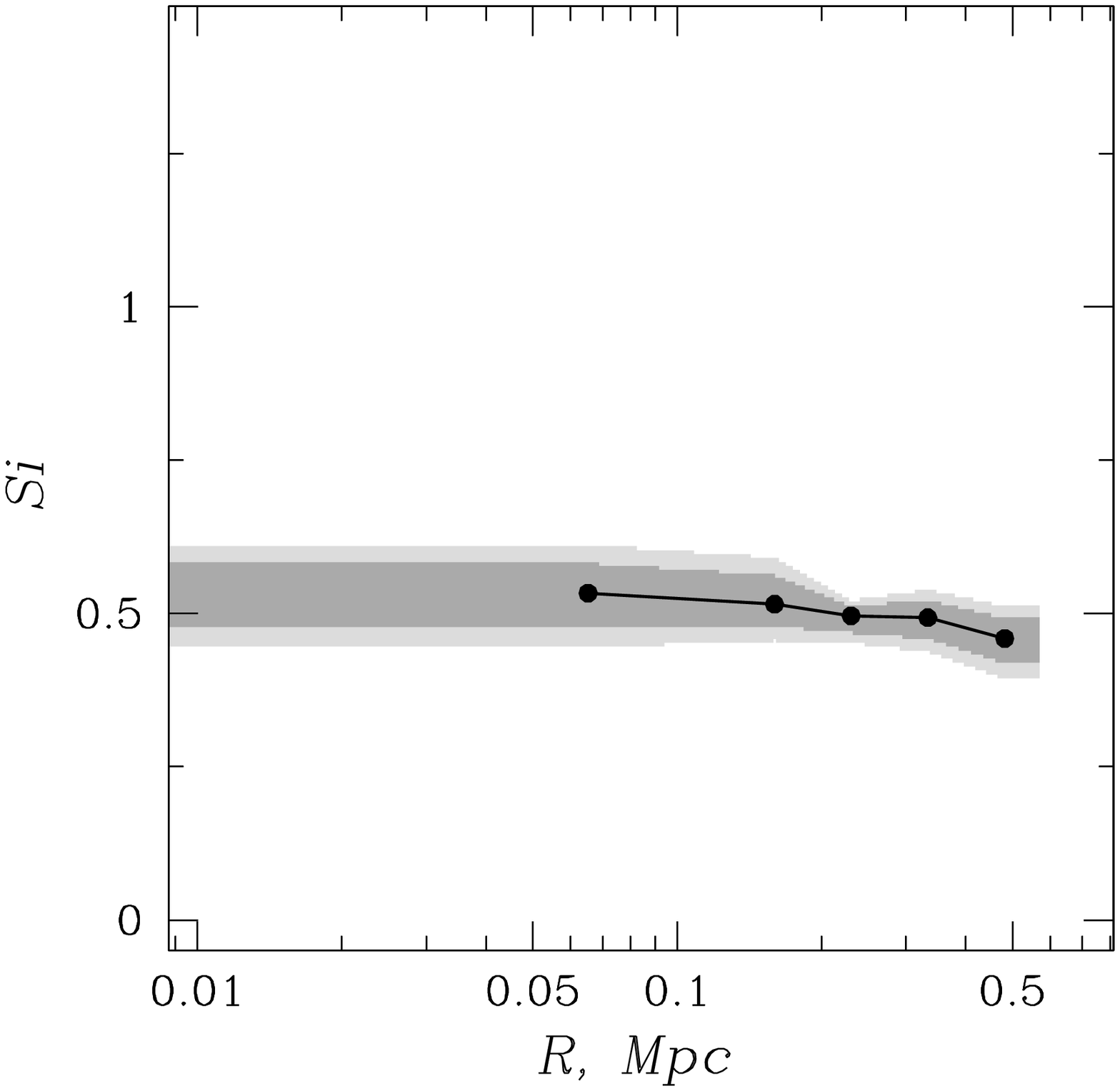}

\figcaption{Derived silicon abundances. Data representation is similar to
Fig.\ref{fe-fig}.
\label{si-fig}}
\vspace*{-15.cm}

{\it \hspace*{3.cm} NGC5129 \hspace*{5.2cm} IC4296 \hspace*{5.6cm} NGC4325}

\vspace*{4.5cm}

{\it \hspace*{3.cm} NGC7619 \hspace*{5.2cm} NGC2300 \hspace*{5.2cm} NGC507 }

\vspace*{4.5cm}

{\it \hspace*{3.cm} NGC2563 \hspace*{5.2cm}  NGC6329 \hspace*{5.2cm}  NGC3268}

\vspace*{5.cm}

\end{figure*}

\begin{figure*}

  \includegraphics[width=2.0in]{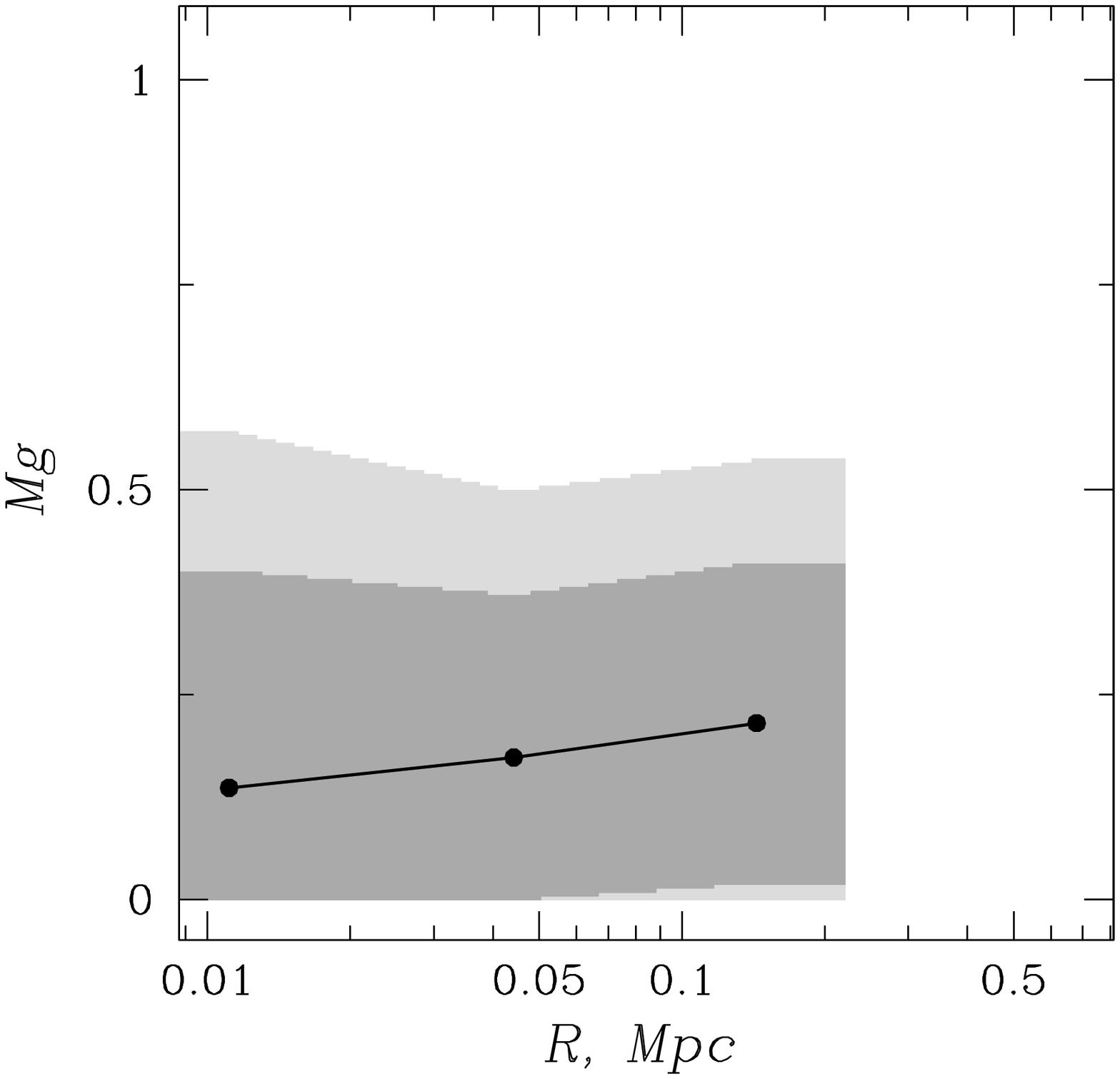} \hfill 
  \includegraphics[width=2.0in]{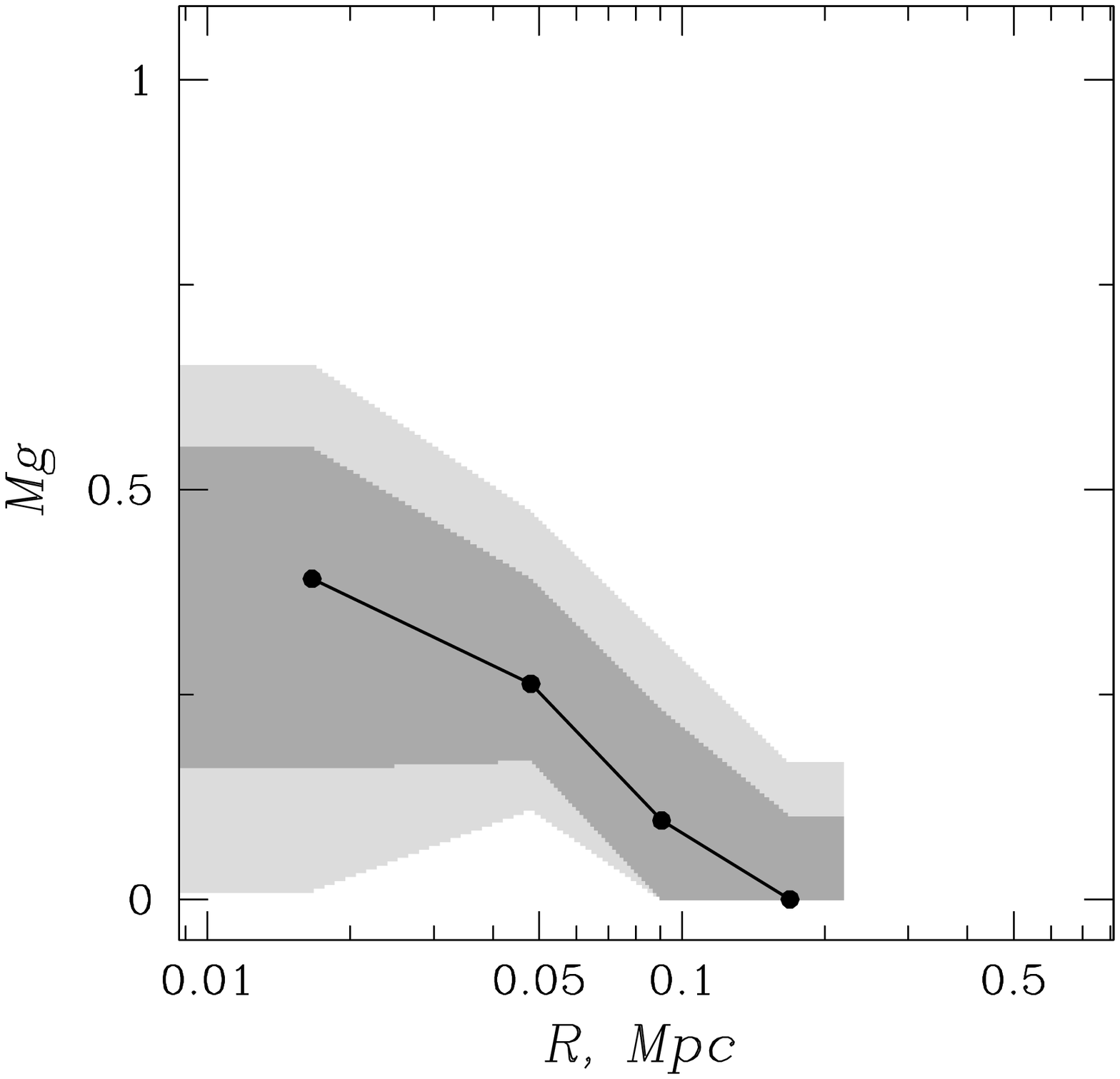} \hfill 
  \includegraphics[width=2.0in]{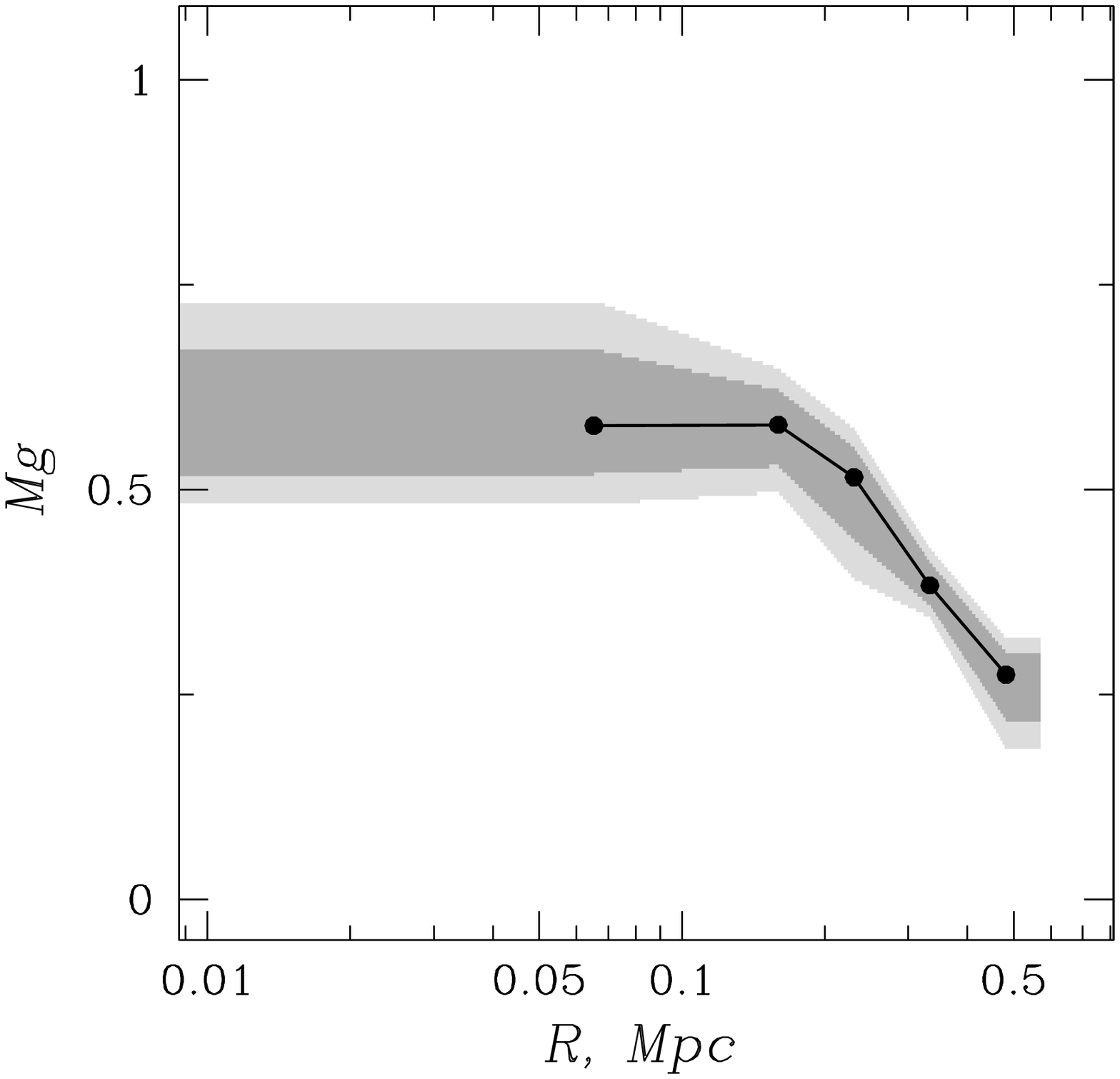}

\figcaption{Derived magnisium abundances.  Data representation is similar to
Fig.\ref{fe-fig}.
\label{mg-fig}}
\vspace*{-5.cm}

{\it \hspace*{3.cm}  NGC2300 \hspace*{5.2cm} NGC507 \hspace*{5.3cm}  NGC3268}

\vspace*{4.90cm}

  \includegraphics[width=2.0in]{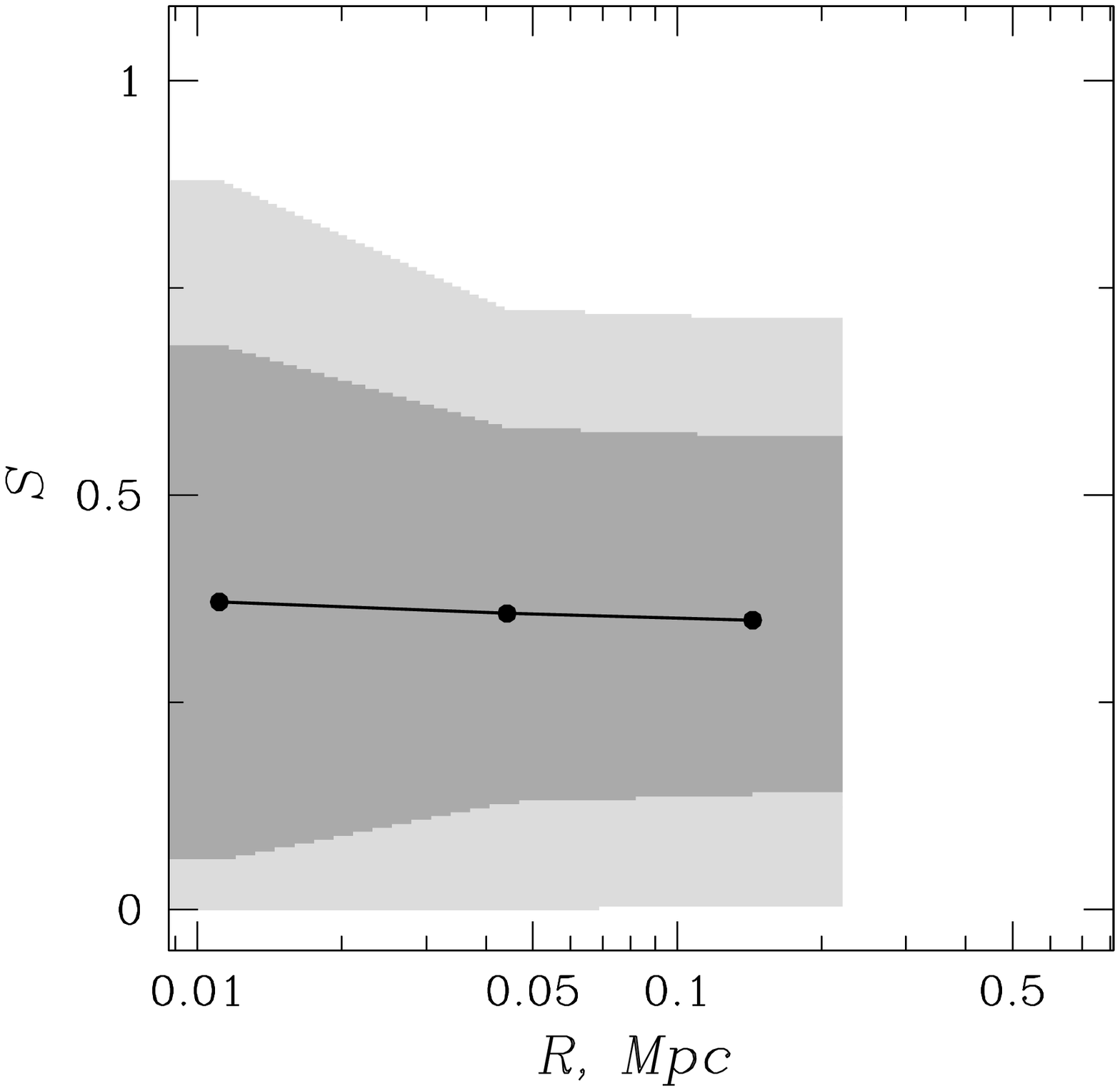} \hfill 
  \includegraphics[width=2.0in]{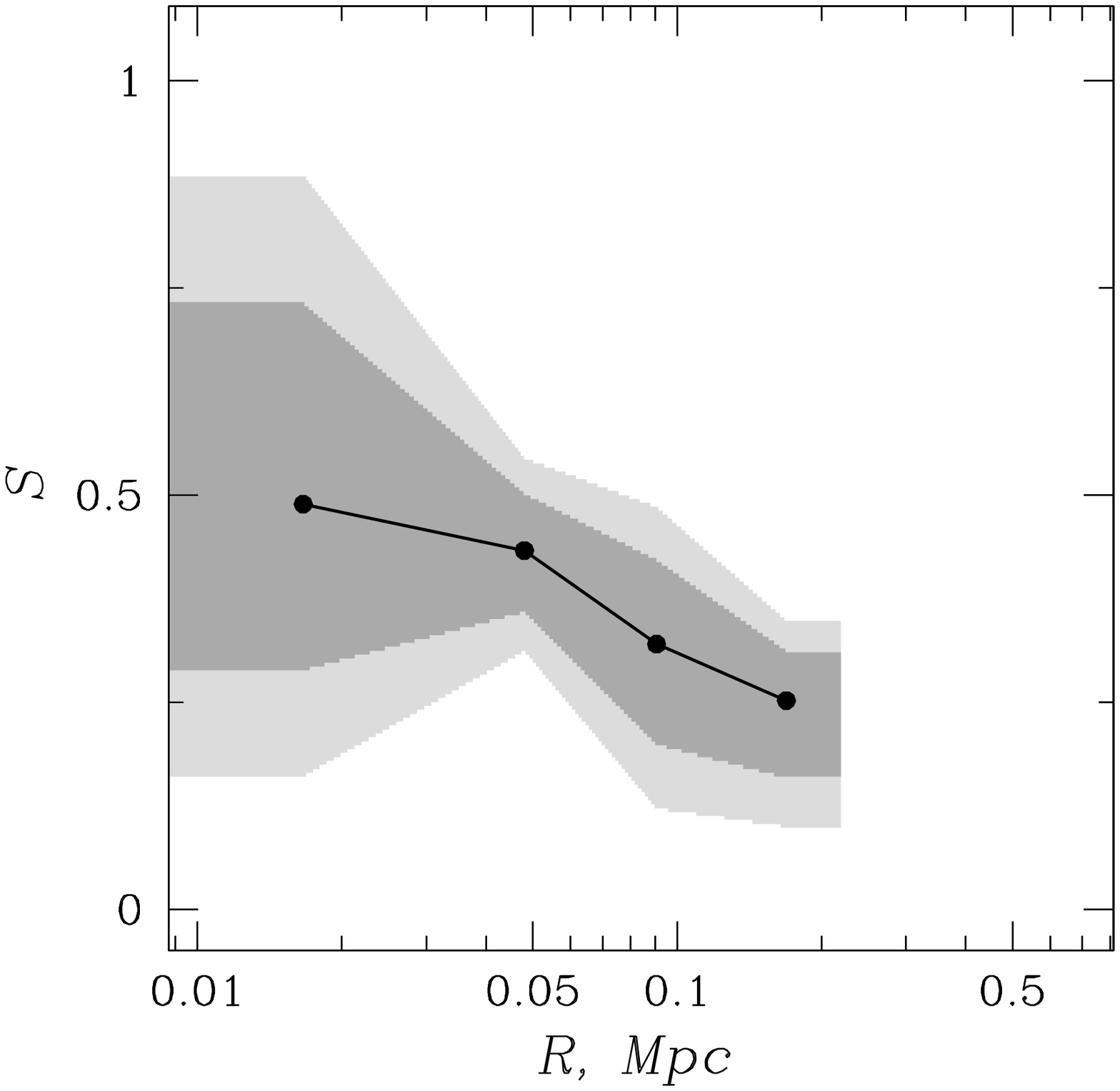} \hfill 
  \includegraphics[width=2.0in]{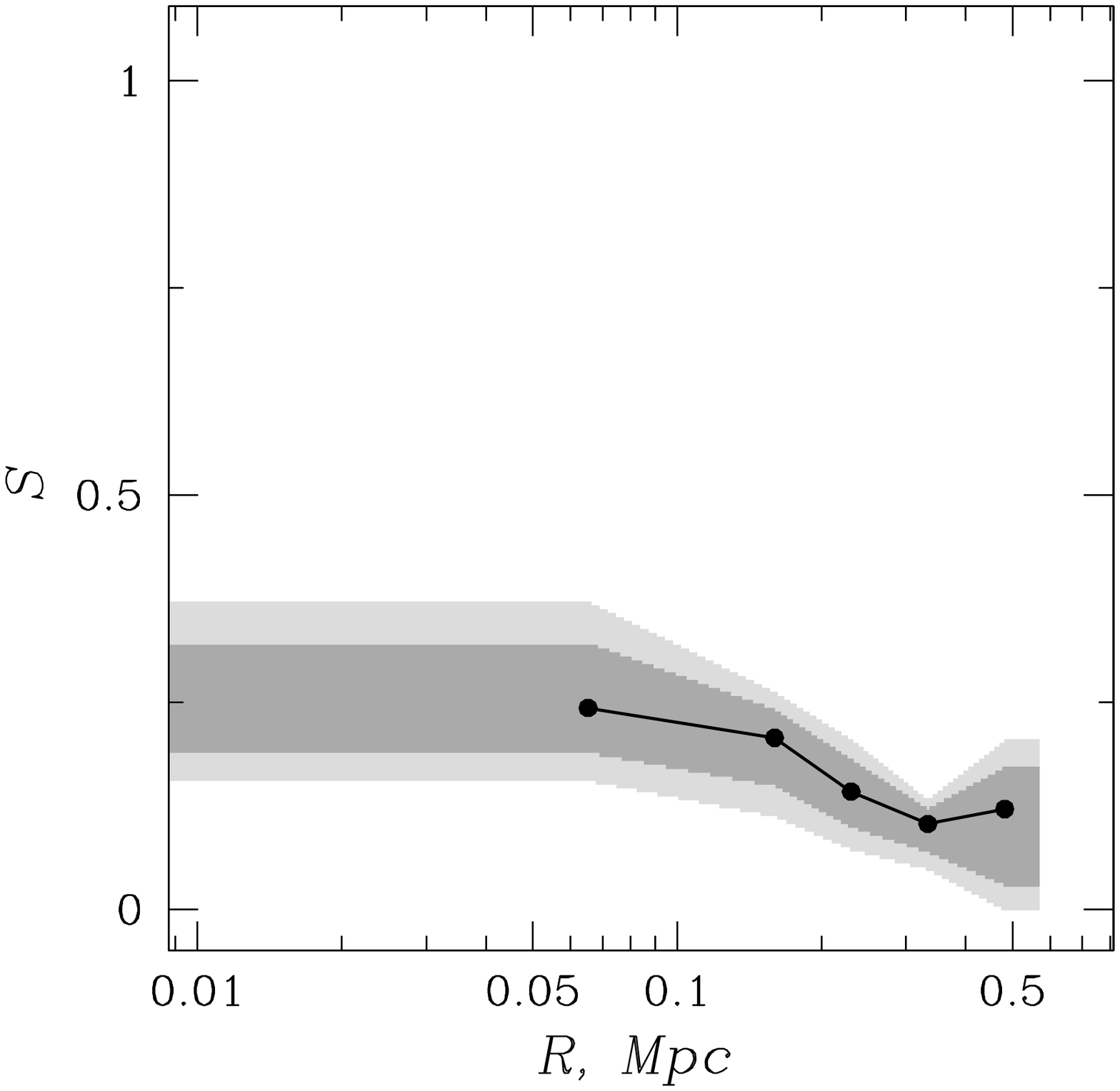}

\figcaption{Derived sulfur abundances.  Data representation is similar to
Fig.\ref{fe-fig}.
\label{s-fig}}
\vspace*{-5.cm}

{\it \hspace*{3.cm}  NGC2300 \hspace*{5.2cm} NGC507 \hspace*{5.2cm}  NGC3268}

\vspace*{4.90cm}

\end{figure*}

Heavy element abundance profiles are shown in Figs.\ref{fe-fig}--\ref{s-fig}. 
Like the temperature, elemental abundance profiles evolve with radius from
stellar mass loss values to intragroup medium values, as indicated by an
increase with radius in the gas-to-light ratio. Later types of the chemical
enrichment are given preference in the (central) zones with low
gas-to-light ratios, characterized by the dominance of SNe Ia in the Fe
production.


With the possible exception of the Mg abundance in NGC3268, the observed
heavy element abundance profiles can be understood within the framework
developed in FDP, FAD and Finoguenov et al. (2002). Groups are expected to
have a lower abundance of SN II products as well as prevalence of SN Ia in
the Fe production, related to low gas fraction in groups. The high gas
temperature in NGC3268 causes the Mg measurements to be compromised
(Mushotzky et al. 1996). This is further complicated by the strong
temperature gradient in NGC3268, which, for the abundance determinations of
light elements could not be sufficiently resolved in our data. Thus a
confirmation of the Mg abundance profile requires observations with higher
spatial resolution, such as XMM-Newton. Given the low X-ray luminosities of
the systems in this study, determination of the central elemental abundance
needs much higher spatial resolution in order to illuminate contributions
from galactic point sources, a central AGN and a possible diffuse hard
source (Finoguenov, Jones 2001). We therefore caution the reader about the
limitations of our abundance results in the central spatial bins.

Determination of the element abundance, characteristic of the
elliptical galaxies has raised a problem when compared to optical data
and raised controversy among the X-ray studies (Buote 2000, FDP).
Element abundance gradients, reported here, are at least partly due to
transition between the galaxy ISM and intragroup gas. While a number
of pitfalls exist in determining of the element abundance for
ellipticals from X-rays (Finoguenov et al. 2002), abundance
determination, corresponding to the intragroup gas is robustly
determined with ASCA data.

\clearpage
\begin{table*}
{
\footnotesize
\centering
\tabcaption{\footnotesize
\centerline{ASCA SIS measurements of temperature, entropy and element abundance$^{\dag}$}
\label{t:data}}

\begin{tabular}{clcll}
\hline
\hline
Annulus (\amin) & kT, keV & S, keV cm$^2$  & $Si/Si_{\odot}$& $Fe/Fe_{\odot}$ \\
\hline
 & & & & \\
\multicolumn{2}{c}{NGC5129} & & & \\
7---15    &0.620  (0.58--0.65)&$280\pm19$&0.507 (0.31--0.69)& 0.193 (0.17--0.21)\\
3.2---7   &1.120  (1.00--1.23)&$423\pm69$&0.465 (0.37--0.54)& 0.240 (0.15--0.33)\\
1.5---3.2 &0.993  (0.94--1.03)&$162\pm17$&0.436 (0.32--0.54)& 0.299 (0.26--0.34)\\
0---1.5   &0.914  (0.87--0.94)&$ 77\pm 5$&0.443 (0.25--0.63)& 0.423 (0.37--0.45)\\
\multicolumn{2}{c}{IC4296} & & & \\
7---15   &0.931  (0.83--1.01)&$241\pm33$&0.318 (0.21--0.46)& 0.056 (0.03--0.09)\\
3.2---7  &1.151  (1.03--1.34)&$308\pm83$&0.360 (0.20--0.58)& 0.323 (0.25--0.44)\\
1.5---3.2&1.110  (1.02--1.32)&$177\pm5 $&0.459 (0.17--0.83)& 0.556 (0.42--0.74)\\
0---1.5  &0.669  (0.61--0.72)&$ 56\pm14$&0.543 (0.15--1.06)& 0.687 (0.46--1.03)\\
\multicolumn{2}{c}{NGC4325} & & & \\
4---14 & 0.960  (0.83--1.16)& $435\pm 158$& 0.718 (0.47--0.99)& 0.112 (0.04--0.18)\\
1.5---4& 1.244  (1.12--1.40)& $166\pm  29$& 0.480 (0.33--0.65)& 0.244 (0.16--0.37)\\
0---1.5& 0.901  (0.88--0.93)& $ 39\pm   3$& 0.284 (0.18--0.43)& 0.296 (0.25--0.38)\\
\multicolumn{2}{c}{NGC7619} & & & \\
5---10& 1.023  (0.98--1.07)& $200\pm  16$& 0.480 (0.36--0.63)& 0.209 (0.18--0.25)\\
2---5 & 1.007  (0.71--1.28)& $336\pm 327$& 0.409 (0.26--0.55)& 0.212 (0.08--0.34)\\
0---2 & 0.806  (0.77--0.84)& $ 50\pm   6$& 0.339 (0.20--0.50)& 0.241 (0.20--0.28)\\
\multicolumn{2}{c}{NGC2300} & & & \\
6---20& 1.215  (1.11--1.35)& $364\pm 47$& 0.390 (0.26--0.54)& 0.061 (0.03--0.10)\\
2---6 & 1.143  (0.92--1.38)& $215\pm 73$& 0.246 (0.10--0.40)& 0.101 (0.02--0.20)\\
0---2 & 0.819  (0.75--0.88)& $ 45\pm 10$& 0.152 (0.00--0.37)& 0.139 (0.09--0.21)\\
\multicolumn{2}{c}{NGC507} & & & \\
4.3---8  &1.480 (1.36--1.63) & $194\pm18$&0.373 (0.31--0.44)& 0.271 (0.24--0.30)\\
2.3---4.3&1.293 (1.11--1.55) & $130\pm23$&0.563 (0.45--0.66)& 0.251 (0.21--0.28)\\
1.2---2.3&1.173 (1.07--1.33) & $ 62\pm7$ &0.632 (0.57--0.70)& 0.258 (0.23--0.28)\\
0---1.2  &0.772 (0.68--0.85) & $ 34\pm5$ &0.599 (0.41--0.77)& 0.322 (0.28--0.36)\\
\multicolumn{2}{c}{NGC2563} & & & \\
4.2---9& 1.441  (1.34--1.60)& $350\pm 48$& 0.670 (0.46--0.92)& 0.285 (0.20--0.39)\\
2---4.2& 1.367  (1.22--1.58)& $231\pm 53$& 0.751 (0.59--0.97)& 0.291 (0.19--0.39)\\
0---2  & 0.983  (0.88--1.06)& $100\pm 27$& 0.854 (0.66--1.20)& 0.265 (0.20--0.44)\\
\multicolumn{2}{c}{NGC6329} & & & \\
5.3---14&  1.485  (1.30--1.72)& $638\pm 128$& 0.496 (0.34--0.70)& 0.275 (0.21--0.37)\\
2---5.3 &  1.600  (1.36--1.86)& $ 35\pm  66$& 0.327 (0.21--0.45)& 0.156 (0.10--0.24)\\
0---2   &  1.041  (0.98--1.09)& $ 82\pm   9$& 0.113 (0.00--0.23)& 0.140 (0.10--0.18)\\
\multicolumn{2}{c}{NGC3268} & & & \\
24.2---35  & 1.721  (1.60--1.77)& $445\pm 22$& 0.459 (0.42--0.49)& 0.100 (0.08--0.14)\\
16.7---24.2& 2.195  (2.09--2.28)& $486\pm 28$& 0.493 (0.46--0.52)& 0.344 (0.31--0.39)\\
11.6---16.7& 2.588  (2.53--2.70)& $465\pm 21$& 0.496 (0.47--0.51)& 0.619 (0.57--0.65)\\
8---11.6   & 2.700  (2.60--2.79)& $493\pm 39$& 0.515 (0.48--0.56)& 0.886 (0.84--0.92)\\
0---8      & 2.625  (2.53--2.71)& $351\pm 14$& 0.533 (0.48--0.58)& 0.960 (0.91--1.00)\\
\hline                                       
\end{tabular}

\begin{enumerate}
\item[{$^{\dag}$}]{\footnotesize ~ Definition of the solar abundance units
is 3.55 and 4.68 $\times10^{-5}$ for the number abundance of Si and Fe
relative to H, respectively. Errors are given at 68\% confidence level (see
text for further details). MEKAL plasma code is used for spectral fitting.}
\end{enumerate}
}
\end{table*}

\end{document}